\documentclass[twocolumn]{aastex631}

\usepackage{amsmath}
\usepackage{natbib}

\newcommand\brantspaper{R26}
\newcommand{\pipe}{\texttt{jwst}}
\newcommand{\pmap}{\texttt{jwst\_1228.pmap}}
\newcommand{\webbpsf}{\texttt{STPSF}~}

\begin{document}

\title{JWST Advanced Deep Extragalactic Survey (JADES) Data Release 5: NIRCam Imaging in GOODS-S and GOODS-N}

\author[0000-0002-9280-7594]{Benjamin D.\ Johnson}
\affiliation{Center for Astrophysics $|$ Harvard \& Smithsonian, 60 Garden St., Cambridge MA 02138 USA}
\email{benjamin.johnson@cfa.harvard.edu}  

\author[0000-0002-4271-0364]{Brant E. Robertson}
\affiliation{Department of Astronomy and Astrophysics, University of California, Santa Cruz, 1156 High Street, Santa Cruz, CA 95064, USA}

\author[0000-0002-2929-3121]{Daniel J.\ Eisenstein}
\affiliation{Center for Astrophysics $|$ Harvard \& Smithsonian, 60 Garden St., Cambridge MA 02138 USA}

\author[0000-0002-8224-4505]{Sandro Tacchella}
\affiliation{Kavli Institute for Cosmology, University of Cambridge, Madingley Road, Cambridge, CB3 0HA, UK}
\affiliation{Cavendish Laboratory, University of Cambridge, 19 JJ Thomson Avenue, Cambridge, CB3 0HE, UK}

\author[0000-0001-8630-2031]{D\'{a}vid Pusk\'{a}s}
\affiliation{Kavli Institute for Cosmology, University of Cambridge, Madingley Road, Cambridge, CB3 0HA, UK}
\affiliation{Cavendish Laboratory, University of Cambridge, 19 JJ Thomson Avenue, Cambridge, CB3 0HE, UK}

\author[0009-0009-8105-4564]{Qiao Duan}
\affiliation{Kavli Institute for Cosmology, University of Cambridge, Madingley Road, Cambridge, CB3 0HA, UK}
\affiliation{Cavendish Laboratory, University of Cambridge, 19 JJ Thomson Avenue, Cambridge, CB3 0HE, UK}

\author[0000-0002-8876-5248]{Zihao Wu}
\affiliation{Center for Astrophysics $|$ Harvard \& Smithsonian, 60 Garden St., Cambridge MA 02138 USA}

\author[0000-0003-4565-8239]{Kevin Hainline}
\affiliation{Steward Observatory, University of Arizona, 933 N. Cherry Avenue, Tucson, AZ 85721, USA}

\author[0000-0002-7893-6170]{Marcia Rieke}
\affiliation{Steward Observatory, University of Arizona, 933 N. Cherry Avenue, Tucson, AZ 85721, USA}

\author[0000-0002-4201-7367]{Chris Willott}
\affiliation{NRC Herzberg, 5071 West Saanich Rd, Victoria, BC V9E 2E7, Canada}

\author[0000-0001-9262-9997]{Christopher N. A. Willmer}
\affiliation{Steward Observatory, University of Arizona, 933 N. Cherry Avenue, Tucson, AZ 85721, USA}

\author[0000-0002-9081-2111]{James A.\ A.\ Trussler}
\affiliation{Center for Astrophysics $|$ Harvard \& Smithsonian, 60 Garden St., Cambridge MA 02138 USA}


\author[0000-0002-8909-8782]{Stacey Alberts}
\affiliation{Steward Observatory, University of Arizona, 933 N. Cherry Avenue, Tucson, AZ 85721, USA}
\affiliation{AURA for the European Space Agency (ESA), Space Telescope Science Institute, 3700 San Martin Dr., Baltimore, MD 21218, USA}

\author[0000-0001-7997-1640]{Santiago Arribas}
\affiliation{Centro de Astrobiolog\'ia (CAB), CSIC–INTA, Cra. de Ajalvir Km.~4, 28850- Torrej\'on de Ardoz, Madrid, Spain}

\author[0000-0003-0215-1104]{William M.\ Baker}
\affiliation{DARK, Niels Bohr Institute, University of Copenhagen, Jagtvej 155A, DK-2200 Copenhagen, Denmark}

\author[0000-0002-8651-9879]{Andrew J.\ Bunker}
\affiliation{Department of Physics, University of Oxford, Denys Wilkinson Building, Keble Road, Oxford OX1 3RH, UK}

\author[0000-0002-0450-7306]{Alex J.\ Cameron}
\affiliation{Cosmic Dawn Center (DAWN), Copenhagen, Denmark}
\affiliation{Niels Bohr Institute, University of Copenhagen, Jagtvej 128, DK-2200, Copenhagen, Denmark}

\author[0000-0002-6719-380X]{Stefano Carniani}
\affiliation{Scuola Normale Superiore, Piazza dei Cavalieri 7, I-56126 Pisa, Italy}

\author[0000-0001-6301-3667]{Courtney Carreira}
\affiliation{Department of Astronomy and Astrophysics, University of California, Santa Cruz, 1156 High Street, Santa Cruz, CA 95064, USA}

\author[0000-0002-1617-8917]{Phillip A. Cargile}
\affiliation{Center for Astrophysics $|$ Harvard \& Smithsonian, 60 Garden St., Cambridge MA 02138 USA}

\author[0000-0002-9551-0534]{Emma Curtis-Lake}
\affiliation{Centre for Astrophysics Research, Department of Physics, Astronomy and Mathematics, University of Hertfordshire, Hatfield AL10 9AB, UK}

\author[0000-0003-1344-9475]{Eiichi Egami}
\affiliation{Steward Observatory, University of Arizona, 933 N. Cherry Avenue, Tucson, AZ 85721, USA}

\author[0000-0002-8543-761X]{Ryan Hausen}
\affiliation{Department of Physics and Astronomy, The Johns Hopkins University, 3400 N. Charles St., Baltimore, MD 21218}

\author[0000-0003-4337-6211]{Jakob M.\ Helton}
\affiliation{The Department of Astronomy \& Astrophysics, The Pennsylvania State University, 525 Davey Lab, University Park, PA 16802}

\author[0000-0001-7673-2257]{Zhiyuan Ji}
\affiliation{Steward Observatory, University of Arizona, 933 N. Cherry Avenue, Tucson, AZ 85721, USA}

\author[0000-0002-4985-3819]{Roberto Maiolino}
\affiliation{Kavli Institute for Cosmology, University of Cambridge, Madingley Road, Cambridge, CB3 0HA, UK.}
\affiliation{Cavendish Laboratory - Astrophysics Group, University of Cambridge, 19 JJ Thomson Avenue, Cambridge, CB3 0HE, UK.}
\affiliation{Department of Physics and Astronomy, University College London, Gower Street, London WC1E 6BT, UK}

\author[0000-0003-4528-5639]{Pablo G. P\'erez-Gonz\'alez}
\affiliation{Centro de Astrobiolog\'ia (CAB), CSIC–INTA, Cra. de Ajalvir Km.~4, 28850- Torrej\'on de Ardoz, Madrid, Spain}

\author[0000-0002-5104-8245]{Pierluigi Rinaldi}
\affiliation{Space Telescope Science Institute, 3700 San Martin Drive, Baltimore, Maryland 21218, USA}

\author[0000-0002-4622-6617]{Fengwu Sun}
\affiliation{Center for Astrophysics $|$ Harvard \& Smithsonian, 60 Garden St., Cambridge MA 02138 USA}

\author[0000-0001-6561-9443]{Yang Sun}
\affiliation{Steward Observatory, University of Arizona, 933 N. Cherry Avenue, Tucson, AZ 85721, USA}

\author[0000-0001-6917-4656]{Natalia C. Villanueva}
\affiliation{Department of Astronomy, The University of Texas at Austin, Austin, TX, USA}

\author[0000-0003-2919-7495]{Christina C. Williams}
\affiliation{NSF National Optical-Infrared Astronomy Research Laboratory, 950 North Cherry Avenue, Tucson, AZ 85719, USA}

\author[0000-0003-3307-7525]{Yongda Zhu}
\affiliation{Steward Observatory, University of Arizona, 933 N. Cherry Avenue, Tucson, AZ 85721, USA}

\begin{abstract}
We present the Near Infrared Camera (NIRCam) imaging products of the fifth data release (DR5) of the
James Webb Space Telescope (JWST) Advanced Deep Extragalactic Survey (JADES).  The JADES survey is one
of the most ambitious programs yet conducted on JWST, producing deep infrared imaging and multiobject
spectroscopy on the GOODS-S and GOODS-N extragalactic deep fields in order to explore galaxies to the earliest epoch.
Here we describe the NIRCam data reduction procedures that result in deep and well-characterized mosaics in up to
18 filters covering 469 arcmin$^2$, with 250 arcmin$^2$ having at least 8 filters of coverage.
This release contains the full NIRCam imaging of JADES, over 800 JWST mission hours, as well as
co-reductions of 19 other programs in these two premier deep fields.
We perform detailed tests on the final data products, thereby characterizing the
photometric properties,
point-spread function,
and astrometric alignment.
We release mosaics for individual programs (or epochs, depending on scheduling) and the mosaics combining data from all programs in order to facilitate photometric variability studies and the deepest possible photometry.
\end{abstract}



\section{Introduction}
\label{sec:intro}

The James Webb Space Telescope (JWST) has opened a fantastic view of galaxies across cosmic time.  Deep and sharp infrared many-band imaging and spectroscopy \citep{gardner23a,rigby23a} reveals exquisite detail in populations heretofore too faint and remote to be explored.  The redshift frontier has been pushed into the first 300 million years after the Big Bang \citep[][among many others]{arrabalharo2023,castellano2022,finkelstein2022,finkelstein2023,adams2023,bunker2023a,curtislake2023,2023ApJ...955L..24G,harikane2023,robertson2023,perezgonzalez2023,atek2023,wang2023,whitler2023,yan2023,hainline2024a,deugenio2024a}, and we see unexpectedly mature galaxies across the first billion years \citep{carniani2024,helton2025,schouws2025,naidu2025}.  We see surprises such as a new manifestation of supermassive black holes \citep[][and many others]{matthee2024}, unusual chemical abundance patterns \citep[e.g.][]{bunker2023a,cameron2023,jones2023,deugenio2024a}, and an overabundance of massive quiescent galaxies at high redshift \citep[e.g.,][]{alberts2024massive,baker2025a,baker2025b,stevenson2026}.
The torrent of results from the last three years will surely continue as the community deepens its study of these fascinating directions.

Multi-observatory deep field programs have been a linchpin of galaxy evolution for the past 30 years, and JWST is rapidly adding to this legacy.  The two premier fields on the sky, GOODS-S and GOODS-N, have been the subject of thousands of nights and hours of time with nearly every narrow-field telescope.  These fields began with the Hubble Deep Field (HDF) in the north \citep{williams96,ferguson2000} and the Chandra Deep Field South \citep{giacconi2002} and were then expanded as the Great Observatory Origins Deep Survey \citep[GOODS;][]{giavalisco2004} to partner Hubble Space Telescope imaging with that from the Spitzer infrared telescope and Chandra X-ray imaging \citep{luo08}.  The Hubble Ultra Deep Field \citep[HUDF;][]{beckwith06} was sited in GOODS-S and has been a magnet for subsequent work.

With JWST, the instrument teams of the Near-Infrared Camera \citep[NIRCam;][]{rieke23a} and Near-Infrared Spectrograph \citep[NIRSpec;][]{jakobsen2022,ferruit22} partnered to form the JWST Advanced Deep Extragalactic Survey \citep[JADES;][]{bunker2020,rieke2020,eisenstein2023}, concentrating nearly 800 hours of guaranteed time on many-band imaging and multi-object spectroscopy in GOODS-N and GOODS-S.  This was complemented by other Cycle 1 programs: the MIDIS \citep{ostlin25,perezgonzalez2024b} and SMILES \citep{rieke2024,alberts2024,lyu2024} programs from the MIRI instrument team provided deep mid-infrared imaging, the NGDEEP program of deep NIRISS slitless spectroscopy and NIRCam parallel imaging \citep{bagley2024}, the FRESCO program of NIRCam wide-field slitless spectroscopy \citep{oesch2023}, and the JEMS program of NIRCam medium-band imaging \citep{williams2023}.  Cycle 2 brought more investment, notably an ultra-deep medium-band imaging program that created the JADES Origins Field \citep{eisenstein2025}.  Extensive targeted spectroscopic programs have been carried out, by JADES
and several integral-field programs including GA-NIFS \citep[e.g.,][]{ubler2023, perna2023}.  Pure parallel imaging and slitless spectroscopy has further expanded the imaging resource.  As we enter into the 4th cycle of JWST, a very rich array of data is now available in the two GOODS fields, totaling over 2000 hours of mission time.

In this paper, we present the NIRCam mosaics of the GOODS-S and GOODS-N fields as part of JADES Data Release 5, the fourth JADES release to include NIRCam data.
This builds on previous JADES data releases \citep{rieke23r,bunker2024,eisenstein2025,deugenio2025,curtislake2025,scholtz2025}.
We provide a detailed description of our processing steps with the JWST Calibration Pipeline and custom processing steps from JADES that improve on aspects such as imaging noise, detector features, alignment and global background modeling.  While optimized for JADES imaging data, we include here nearly all of the extant NIRCam imaging data in these fields, expanding the footprint and putting the whole field onto a common setting.
The resulting NIRCam images are available on MAST as High Level Science Products via \dataset[10.17909/5kka-ms10]{\doi{10.17909/ 8tdj-8n28}}.

This paper is one of four detailing DR5.  \citet{Wu26} presents methodology for template removal of wisp artifacts from the NIRCam images.
\citet{Alberts26} presents the reductions and mosaicking of JADES imaging from the Mid-Infrared Imager (MIRI).
Finally, \citet[][hereafter \brantspaper]{Robertson26} presents the object catalogs built from the NIRCam, MIRI, and archival Hubble Space Telescope images.

This paper is organized as follows.  \S\ref{sec:observations} summarizes the observations from JADES and other programs.
\S\ref{sec:pipeline} presents the JADES NIRCam pipeline, starting with ramp fitting, then describing the calibration of individual exposures, and finally mosaicking and point spread function (PSF) modeling.
Appendices \ref{sec:refcat} and \ref{sec:sky-flats} present supporting data products of an astrometric solution for the two fields and custom derived on-sky flat fields.
\S\ref{sec:image_quality} presents assessments and validations of the resulting mosaics, as well as remaining liens and caveats.
Appendix \ref{app:data_release} presents the file structure of the resulting mosaics.
We conclude in \S~\ref{sec:conclusions}.

\section{Observations}
\label{sec:observations}

\begin{figure*}[tb]
\includegraphics[width=\textwidth]{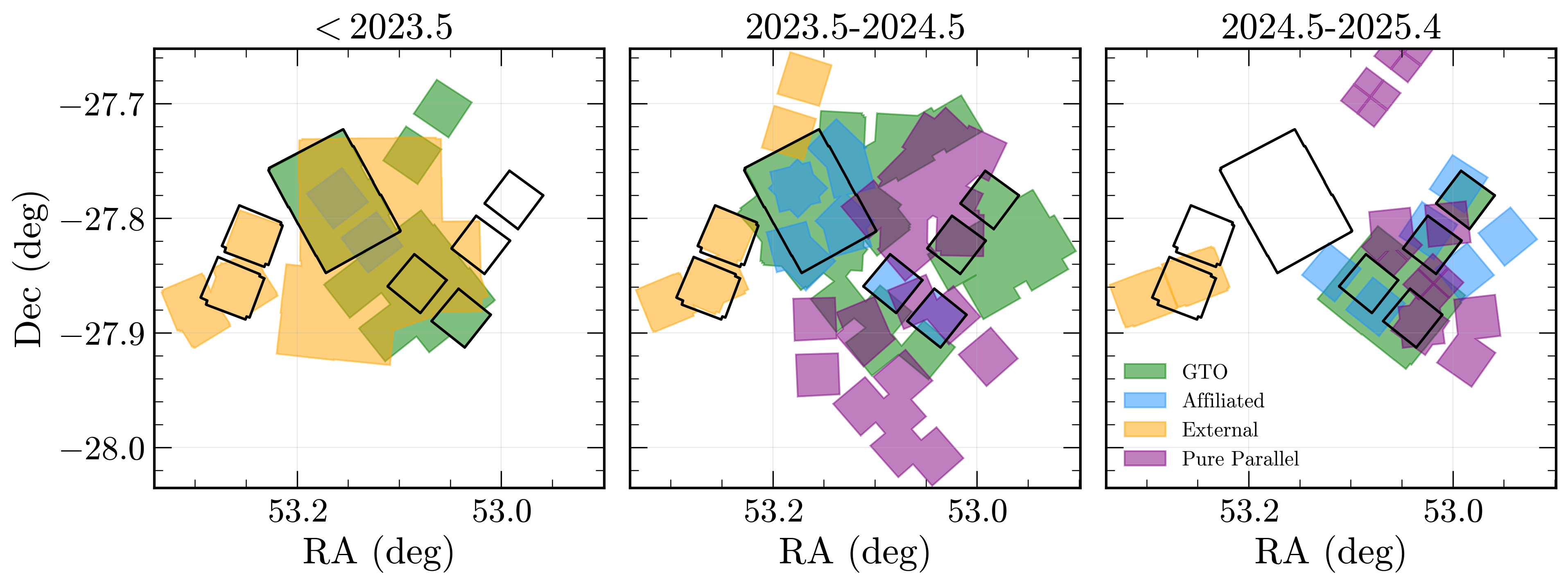}
\caption{Footprint of the GOODS-S NIRCam LW images in different years, colored by the program kind (\S\ref{sec:observations}).  We separate the pure parallel programs, all external to JADES, to highlight the important role these programs have played in expanding the coverage.  The heavy black outlines in each panel indicate regions of particularly deep imaging: from east to west these are NGDEEP NIRCam parallel imaging, the 1180/Deep area, the JADES Origins Field (JOF), and the 1287 Parallel area.  These footprints do not account for area lost to artifact masking. When LW imaging was not obtained (i.e. for some NIRCam/Grism observations without pre-imaging) we show the SW footprint instead. \label{fig:footprint_GS}}
\end{figure*}

\subsection{JADES Guaranteed Time Observations}

The core of this data realease is comprised of NIRCam observations from JADES programs 1180, 1181, 1210, 1286, 1287, and 4540, totalling about 800 hours of guaranteed time observations (GTO) from JWST instrument teams.  The planning of these observations is described in detail in \citet{eisenstein2023} and \citet{eisenstein2025}; we give a brief summary here.

In both GOODS fields, JADES NIRCam data are organized into a filled mosaic where NIRCam was effectively Prime and a set of Parallel fields where NIRSpec was pointing at the Prime field with only small dithers.  In practice, the parallel fields have some overlap, so some of the gaps between short-wave (SW) chips and those between the two NIRCam modules are filled.
For each of these two types, JADES distinguishes between Deep and Medium regions.  The Deep region of 42 arcmin$^2$ is entirely in GOODS-S and consists of 4 overlapping long pointings covering the HUDF (Deep Prime) and 2 separate single pointings that were parallels to long NIRSpec pointings (Deep Parallel); summing over the filters, this averages about 100 hours of exposure time.  The Medium region of 167 arcmin$^2$ surrounds this region in GOODS-S and provides core coverage in GOODS-N and HDF; it averages about 25 hours of exposure time summing over filters.  In practice, there are inescapable variations in depth due to the complex footprint of NIRCam and the overlapping of pointings (\ref{sec:depth}).  GOODS-N Medium is somewhat shallower on average than GOODS-S Medium.

JADES always observed with at least 8 filters: F090W, F115W, F150W, F200W, F277W, F356W, F410M, and F444W.  In most cases a 9th filter, F335M, was included.  We found that the medium-band F335M and F410M filters were extremely helpful in identifying the strong rest-optical emission lines common in $z>3$ galaxies \citep[e.g.][]{simmonds24,zhu25}.  The Medium Parallel fields included a 10th filter, F070W, because the coverage from HST optical imaging was not as deep away from the heart of the GOODS fields.  Program 4540 added F070W and NIRCam wide-field grism spectroscopy to the JADES Origins Field region.   In addition, the last of the Deep Parallel fields (program 1287) included F162M, F182M, and F300M as well, totalling 12 filters in that pointing to build on the practice of using a medium filter to divide a wide filter in two.

JADES utilized long integrations, typically 900-1375 sec, to suppress detector noise, and nearly always observed with at least 6 dither positions per pointing, to guard against bad pixels and other image flaws.  Boundary and gap areas can have fewer exposures, of course.  JADES used the DEEP8 readout pattern where possible, to economize on telemetry; some Medium fields needed to use the MEDIUM8 pattern to accommodate shorter exposures.  Dither patterns varied: subpixel dithers used long-wavelength MIRI patterns to get arcsecond-level steps, while the larger steps were arranged in different ways: APT mosaics, multiple MSA configurations, and multiple explicit observations.
As described in \citet{eisenstein2023}, occasional issues with NIRSpec multi-shutter array (MSA) short circuits and telescope guide star acquisition did impact the final geometry of the JADES observations, resulting in re-plans that sought to fill these gaps, sometimes with modest compromises on depth or geometrical cleanliness.

The JADES NIRCam program was scheduled over the first two years of the mission, with some last portions of GOODS-S not happening until the third year.  Some of this sequencing resulted in repeated area between years, opening time-domain opportunities \citep{decoursey2025}.  In particular, the Deep Prime program was split nearly evenly between the first two years, and the Medium Prime data that overlap it were observed in year 2.  Parts of the southern portion of the Prime mosaic were covered twice due to parallel exposures and the repetition of one failed pointing.  Finally, two of the parallel pointings to the west have repeated imaging.

Within a visit, we usually began with F090W as the SW filter.  Occasionally F115W was first.  This proved to be fortuitous, as it meant that persistence from the prior visit would primarily affect F090W, which is less affected by the wisp and claw stray light artifacts that can be prominent in F150W and F200W.  This helped to separate these issues.  In contrast, the long-wave (LW) images tended to be clean, save for some rare scattered light discussed in \citet{eisenstein2023} and later in this paper.

In total, the JADES GTO NIRCam observations comprise 486.7 hours of SW integration in GOODS-S and 90.9 hours in GOODS-N (Tables \ref{tbl:gs_subregions} and \ref{tbl:gn_subregions}).  LW integration time is mildly shorter, 457.4 and 89.3 respectively, due to observations with the NIRCam grism and two cases of LW imaging that had to be discarded due to glowing short-circuits in NIRSpec.

We note that all JADES observations were designed with coordinated parallels involving either MIRI imaging or NIRSpec multi-object spectroscopy.  The NIRSpec spectroscopy was fully included in JADES Data Release 4 \citep{curtislake2025,scholtz2025}, building on prior releases \citep{bunker2024,deugenio2025}.  The MIRI imaging is being released in a companion paper (Alberts et., to be submitted), building on the pipeline described in \citet{alberts2024}.

\begin{figure*}[tb]
\includegraphics[width=\textwidth]{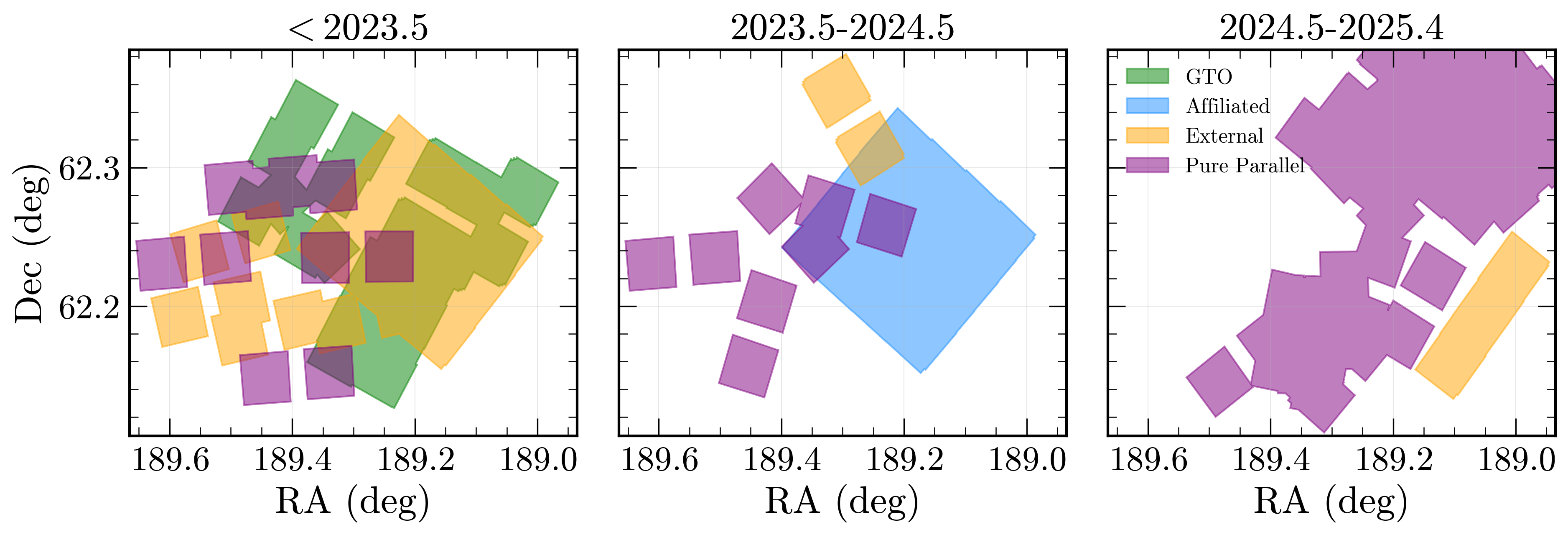}
\caption{Footprint of the GOODS-N NIRCam LW images in three separate years, colored by the program kind. These footprints do not account for area lost to artifact masking.  We note that shallower and SW-only data taken in pure-parallel mode extends off the northwest edge of the full JADES mosaic.
\label{fig:footprint_GN}}
\end{figure*}

\subsection{JADES Affiliated Guest Observer Programs}
\newcommand{\thissect}[1]{{$\bullet$ \bf #1:}}

Members of the JADES team led successful guest observer (GO) programs that produced additional NIRCam data included in this release.  These programs were substantially coordinated with JADES and its ongoing data reduction efforts.

\thissect{Program 1963}
The Cycle 1 JWST Extragalactic Medium-band Survey program \citep[JEMS, PIs: Williams, Tacchella, \& Maseda;][]{williams2023} observed one pointing in the JADES Deep Prime region with one NIRCam module on the HUDF.  This provides deep coverage in F182M, F210M, F430M, F460M, and F480M.

\thissect{Program 3215}
The Cycle 2 program 3215 (PI: Eisenstein) carried out ultradeep medium-band imaging and spectroscopy in a single pointing at the same location as the Deep Parallel program 1210, producing the JADES Origins Field (JOF).
Program 3215 provides very long ($\sim 30$ hour per filter)  F162M, F182M, F210M, F250M, F300M, and F335M observations on this field, bringing the total exposure time to about 350 hrs across 15 filters.

As described in \citet{eisenstein2025}, a portion of these observations were impacted by a glowing NIRSpec MSA short circuit, and we were permitted to replan these to shift the pointing to place NIRSpec on the JOF while imaging the 6 medium-bands on a part of the JADES Medium Parallel, creating another 15-filter region.

\thissect{Program 3577}
The Cycle 2 program 3577, the Complete NIRCam Grism Redshift Survey (CONGRESS, PI: Egami), conducted NIRCam F356W wide-field slitless spectroscopy in GOODS-N, returning to the FRESCO field.  This yields medium-depth F090W and F115W imaging as well as shallower F356W direct imaging.

\thissect{Program 5997}
The Cycle 3 program Observing All phases of StochastIc Star formation (OASIS, PIs: Looser \& D'Eugenio) conducted two deep NIRSpec MOS pointings, which produced two NIRCam coordinated parallel fields.  One falls near the JOF and within the JADES Medium Prime mosaic; this used F150W, F162M, F182M, F210M, F250M, F300M, F335M, and F480M, each of 7 hr depth.
The other produces a new field partially overlapping the Deep Parallel 1287 field; this used the 8 base JADES filters.

\thissect{Program 6541}
The Cycle 2 Director Discretionary Program 6541 (PI: Egami) provides spectroscopic follow-up and two additional epochs of wide-band imaging on transients in the Deep Prime region.  This provides shallow coverage in F115W, F150W, F200W, F277W, F356W, and F444W.

\subsection{NIRCam Imaging External to JADES}
\label{subsec:ancillary_data}

We also include in these DR5 reductions a wide array of now-public NIRCam imaging from 14 other programs.  We briefly describe these here.

\thissect{Program 1176}
The Cycle 1 guaranteed time program 1176, the Prime Extragalactic Areas for Reionization and Lensing Science \citep[PEARLS, PI: Windhorst][]{windhorst2023}, performed one pointing in GOODS-S to the north of JADES, yielding medium depth in the same 8 filters that JADES uses as its base.

\thissect{Programs 1283 and 6511}
The Cycle 1 guaranteed time program 1283, the MIRI Deep Imaging Survey \citep[MIDIS, PI: \"Ostlin;][]{ostlin25,perezgonzalez2024b}, conducted very deep MIRI imaging in the HUDF, producing a NIRCam deep field in GOODS-S to the east of JADES.
These were returned to in Cycle 3 GO program 6511 \citep[PI: \"Ostlin;][]{perezgonzalex25midis}, adding additional depth and filters.  We note that we only include the 2024 data from 6511, which comprise observations using F200W and F444W.

\thissect{Program 1264}
The Cycle 1 program 1264 (PI: Colina) conducted one NIRCam pointing in GOODS-N, yielding half-hour depth in six wide filters to the north of JADES.

\thissect{Program 1895}
The Cycle 1 program 1895, the First Reionization Epoch Spectroscopic Complete survey \citep[FRESCO, PI: Oesch;][]{oesch2023}, performed NIRCam wide-field slitless spectroscopy in F444W, yielding SW imaging in F182M and F210M in both GOODS-S and GOODS-N, and shallow direct imaging in F444W.

\thissect{Program 2079}
The Cycle 1 program 2079, the Next Generation Deep Extragalactic Exploratory Public survey \citep[NGDEEP, PI: Finkelstein;][]{bagley2024}, performed NIRISS wide-field slitless spectroscopy on the HUDF, yielding a deep parallel NIRCam field in GOODS-S east of JADES \citep{bagley2024} in F115W, F150W, F200W, F277W, F356W, and F444W.  These observations were split over two years of the mission.  We omitted the shorter exposures taken parallel to direct NIRISS images as their noise properties were substantially different than the bulk of the data.

\thissect{Program 2198}
The Cycle 1 program 2198 (PI: Barrufet) conducted shallow NIRCam pre-imaging in F200W and F444W in GOODS-S south of the HUDF.

\thissect{Program 2514}
The Cycle 1 pure-parallel program 2514, Parallel wide-Area Nircam Observations to Reveal And Measure the
Invisible Cosmos \citep[PANORAMIC, PI: Williams;][]{williams2025}) conducted NIRCam imaging, attaching to a number of pointings in both GOODS-S and GOODS-N.  Most of these provide 6-band coverage in F115W, F150W, F200W, F277W, F356W, and F444W.  One field in GOODS-S is particularly deep, adding F410M and providing 22 hr depth in F115W.  One field in GOODS-N duplicated a JADES Medium Parallel pointing and provides novel filters: F162M, F182M, F210M, F300M, F430M, and F460M.

\thissect{Program 2516}
The Cycle 1 program 2516 (PI: Hodge) conducted NIRCam imaging in GOODS-S in F200W, F356W, and F444W.  We use one pointing, observation 3, that overlaps other imaging in the release.

\thissect{Program 2674}
The Cycle 1 program 2674 (PI: Arrabal Haro) conducted F182M, F187N, F405N, and F410M imaging in GOODS-N.  We include four of these pointings, using only the medium bands.

\thissect{Program 3990}
The Cycle 2 pure-parallel program 3990, Bias-free Extragalactic Analysis for Cosmic Origins with NIRCam \citep[BEACON, PI: Morishita;][]{morishita2025}, attached to several pointings in GOODS-S, yielding wide-band coverage to the west of the HUDF.  Another pointing provides F140M and F250M near the JOF.

\thissect{Program 4762}
The Cycle 2 program 4762 (PI: Fujimoto) performed NIRCam wide-field slitless spectroscopy in GOODS-N, west of JADES and adjacent to CONGRESS and FRESCO, producing medium-depth images in F182M and F210M and shallow direct images in F150W, F356W, and F444W.

\thissect{Program 5398}
The Cycle 3 pure-parallel program 5398, the Public Observation Pure Parallel Infrared Emission-Line Survey (POPPIES, PI: Kartaltepe), is conducting NIRCam wide-field slitless spectroscopy. Direct images are also obtained in the SW channel and for pre-imaging in the LW channel. This program has attached to a series of observations in GOODS-N, particularly those from program 5407 (PI: Leung), which leads to a large but shallow imaging mosaic in F115W, F200W, and F444W to the northwest of JADES.

\thissect{Program 6434}
The Cycle 3 pure-parallel program 6434, the Slitless Areal Pure-Parallel High-Redshift Emission Survey \citep[SAPPHIRES, PI: Egami;][]{Sun_sapphire}, is conducting NIRCam wide-field slitless spectroscopy. Direct images are also obtained in the SW channel and sometimes in the LW channel for pre-imaging. This program attached to a set of observations in both GOODS-S and GOODS-N, and it coordinated those 
designs and image reductions with JADES.  The exact filter choices vary with the opportunity.  There are two long observations to particularly mention.  One pointing in GOODS-N was observed in F162M, F200W, and F277W for 16-21 hrs/filter, substantially overlapping JADES many-band imaging.
Another pointing in GOODS-S was observed for 12 hrs in F200W.

\begin{figure*}[t]
\includegraphics[width=\textwidth]{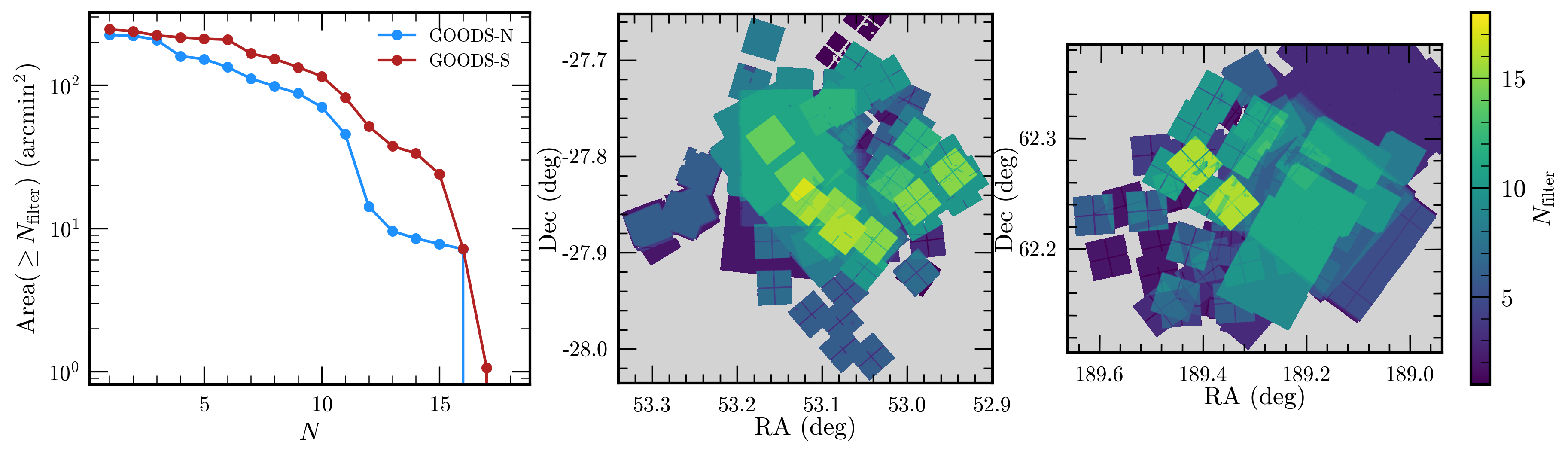}
\caption{Filter coverage of JADES DR5. {\it Left:} Area as a function of the number of filters.  {\it Middle}: Map of the number of filters available in GOODS-S.  {\it Right}: Same as middle, but for GOODS-N.
\label{fig:nfilter}}
\end{figure*}

\subsection{Notable omissions}

The data assembly for JADES DR5 was closed in May 2025.  This was before the last epoch of the MIDIS program 6511 was observed in late 2025 in GOODS-S, and before the Cycle 3 program 4549 (PI: G. Rieke) of slitless spectroscopy observed the JADES GOODS-S Deep Prime region.
We also do not include the Cycle 4 medium-band MINERVA program 7814 (PI: A.Muzzin) in GOODS-N nor the imaging collected by the JWST Multi-Cycle Transient survey (program 8060, PI: E. Egami) in late 2025.  We did not include a few shallow pointings of F140M, F187N, and F405N imaging in the GOODS fields. Finally, we have not included direct imaging from the NIRISS instrument, several fields of which exist in GOODS-S.

\subsection{Summary of the data}

As is clear from this listing, the JWST archival imaging in GOODS-S and GOODS-N is now impressively large.
The total exposure time in the SW filters is 985 hours in GOODS-S and 268 hours in GOODS-N, 1253 total.  The exposure time in the LW is 913 hours in GOODS-S and 157 hours in GOODS-N; the remaining 183 hrs were mostly spent performing slitless spectroscopy with the LW arm.  The charged mission time is typically 40-50\% larger, so this implies around 1800 hours of mission time.  About 63\% of the 1253 SW hrs come from JADES or JADES-affiliated projects.

For reasons detailed in \S\ref{sec:subregion}, we group the NIRCam exposures into sets, called subregions, each containing the data from one program at nearly the same position angle (PA) and epoch.  As the telescope can only hold a given PA for a couple weeks, in practical terms one epoch means all of the data collected in a given year at that PA.  There are 45 such subregions in GOODS-S and 29 in GOODS-N.  The basic descriptions of each subregion are given in Tables \ref{tbl:gs_subregions} and \ref{tbl:gn_subregions}.  Tables \ref{tbl:gs_nim} and \ref{tbl:gn_nim} then list the number of individual Sensor Chip Assembly (SCA) exposures in each.  There are a total of 22,202 SCA exposures in GOODS-S and 10,125 in GOODS-N, each a 4 megapixel image sampled multiple times per integration.

The data span 3 years of collection in each field.  We present the footprints covered in each of the 3 years in  Figure \ref{fig:footprint_GS} for GOODS-S and Figure \ref{fig:footprint_GN} for GOODS-N.  Of course, the real situation is yet more complicated, as each year observed different sets of filters.  This multi-cycle coverage is of course a great opportunity for studies of infrared variability at very faint flux levels, such as highlighted in \citet{decoursey2025} for transients and \citet{hainline2024b, hainline2026a} for proper motions of brown dwarfs.

A total of 469 arcmin$^2$, 245 in GOODS-S and 224 in GOODS-N, have at least one filter of JWST NIRCam imaging in DR5.  However, the coverage is heterogeneous; Figure \ref{fig:nfilter} shows the number of distinct filters available at each location in the field.
The area covered by the 8 core JADES filters, enabling accurate photometric redshifts over a wide redshift range, is 151 arcmin$^2$ in GOODS-S and 89 arcmin$^2$ in GOODS-N, 240 arcmin$^2$ total, of which 91\% and 100\% are provided by JADES GTO imaging respectively.  For coverage by the 6 wide band filters F115W, F150W, F200W, F277W, F356W, and F444W --- a set that allows some confidence in photometric redshift estimates --- the total area is 209 arcmin$^2$ in GOODS-S and 114 arcmin$^2$ in GOODS-N, for a total of 323 arcmin$^2$.
There is 65 arcmin$^2$ with more than 12 filters.

\begin{deluxetable*}{lchccccccccc}
\tabletypesize{\footnotesize}
\tablecaption{GOODS-S Subregions\label{tbl:gs_subregions}}
\tablehead{
\colhead{Subregion} & \colhead{PID} & \nocolhead{ID} & \colhead{Bit\tablenotemark{a}} & \colhead{$N_{\rm obs}$} & \colhead{$N_{\rm band}$} & \colhead{$N_{\rm band, LW}$} & \colhead{$t_{\rm exp, SW}$\tablenotemark{b}} & \colhead{$t_{\rm exp, LW}$\tablenotemark{b}} & \colhead{Area\tablenotemark{c}} & \colhead{Epoch\tablenotemark{d}} & \colhead{PA\tablenotemark{d}} \\ 
\colhead{} & \colhead{} & \colhead{} & \nocolhead{} & \colhead{} & \colhead{} & \colhead{} & \colhead{(hr)} & \colhead{(hr)} & \colhead{(arcmin$^2$)} & \colhead{(jyear)} & \colhead{ (deg)} \\
}
\startdata
\texttt{jw011760\_gs} & 01176 & 0 & 13 & 1 & 8 & 4 & 3.5 & 3.5 & 10.2 & 2023.61 & 253.1 \\
\texttt{jw011800\_deep} & 01180 & 1 & 0 & 6 & 9 & 5 & 87.0 & 87.0 & 27.0 & 2022.75 & 298.5 \\
\texttt{jw011800\_deep23} & 01180 & 2 & 0 & 6 & 9 & 5 & 87.0 & 87.0 & 27.4 & 2023.75 & 298.5 \\
\texttt{jw011800\_medium} & 01180 & 3 & 1 & 6 & 8 & 4 & 36.6 & 30.0 & 35.8 & 2022.77 & 308.0 \\
\texttt{jw011800\_medium23} & 01180 & 4 & 2 & 2 & 9 & 5 & 15.6 & 15.6 & 22.4 & 2023.77 & 308.0 \\
\texttt{jw011800\_medium\_obs022} & 01180 & 5 & 2 & 1 & 9 & 5 & 7.7 & 7.7 & 11.2 & 2023.87 & 356.7 \\
\texttt{jw011800\_medium\_obs219} & 01180 & 6 & 2 & 1 & 9 & 5 & 7.7 & 7.7 & 11.2 & 2023.87 & 356.1 \\
\texttt{jw011800\_medium\_obs220+222} & 01180 & 7 & 2 & 2 & 2 & 1 & 2.8 & 2.8 & 22.4 & 2024.00 & 43.0 \\
\texttt{jw011800\_medium\_obs223} & 01180 & 8 & 2 & 1 & 9 & 5 & 7.7 & 7.7 & 11.2 & 2024.00 & 52.1 \\
\texttt{jw011800\_medium\_redo} & 01180 & 9 & 2 & 1 & 8 & 4 & 6.7 & 6.7 & 9.5 & 2023.77 & 308.0 \\
\texttt{jw012100} & 01210 & 10 & 5 & 3 & 9 & 5 & 55.0 & 55.0 & 9.5 & 2022.81 & 321.0 \\
\texttt{jw012830} & 01283 & 11 & 14 & 3 & 4 & 2 & 22.9 & 22.9 & 14.6 & 2022.93 & 23.2 \\
\texttt{jw012830\_obs007} & 01283 & 12 & 14 & 1 & 2 & 1 & 8.4 & 8.4 & 11.9 & 2023.93 & 22.8 \\
\texttt{jw012860} & 01286 & 13 & 6 & 1 & 10 & 5 & 11.0 & 11.0 & 9.5 & 2023.03 & 56.2 \\
\texttt{jw012860\_dec23} & 01286 & 14 & 6 & 6 & 10 & 5 & 70.9 & 70.9 & 42.4 & 2023.96 & 30.2 \\
\texttt{jw012860\_oct23} & 01286 & 15 & 6 & 1 & 10 & 5 & 11.8 & 11.8 & 9.5 & 2023.80 & 319.9 \\
\texttt{jw012870} & 01287 & 16 & 7 & 2 & 12 & 6 & 36.6 & 36.6 & 9.5 & 2024.03 & 53.0 \\
\texttt{jw012870\_obs003} & 01287 & 17 & 7 & 1 & 12 & 6 & 18.3 & 18.3 & 9.4 & 2025.03 & 53.0 \\
\texttt{jw018950\_gs} & 01895 & 18 & 15 & 8 & 3 & 1 & 17.7 & 2.1 & 62.7 & 2022.88 & 180.2 \\
\texttt{jw019630} & 01963 & 19 & 16 & 1 & 5 & 3 & 15.5 & 15.5 & 10.4 & 2022.78 & 307.2 \\
\texttt{jw020790\_obs001} & 02079 & 20 & 17 & 3 & 6 & 3 & 47.4 & 47.4 & 11.0 & 2024.08 & 66.9 \\
\texttt{jw020790\_obs004} & 02079 & 21 & 17 & 3 & 6 & 3 & 46.8 & 46.8 & 11.1 & 2023.08 & 69.9 \\
\texttt{jw021980\_pa001} & 02198 & 22 & 18 & 4 & 2 & 1 & 0.5 & 0.5 & 27.8 & 2022.88 & 0.6 \\
\texttt{jw021980\_pa353} & 02198 & 23 & 18 & 4 & 2 & 1 & 0.5 & 0.5 & 27.9 & 2022.86 & 353.9 \\
\texttt{jw025140\_gs\_pa064} & 02514 & 24 & 19 & 3 & 6 & 3 & 1.5 & 1.5 & 9.4 & 2024.07 & 64.2 \\
\texttt{jw025140\_gs\_pa272} & 02514 & 25 & 19 & 6 & 7 & 4 & 27.5 & 27.5 & 9.5 & 2023.67 & 272.3 \\
\texttt{jw025140\_gs\_pa311} & 02514 & 26 & 19 & 15 & 6 & 3 & 7.7 & 7.7 & 44.0 & 2023.78 & 310.8 \\
\texttt{jw025160} & 02516 & 27 & 20 & 1 & 3 & 2 & 1.0 & 1.0 & 10.2 & 2022.91 & 19.6 \\
\texttt{jw032150} & 03215 & 28 & 8 & 5 & 6 & 3 & 103.1 & 100.8 & 9.6 & 2023.80 & 321.1 \\
\texttt{jw032150\_obs901} & 03215 & 29 & 8 & 1 & 8 & 4 & 18.8 & 18.8 & 9.4 & 2024.96 & 39.0 \\
\texttt{jw039900\_obs092} & 03990 & 30 & 23 & 1 & 2 & 1 & 3.5 & 3.5 & 9.4 & 2023.98 & 40.9 \\
\texttt{jw039900\_obs093} & 03990 & 31 & 23 & 1 & 1 & 1 & 0.0 & 0.9 & 9.4 & 2023.93 & 22.6 \\
\texttt{jw039900\_obs563} & 03990 & 32 & 23 & 1 & 1 & 1 & 0.0 & 4.1 & 9.4 & 2024.81 & 324.5 \\
\texttt{jw039900\_pa045} & 03990 & 33 & 23 & 3 & 6 & 3 & 6.3 & 6.3 & 9.5 & 2024.00 & 45.5 \\
\texttt{jw039900\_pa058} & 03990 & 34 & 23 & 2 & 4 & 2 & 14.2 & 14.2 & 9.4 & 2024.05 & 58.0 \\
\texttt{jw039900\_pa358} & 03990 & 35 & 23 & 4 & 8 & 4 & 13.8 & 13.8 & 11.5 & 2023.88 & 358.4 \\
\texttt{jw045400} & 04540 & 36 & 9 & 18 & 5 & 2 & 24.3 & 1.6 & 29.3 & 2024.80 & 321.0 \\
\texttt{jw059970\_jan25} & 05997 & 37 & 10 & 2 & 8 & 4 & 27.5 & 27.5 & 9.5 & 2025.01 & 56.4 \\
\texttt{jw059970\_oct24} & 05997 & 38 & 10 & 2 & 8 & 4 & 27.5 & 27.5 & 9.5 & 2024.80 & 321.0 \\
\texttt{jw064340\_obs041} & 06434 & 39 & 11 & 1 & 1 & 0 & 11.8 & 0.0 & 9.4 & 2024.78 & 316.8 \\
\texttt{jw064340\_obs072} & 06434 & 40 & 11 & 1 & 1 & 0 & 6.1 & 0.0 & 9.2 & 2025.02 & 52.4 \\
\texttt{jw064340\_pa006} & 06434 & 41 & 11 & 4 & 3 & 1 & 15.1 & 3.8 & 18.8 & 2024.89 & 6.3 \\
\texttt{jw065110} & 06511 & 42 & 24 & 6 & 2 & 1 & 45.8 & 45.8 & 14.2 & 2024.92 & 20.9 \\
\texttt{jw065410\_pa015} & 06541 & 43 & 12 & 2 & 6 & 3 & 2.1 & 2.1 & 18.7 & 2023.91 & 14.5 \\
\texttt{jw065410\_pa045} & 06541 & 44 & 12 & 2 & 5 & 3 & 2.1 & 2.1 & 18.7 & 2024.00 & 45.4 \\
\enddata
\tablenotetext{a}{Bit corresponding to this subregion in the bithash image (\S\ref{sec:bithash}).}
\tablenotetext{b}{Exposure times are computed by summing the exposure time in every image in a channel and dividing by the number of detectors in that channel, 8 for SW or 2 for LW.}
\tablenotetext{c}{Area corresponds to the LW imaging when available, or SW if not.  Does not account for area lost to artifacts.}
\tablenotetext{d}{These are means of the epoch and PA of the constituent exposures}
\end{deluxetable*}

\begin{deluxetable*}{lchccccccccc}
\tabletypesize{\footnotesize}
\tablecaption{GOODS-N Subregions\label{tbl:gn_subregions}}
\tablehead{
\colhead{Subregion} & \colhead{PID} & \nocolhead{ID} & \colhead{Bit\tablenotemark{a}} & \colhead{$N_{\rm obs}$} & \colhead{$N_{\rm band}$} & \colhead{$N_{\rm band, LW}$} & \colhead{$t_{\rm exp, SW}$\tablenotemark{b}} & \colhead{$t_{\rm exp, LW}$\tablenotemark{b}} & \colhead{Area\tablenotemark{c}} & \colhead{Epoch\tablenotemark{d}} & \colhead{PA\tablenotemark{d}} \\ 
\colhead{} & \colhead{} & \colhead{} & \nocolhead{} & \colhead{} & \colhead{} & \colhead{} & \colhead{(hr)} & \colhead{(hr)} & \colhead{(arcmin$^2$)} & \colhead{(jyear)} & \colhead{ (deg)} \\
}
\startdata
\texttt{jw011810\_hst} & 01181 & 0 & 3 & 4 & 9 & 5 & 31.5 & 29.9 & 27.4 & 2023.10 & 241.1 \\
\texttt{jw011810\_jwst} & 01181 & 1 & 4 & 3 & 10 & 5 & 35.4 & 35.4 & 27.7 & 2023.33 & 150.6 \\
\texttt{jw011810\_miri} & 01181 & 2 & 3 & 3 & 9 & 5 & 12.2 & 12.2 & 29.7 & 2023.09 & 241.0 \\
\texttt{jw011810\_obs098} & 01181 & 3 & 4 & 1 & 10 & 5 & 11.8 & 11.8 & 9.5 & 2023.40 & 133.1 \\
\texttt{jw012640} & 01264 & 4 & 29 & 1 & 6 & 3 & 1.9 & 1.9 & 11.0 & 2023.93 & 300.8 \\
\texttt{jw018950\_gn} & 01895 & 5 & 15 & 8 & 3 & 1 & 17.7 & 2.1 & 62.7 & 2023.12 & 230.6 \\
\texttt{jw025140\_gn24\_pa184} & 02514 & 6 & 19 & 2 & 4 & 2 & 0.9 & 0.9 & 9.4 & 2024.23 & 184.4 \\
\texttt{jw025140\_gn\_pa133} & 02514 & 7 & 19 & 6 & 6 & 3 & 4.7 & 4.7 & 9.4 & 2024.38 & 133.0 \\
\texttt{jw025140\_gn\_pa163} & 02514 & 8 & 19 & 2 & 4 & 2 & 7.5 & 7.5 & 9.4 & 2024.29 & 163.1 \\
\texttt{jw025140\_gn\_pa180} & 02514 & 9 & 19 & 2 & 4 & 2 & 1.1 & 1.1 & 9.4 & 2023.23 & 180.3 \\
\texttt{jw025140\_gn\_pa184} & 02514 & 10 & 19 & 8 & 4 & 2 & 4.4 & 4.0 & 34.2 & 2023.23 & 184.3 \\
\texttt{jw025140\_gn\_pa253} & 02514 & 11 & 19 & 3 & 6 & 3 & 1.5 & 1.5 & 9.4 & 2024.08 & 252.7 \\
\texttt{jw026740\_obs001} & 02674 & 12 & 21 & 1 & 3 & 2 & 0.9 & 1.8 & 9.4 & 2023.21 & 195.6 \\
\texttt{jw026740\_pa193} & 02674 & 13 & 21 & 3 & 3 & 2 & 2.6 & 5.3 & 25.9 & 2023.21 & 192.7 \\
\texttt{jw035770} & 03577 & 14 & 22 & 12 & 3 & 1 & 15.7 & 3.1 & 67.2 & 2024.13 & 228.3 \\
\texttt{jw047620} & 04762 & 15 & 30 & 2 & 5 & 2 & 3.6 & 0.7 & 16.5 & 2025.10 & 232.9 \\
\texttt{jw053980\_gn\_obs290\_291} & 05398 & 16 & 28 & 2 & 3 & 1 & 3.9 & 1.9 & 9.4 & 2025.37 & 135.0 \\
\texttt{jw053980\_gn\_pa220} & 05398 & 17 & 28 & 22 & 3 & 1 & 10.8 & 1.8 & 60.1 & 2025.15 & 219.2 \\
\texttt{jw053980\_gn\_pa227} & 05398 & 18 & 28 & 2 & 3 & 1 & 1.0 & 0.2 & 10.9 & 2025.12 & 226.2 \\
\texttt{jw053980\_gn\_pa230} & 05398 & 19 & 28 & 6 & 3 & 1 & 2.9 & 0.5 & 30.9 & 2025.11 & 230.3 \\
\texttt{jw064340\_gn\_obs215} & 06434 & 20 & 11 & 1 & 1 & 0 & 4.2 & 0.0 & 9.2 & 2025.03 & 268.1 \\
\texttt{jw064340\_gn\_pa217} & 06434 & 21 & 11 & 2 & 3 & 1 & 5.1 & 0.9 & 9.4 & 2025.15 & 217.1 \\
\texttt{jw064340\_gn\_pa228} & 06434 & 22 & 11 & 4 & 3 & 1 & 17.2 & 3.5 & 12.9 & 2025.12 & 228.3 \\
\texttt{jw064340\_gn\_pa239} & 06434 & 23 & 11 & 2 & 3 & 1 & 5.1 & 0.9 & 9.4 & 2025.11 & 239.7 \\
\texttt{jw064340\_gn\_pa249} & 06434 & 24 & 11 & 2 & 3 & 1 & 5.8 & 1.6 & 9.4 & 2025.07 & 249.9 \\
\texttt{jw064340\_gn\_pa258} & 06434 & 25 & 11 & 3 & 3 & 1 & 7.4 & 3.2 & 18.1 & 2025.05 & 258.5 \\
\texttt{jw064340\_gn\_pa271} & 06434 & 26 & 11 & 2 & 3 & 1 & 5.8 & 1.6 & 9.4 & 2025.02 & 271.4 \\
\texttt{jw064340\_obs381\_382} & 06434 & 27 & 11 & 2 & 3 & 1 & 37.7 & 16.4 & 9.4 & 2025.37 & 140.7 \\
\texttt{jw064340\_obs413\_414} & 06434 & 28 & 11 & 2 & 3 & 1 & 2.6 & 0.5 & 9.4 & 2025.36 & 142.0 \\
\enddata
\tablenotetext{a}{Bit corresponding to this subregion in the bithash image (\S\ref{sec:bithash}).}
\tablenotetext{b}{Exposure times are computed by summing the exposure time in every image and dividing by the number of detectors, 8 for SW or 2 for LW..}
\tablenotetext{c}{Area corresponds to the LW imaging when available, or SW if not.  Does not account for area lost to artifacts.}
\tablenotetext{d}{These are means of the epoch and PA of the constituent exposures}
\end{deluxetable*}

\begin{deluxetable*}{lcccccccccccccccccccc}
\tabletypesize{\footnotesize}
\tablecaption{GOODS-S SCA Exposures \label{tbl:gs_nim}}
\tablehead{
\colhead{ID} & \multicolumn{8}{c}{Number of Wide-band SCA Exposures} & \hspace*{10pt} & \multicolumn{10}{c}{Number of Medium-band SCA Exposures} \\[-6pt]
\\[-6pt]
\colhead{} & \colhead{${\rm 070}$} & \colhead{${\rm 090}$} & \colhead{${\rm 115}$} & \colhead{${\rm 150}$} & \colhead{${\rm 200}$} & \colhead{${\rm 277}$} & \colhead{${\rm 356}$} & \colhead{${\rm 444}$} & \colhead{} & \colhead{${\rm 162}$} & \colhead{${\rm 182}$} & \colhead{${\rm 210}$} & \colhead{${\rm 250}$} & \colhead{${\rm 300}$} & \colhead{${\rm 335}$} & \colhead{${\rm 410}$} & \colhead{${\rm 430}$} & \colhead{${\rm 460}$} & \colhead{${\rm 480}$} \\
}
\startdata
\texttt{jw011760\_gs} &  -  & $32$ & $32$ & $32$ & $32$ & $8$ & $8$ & $8$ &  &  -  &  -  &  -  &  -  &  -  &  -  & $8$ &  -  &  -  &  -  \\
\texttt{jw011800\_deep} &  -  & $416$ & $704$ & $416$ & $288$ & $104$ & $72$ & $104$ &  &  -  &  -  &  -  &  -  &  -  & $72$ & $104$ &  -  &  -  &  -  \\
\texttt{jw011800\_deep23} &  -  & $416$ & $704$ & $416$ & $288$ & $104$ & $72$ & $104$ &  &  -  &  -  &  -  &  -  &  -  & $72$ & $104$ &  -  &  -  &  -  \\
\texttt{jw011800\_medium} &  -  & $264$ & $264$ & $264$ & $264$ & $54$ & $54$ & $54$ &  &  -  &  -  &  -  &  -  &  -  &  -  & $54$ &  -  &  -  &  -  \\
\texttt{jw011800\_medium23} &  -  & $96$ & $192$ & $96$ & $48$ & $12$ & $24$ & $24$ &  &  -  &  -  &  -  &  -  &  -  & $24$ & $24$ &  -  &  -  &  -  \\
\texttt{jw011800\_medium\_obs022} &  -  & $48$ & $96$ & $48$ & $48$ & $12$ & $12$ & $12$ &  &  -  &  -  &  -  &  -  &  -  & $12$ & $12$ &  -  &  -  &  -  \\
\texttt{jw011800\_medium\_obs219} &  -  & $48$ & $96$ & $48$ & $48$ & $12$ & $12$ & $12$ &  &  -  &  -  &  -  &  -  &  -  & $12$ & $12$ &  -  &  -  &  -  \\
\texttt{jw011800\_medium\_obs220+222} &  -  &  -  &  -  &  -  & $96$ & $24$ &  -  &  -  &  &  -  &  -  &  -  &  -  &  -  &  -  &  -  &  -  &  -  &  -  \\
\texttt{jw011800\_medium\_obs223} &  -  & $48$ & $96$ & $48$ & $48$ & $12$ & $12$ & $12$ &  &  -  &  -  &  -  &  -  &  -  & $12$ & $12$ &  -  &  -  &  -  \\
\texttt{jw011800\_medium\_redo} &  -  & $48$ & $48$ & $48$ & $48$ & $12$ & $12$ & $12$ &  &  -  &  -  &  -  &  -  &  -  &  -  & $12$ &  -  &  -  &  -  \\
\texttt{jw012100} &  -  & $144$ & $192$ & $144$ & $96$ & $30$ & $24$ & $36$ &  &  -  &  -  &  -  &  -  &  -  & $18$ & $36$ &  -  &  -  &  -  \\
\texttt{jw012830} &  -  &  -  & $160$ & $80$ &  -  & $40$ & $20$ &  -  &  &  -  &  -  &  -  &  -  &  -  &  -  &  -  &  -  &  -  &  -  \\
\texttt{jw012830\_obs007} &  -  &  -  &  -  & $88$ &  -  &  -  & $22$ &  -  &  &  -  &  -  &  -  &  -  &  -  &  -  &  -  &  -  &  -  &  -  \\
\texttt{jw012860} & $48$ & $72$ & $96$ & $72$ & $72$ & $18$ & $18$ & $24$ &  &  -  &  -  &  -  &  -  &  -  & $12$ & $18$ &  -  &  -  &  -  \\
\texttt{jw012860\_dec23} & $288$ & $576$ & $576$ & $432$ & $288$ & $108$ & $72$ & $144$ &  &  -  &  -  &  -  &  -  &  -  & $72$ & $144$ &  -  &  -  &  -  \\
\texttt{jw012860\_oct23} & $48$ & $72$ & $96$ & $72$ & $72$ & $18$ & $18$ & $24$ &  &  -  &  -  &  -  &  -  &  -  & $12$ & $18$ &  -  &  -  &  -  \\
\texttt{jw012870} &  -  & $72$ & $96$ & $72$ & $48$ & $12$ & $18$ & $24$ &  & $48$ & $48$ &  -  &  -  & $12$ & $12$ & $18$ &  -  &  -  &  -  \\
\texttt{jw012870\_obs003} &  -  & $24$ & $72$ & $24$ & $24$ & $6$ & $6$ & $12$ &  & $24$ & $24$ &  -  &  -  & $6$ & $6$ & $12$ &  -  &  -  &  -  \\
\texttt{jw018950\_gs} &  -  &  -  &  -  &  -  &  -  &  -  &  -  & $48$ &  &  -  & $448$ & $256$ &  -  &  -  &  -  &  -  &  -  &  -  &  -  \\
\texttt{jw019630} &  -  &  -  &  -  &  -  &  -  &  -  &  -  &  -  &  &  -  & $192$ & $192$ &  -  &  -  &  -  &  -  & $24$ & $24$ & $48$ \\
\texttt{jw020790\_obs001} &  -  &  -  & $336$ & $112$ & $112$ & $12$ & $92$ & $36$ &  &  -  &  -  &  -  &  -  &  -  &  -  &  -  &  -  &  -  &  -  \\
\texttt{jw020790\_obs004} &  -  &  -  & $336$ & $112$ & $112$ & $12$ & $92$ & $36$ &  &  -  &  -  &  -  &  -  &  -  &  -  &  -  &  -  &  -  &  -  \\
\texttt{jw021980\_pa001} &  -  &  -  &  -  &  -  & $96$ &  -  &  -  & $24$ &  &  -  &  -  &  -  &  -  &  -  &  -  &  -  &  -  &  -  &  -  \\
\texttt{jw021980\_pa353} &  -  &  -  &  -  &  -  & $96$ &  -  &  -  & $24$ &  &  -  &  -  &  -  &  -  &  -  &  -  &  -  &  -  &  -  &  -  \\
\texttt{jw025140\_gs\_pa064} &  -  &  -  & $24$ & $16$ & $16$ & $4$ & $4$ & $6$ &  &  -  &  -  &  -  &  -  &  -  &  -  &  -  &  -  &  -  &  -  \\
\texttt{jw025140\_gs\_pa272} &  -  &  -  & $112$ & $24$ & $8$ & $6$ & $6$ & $8$ &  &  -  &  -  &  -  &  -  &  -  &  -  & $16$ &  -  &  -  &  -  \\
\texttt{jw025140\_gs\_pa311} &  -  &  -  & $120$ & $80$ & $80$ & $20$ & $20$ & $30$ &  &  -  &  -  &  -  &  -  &  -  &  -  &  -  &  -  &  -  &  -  \\
\texttt{jw025160} &  -  &  -  &  -  &  -  & $64$ &  -  & $8$ & $8$ &  &  -  &  -  &  -  &  -  &  -  &  -  &  -  &  -  &  -  &  -  \\
\texttt{jw032150} &  -  &  -  &  -  &  -  &  -  &  -  &  -  &  -  &  & $240$ & $480$ & $360$ & $120$ & $90$ & $54$ &  -  &  -  &  -  &  -  \\
\texttt{jw032150\_obs901} &  -  &  -  &  -  & $24$ &  -  &  -  & $6$ &  -  &  & $48$ & $72$ & $72$ & $18$ & $18$ & $12$ &  -  &  -  &  -  &  -  \\
\texttt{jw039900\_obs092} &  -  &  -  &  -  &  -  & $80$ &  -  &  -  & $20$ &  &  -  &  -  &  -  &  -  &  -  &  -  &  -  &  -  &  -  &  -  \\
\texttt{jw039900\_obs093} &  -  &  -  &  -  &  -  &  -  &  -  &  -  &  -  &  &  -  &  -  &  -  & $8$ &  -  &  -  &  -  &  -  &  -  &  -  \\
\texttt{jw039900\_obs563} &  -  &  -  &  -  &  -  &  -  &  -  &  -  &  -  &  &  -  &  -  &  -  & $20$ &  -  &  -  &  -  &  -  &  -  &  -  \\
\texttt{jw039900\_pa045} &  -  & $24$ & $24$ & $24$ &  -  & $6$ & $6$ & $6$ &  &  -  &  -  &  -  &  -  &  -  &  -  &  -  &  -  &  -  &  -  \\
\texttt{jw039900\_pa058} &  -  &  -  & $72$ & $72$ &  -  & $18$ & $18$ &  -  &  &  -  &  -  &  -  &  -  &  -  &  -  &  -  &  -  &  -  &  -  \\
\texttt{jw039900\_pa358} &  -  & $16$ & $16$ & $16$ & $16$ & $4$ & $4$ & $4$ &  &  -  &  -  &  -  &  -  &  -  &  -  & $4$ &  -  &  -  &  -  \\
\texttt{jw045400} & $672$ &  -  &  -  & $48$ & $96$ &  -  & $12$ & $24$ &  &  -  &  -  &  -  &  -  &  -  &  -  &  -  &  -  &  -  &  -  \\
\texttt{jw059970\_jan25} &  -  & $72$ & $72$ & $72$ & $72$ & $18$ & $18$ & $18$ &  &  -  &  -  &  -  &  -  &  -  &  -  & $18$ &  -  &  -  &  -  \\
\texttt{jw059970\_oct24} &  -  &  -  &  -  & $72$ &  -  &  -  &  -  &  -  &  & $72$ & $72$ & $72$ & $18$ & $18$ & $18$ &  -  &  -  &  -  & $18$ \\
\texttt{jw064340\_obs041} &  -  &  -  &  -  &  -  & $40$ &  -  &  -  &  -  &  &  -  &  -  &  -  &  -  &  -  &  -  &  -  &  -  &  -  &  -  \\
\texttt{jw064340\_obs072} &  -  &  -  &  -  &  -  & $128$ &  -  &  -  &  -  &  &  -  &  -  &  -  &  -  &  -  &  -  &  -  &  -  &  -  &  -  \\
\texttt{jw064340\_pa006} &  -  & $224$ &  -  &  -  & $224$ &  -  &  -  & $56$ &  &  -  &  -  &  -  &  -  &  -  &  -  &  -  &  -  &  -  &  -  \\
\texttt{jw065110} &  -  &  -  &  -  &  -  & $480$ &  -  &  -  & $120$ &  &  -  &  -  &  -  &  -  &  -  &  -  &  -  &  -  &  -  &  -  \\
\texttt{jw065410\_pa015} &  -  &  -  & $48$ & $48$ & $48$ & $12$ & $12$ & $12$ &  &  -  &  -  &  -  &  -  &  -  &  -  &  -  &  -  &  -  &  -  \\
\texttt{jw065410\_pa045} &  -  &  -  &  -  & $48$ & $96$ & $12$ & $12$ & $12$ &  &  -  &  -  &  -  &  -  &  -  &  -  &  -  &  -  &  -  &  -  \\
\enddata
\tablecomments{An SCA exposure refers an image from one of the 10 NIRCam detectors in one band; a single NIRCam exposure will consist of 8 SCA exposures in the SW channel and 2 SCA exposures in the LW channel.}
\end{deluxetable*}

\begin{deluxetable*}{lcccccccccccccccccccc}
\tabletypesize{\footnotesize}
\tablecaption{GOODS-N SCA Exposures \label{tbl:gn_nim}}
\tablehead{
\colhead{ID} & \multicolumn{8}{c}{Number of Wide-band SCA Exposures} & \hspace*{10pt} & \multicolumn{10}{c}{Number of Medium-band SCA Exposures} \\[-6pt]
\\[-6pt]
\colhead{} & \colhead{${\rm 070}$} & \colhead{${\rm 090}$} & \colhead{${\rm 115}$} & \colhead{${\rm 150}$} & \colhead{${\rm 200}$} & \colhead{${\rm 277}$} & \colhead{${\rm 356}$} & \colhead{${\rm 444}$} & \colhead{} & \colhead{${\rm 162}$} & \colhead{${\rm 182}$} & \colhead{${\rm 210}$} & \colhead{${\rm 250}$} & \colhead{${\rm 300}$} & \colhead{${\rm 335}$} & \colhead{${\rm 410}$} & \colhead{${\rm 430}$} & \colhead{${\rm 460}$} & \colhead{${\rm 480}$} \\
}
\startdata
\texttt{jw011810\_hst} &  -  & $192$ & $384$ & $192$ & $192$ & $48$ & $48$ & $48$ &  &  -  &  -  &  -  &  -  &  -  & $42$ & $42$ &  -  &  -  &  -  \\
\texttt{jw011810\_jwst} & $144$ & $288$ & $288$ & $216$ & $144$ & $54$ & $36$ & $72$ &  &  -  &  -  &  -  &  -  &  -  & $36$ & $72$ &  -  &  -  &  -  \\
\texttt{jw011810\_miri} &  -  & $144$ & $240$ & $144$ & $144$ & $36$ & $36$ & $36$ &  &  -  &  -  &  -  &  -  &  -  & $24$ & $36$ &  -  &  -  &  -  \\
\texttt{jw011810\_obs098} & $48$ & $96$ & $96$ & $72$ & $48$ & $18$ & $12$ & $24$ &  &  -  &  -  &  -  &  -  &  -  & $12$ & $24$ &  -  &  -  &  -  \\
\texttt{jw012640} &  -  &  -  & $56$ & $56$ & $40$ & $14$ & $10$ & $14$ &  &  -  &  -  &  -  &  -  &  -  &  -  &  -  &  -  &  -  &  -  \\
\texttt{jw018950\_gn} &  -  &  -  &  -  &  -  &  -  &  -  &  -  & $48$ &  &  -  & $448$ & $256$ &  -  &  -  &  -  &  -  &  -  &  -  &  -  \\
\texttt{jw025140\_gn24\_pa184} &  -  &  -  &  -  & $16$ & $16$ & $4$ & $4$ &  -  &  &  -  &  -  &  -  &  -  &  -  &  -  &  -  &  -  &  -  &  -  \\
\texttt{jw025140\_gn\_pa133} &  -  &  -  &  -  &  -  &  -  &  -  &  -  &  -  &  & $48$ & $48$ & $48$ &  -  & $12$ &  -  &  -  & $12$ & $12$ &  -  \\
\texttt{jw025140\_gn\_pa163} &  -  &  -  & $56$ & $56$ &  -  &  -  & $14$ & $14$ &  &  -  &  -  &  -  &  -  &  -  &  -  &  -  &  -  &  -  &  -  \\
\texttt{jw025140\_gn\_pa180} &  -  &  -  & $24$ & $16$ &  -  &  -  & $4$ & $6$ &  &  -  &  -  &  -  &  -  &  -  &  -  &  -  &  -  &  -  &  -  \\
\texttt{jw025140\_gn\_pa184} &  -  &  -  & $96$ & $64$ &  -  &  -  & $12$ & $24$ &  &  -  &  -  &  -  &  -  &  -  &  -  &  -  &  -  &  -  &  -  \\
\texttt{jw025140\_gn\_pa253} &  -  &  -  & $24$ & $16$ & $16$ & $4$ & $4$ & $6$ &  &  -  &  -  &  -  &  -  &  -  &  -  &  -  &  -  &  -  &  -  \\
\texttt{jw026740\_obs001} &  -  &  -  &  -  &  -  &  -  &  -  &  -  & $6$ &  &  -  & $24$ &  -  &  -  &  -  &  -  & $6$ &  -  &  -  &  -  \\
\texttt{jw026740\_pa193} &  -  &  -  &  -  &  -  &  -  &  -  &  -  & $18$ &  &  -  & $72$ &  -  &  -  &  -  &  -  & $18$ &  -  &  -  &  -  \\
\texttt{jw035770} &  -  & $384$ & $672$ &  -  &  -  &  -  & $72$ &  -  &  &  -  &  -  &  -  &  -  &  -  &  -  &  -  &  -  &  -  &  -  \\
\texttt{jw047620} &  -  &  -  &  -  & $16$ &  -  &  -  & $4$ & $12$ &  &  -  & $111$ & $64$ &  -  &  -  &  -  &  -  &  -  &  -  &  -  \\
\texttt{jw053980\_gn\_obs290\_291} &  -  &  -  & $24$ &  -  & $24$ &  -  &  -  & $6$ &  &  -  &  -  &  -  &  -  &  -  &  -  &  -  &  -  &  -  &  -  \\
\texttt{jw053980\_gn\_pa220} &  -  &  -  & $352$ &  -  & $352$ &  -  &  -  & $88$ &  &  -  &  -  &  -  &  -  &  -  &  -  &  -  &  -  &  -  &  -  \\
\texttt{jw053980\_gn\_pa227} &  -  &  -  & $32$ &  -  & $32$ &  -  &  -  & $8$ &  &  -  &  -  &  -  &  -  &  -  &  -  &  -  &  -  &  -  &  -  \\
\texttt{jw053980\_gn\_pa230} &  -  &  -  & $96$ &  -  & $96$ &  -  &  -  & $24$ &  &  -  &  -  &  -  &  -  &  -  &  -  &  -  &  -  &  -  &  -  \\
\texttt{jw064340\_gn\_obs081} &  -  &  -  &  -  &  -  & $112$ &  -  &  -  &  -  &  &  -  &  -  &  -  &  -  &  -  &  -  &  -  &  -  &  -  &  -  \\
\texttt{jw064340\_gn\_obs215} &  -  & $64$ &  -  &  -  &  -  &  -  &  -  &  -  &  &  -  &  -  &  -  &  -  &  -  &  -  &  -  &  -  &  -  &  -  \\
\texttt{jw064340\_gn\_pa217} &  -  & $64$ &  -  &  -  & $64$ &  -  &  -  & $16$ &  &  -  &  -  &  -  &  -  &  -  &  -  &  -  &  -  &  -  &  -  \\
\texttt{jw064340\_gn\_pa228} &  -  & $224$ &  -  &  -  & $224$ &  -  &  -  & $56$ &  &  -  &  -  &  -  &  -  &  -  &  -  &  -  &  -  &  -  &  -  \\
\texttt{jw064340\_gn\_pa239} &  -  & $64$ &  -  &  -  & $64$ &  -  &  -  & $16$ &  &  -  &  -  &  -  &  -  &  -  &  -  &  -  &  -  &  -  &  -  \\
\texttt{jw064340\_gn\_pa249} &  -  & $64$ &  -  &  -  & $64$ &  -  &  -  & $16$ &  &  -  &  -  &  -  &  -  &  -  &  -  &  -  &  -  &  -  &  -  \\
\texttt{jw064340\_gn\_pa258} &  -  & $64$ &  -  &  -  & $128$ &  -  &  -  & $32$ &  &  -  &  -  &  -  &  -  &  -  &  -  &  -  &  -  &  -  &  -  \\
\texttt{jw064340\_gn\_pa271} &  -  & $64$ &  -  &  -  & $64$ &  -  &  -  & $16$ &  &  -  &  -  &  -  &  -  &  -  &  -  &  -  &  -  &  -  &  -  \\
\texttt{jw064340\_obs381\_382} &  -  &  -  &  -  &  -  & $176$ & $34$ &  -  &  -  &  & $136$ &  -  &  -  &  -  &  -  &  -  &  -  &  -  &  -  &  -  \\
\texttt{jw064340\_obs413\_414} &  -  & $64$ &  -  &  -  & $64$ &  -  &  -  & $16$ &  &  -  &  -  &  -  &  -  &  -  &  -  &  -  &  -  &  -  &  -  \\
\enddata
\tablecomments{An SCA exposure refers an image from one of the 10 NIRCam detectors in one band; a single NIRCam exposure will consist of 8 SCA exposures in the SW channel and 2 SCA exposures in the LW channel.}
\end{deluxetable*}

\section{JADES NIRCam Reduction Pipeline}
\label{sec:pipeline}
The JADES NIRCam image reductions rely on the \pipe\ pipeline, which consists of three stages.  Stage 1 fits the raw ramp data to produce count rate images.  Stage 2 goes from the rate images to images that are calibrated astrometrically and in flux and have the background subtracted.  Stage 3 resamples and combines images to produce mosaics.  For this release we use pipeline version 1.14.0 with context map \pmap\ from May 2024. The \pipe\ pipeline is continually evolving. To our knowledge, the updates most relevant for NIRCam image reduction since this pipeline version and context map are updated bad pixel maps, updated flat fields, experimental $1/f$ noise corrections and reference pixel trend fitting (which are disabled by default), and some changes to the treatment of cosmic ray jumps early in ramps.

\begin{figure*}[t!]
\includegraphics[width=\textwidth]{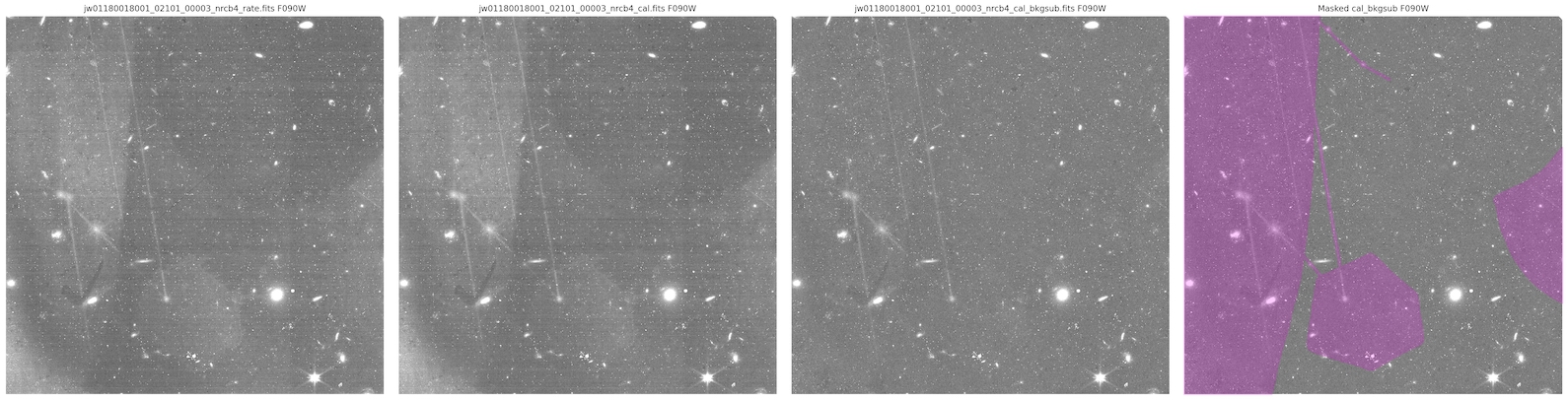}
\caption{Example of a single F090W exposure in the B4 detector at different stages of reduction.  These are: $a$) the rate image, after stage 1; $b$) the flux calibrated, flat-fielded image before subtraction of detector and physical backgrounds; $c$) the flux calibrated, $1/f$ noise corrected, artifact template template and background subtracted image; $d$) the same as $c$) but with masked artifacts indicated with a purple shade.  Note the strong persistence in the left side of the detector including the track of a bright star through a dither pattern.
\label{fig:gif_f090w_b4}}
\end{figure*}

\begin{figure*}[t]
\includegraphics[width=\textwidth]{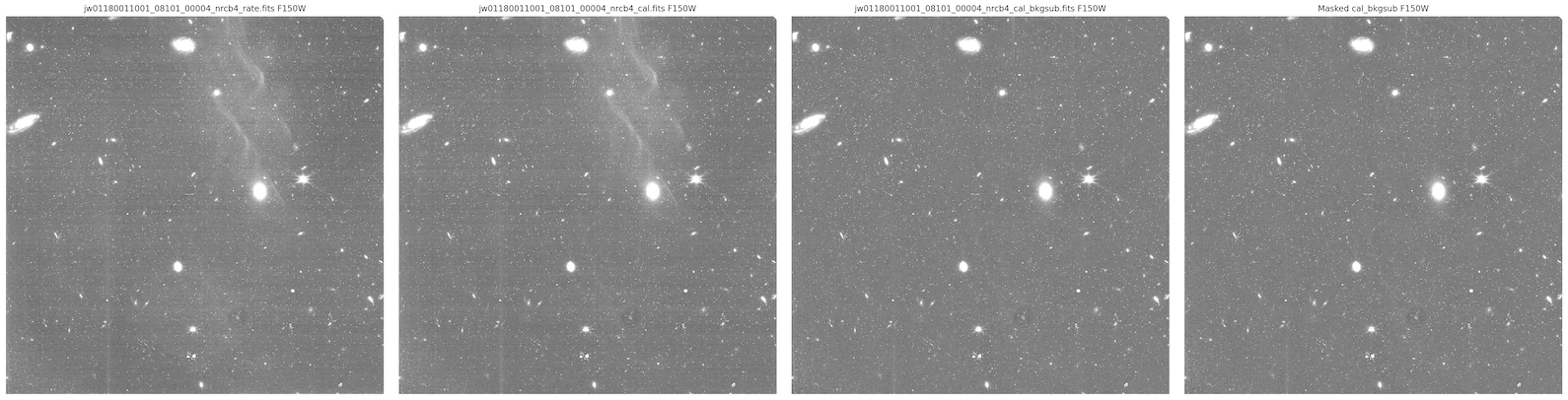}
\caption{Same as Figure \ref{fig:gif_f090w_b4} but for F150W.  Note the strong wisp feature in panels a) and b).
\label{fig:gif_f150w_b4}}
\end{figure*}

\begin{figure*}[t]
\includegraphics[width=\textwidth]{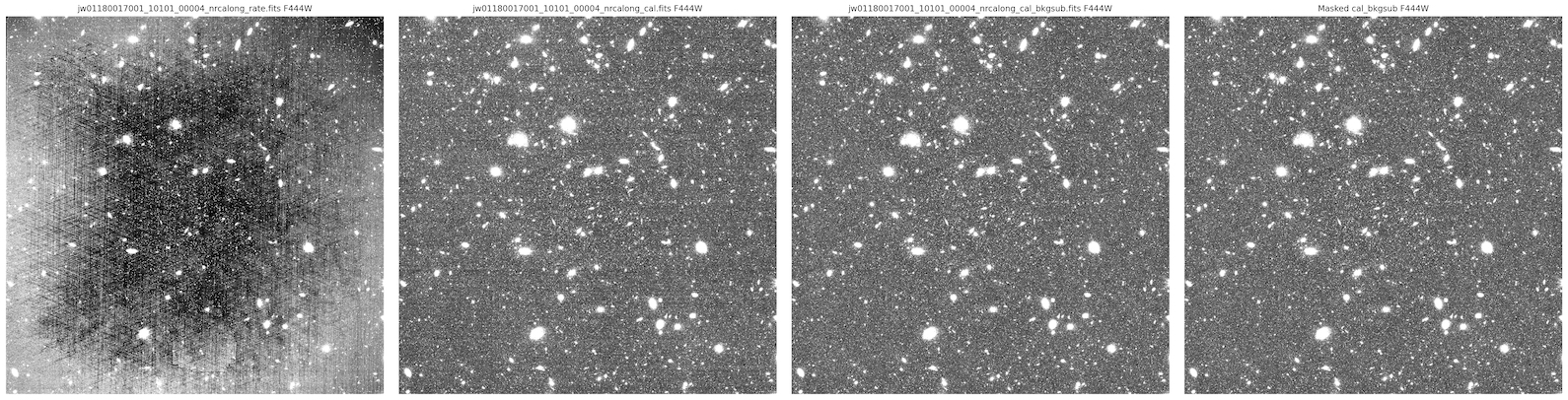}
\caption{Same as Figure \ref{fig:gif_f090w_b4} but for an F444W exposure in ALONG.
\label{fig:gif_f277w_along}}
\end{figure*}

Example images at different parts of the stage 1 and stage 2 reduction are given in Figure \ref{fig:gif_f090w_b4} (F090W in the B4 detector), Figure \ref{fig:gif_f150w_b4} (F150W in the B4 detector) and Figure \ref{fig:gif_f277w_along}.  These show exposures after stage 1 ramp-fitting (\S\ref{sec:stage1}), after flat fielding (\S\ref{sec:flats}), after subtraction of additional additive effects including $1/f$ noise and astrophysical and detector backgrounds (\S\ref{sec:1overf} and \S\ref{sec:exposure_bg}), and with an exposure level artifact mask (\S\ref{sec:exposure-masking}) indicated.

\subsection{Ramps to Rates}
\label{sec:stage1}
In Stage 1 the slopes of the raw data ramps are fit resulting in count-rate images. We make only minor modifications to the default parameters of the \pipe\ pipeline for this stage.  These changes include increasing the threshold for the detection of cosmic ray jumps, changing the requirements for triggering large area `snowball' flagging, and employing an intra-pixel-capacitance (IPC) correction that is normally turned off.
We also generate saturated pixel maps from the DQ layer of the individual ramp images for use in later processing, since the saturation flag is normally reset after jump-detection and rate measurement.

\subsubsection{Crosstalk Correction}
\label{sec:xtalk}
Crosstalk in the readout electronics can cause the electrical signal from pixels read out in one amplifier to be affected by the signal being read out at the same time from a corresponding pixel in another amplifier, leading to artifacts in the images.  This effect was measured on the ground for each amplifier pair of each of the NIRCam detectors\footnote{\url{https://www.stsci.edu/files/live/sites/www/files/home/jwst/documentation/technical-documents/_documents/JWST-STScI-004361.pdf}}
, with the fractional amount of cross talk between amplifiers of order $10^{-5}$ to $10^{-4}$ of the original signal. There is both a prompt component affecting pixels read out at the same time and another component affecting the subsequently read out pixel, usually of approximately equal amplitude and opposite sign.  The small amplitude of this effect and the self-cancelling nature means that it is only noticeable when the dynamic range of an image exceeds 10 magnitudes and when the source of the effect is highly spatially concentrated.  Moreover, the different readout directions of adjacent amplifiers implies that many affected pixels do not remain consistent in sky coordinates between dithers, and may be flagged during mosaic outlier rejection.

Nevertheless, in the deep JADES imaging it became clear that spurious signal in mosaics close to the detection limit with tell-tale adjacent positive and negative components in LW channel mosaics was being induced by crosstalk from the centers of bright (magnitude $\sim 20$) galaxies.  We therefore used the ground-measured crosstalk coefficients to predict and subtract this crosstalk effect on every exposure.

We note that this correction is only approximate; the effect should properly be addressed in the ramp data, before jump detection and rate computation.  Furthermore, the small amplitude of the effect allows us to treat the problem simply, but in general crosstalk is complicated by the lack of knowledge of the true source signal as every pixel is to some degree affected.  For these reasons, and due to missing ground measurements for one detector, we also added an additional uncertainty equal to the predicted crosstalk effect to the read-noise variance layer of the individual exposures. We also masked the sympathetic pixels of saturated pixels. This crosstalk correction was found to remove the spurious signal in the mosaics, without introducing new spurious features.  However, when there are not enough widely separated dithers, areas sympathetic to large groups of saturated pixels may be lost.

\subsubsection{Saturated Pixel Persistence Corrections}
The NIRCam IR detectors suffer from some degree of persistence, charge that is `trapped' in pixels and later slowly deposited as signal, potentially in subsequent exposures.
In the case of cosmic ray persistence, we use the saturated pixel masks to flag pixels in subsequent exposures taken within 3600 or 1800 seconds and in the same visit, with the larger values used for the SW detectors more strongly affected by persistence.

\subsubsection{Bad Pixel Masks}
\label{sec:bpms}
The JWST near-infrared detectors are experiencing a moderate increase over time in the number of pixels with high dark current, so-called hot pixels. The bad pixel masks from \pmap\ are based on data taken during the instrument commissioning period and therefore do not contain all the hot pixels at the time of our observations. We supplement these masks with additional bad pixels identified from NIRCam dark exposures obtained in cycles 1 and 2. The processing to identify hot pixels followed the approach of \citet{cooper2023}.
In addition, we use the `jump' files produced during the ramp-fitting to mask any pixels that saturated within the first 3 groups, as rates derived from such shortened ramps are unreliable.

\subsection{Calibrated exposures}
The next reduction stage involves photometric calibration of the rate images, assignment of world coordinate system (WCS) information, and background subtraction to prepare the individual exposures for combination into a mosaic.  We again use the \pipe\ pipeline (version 1.14.0 with context map \pmap). Here we describe changes from the pipeline and reference file defaults, along with custom steps applied after the stage 2 pipeline is run but before mosaicing.

\subsubsection{Custom LW Channel Flat Fields}
\label{sec:flats}
The NIRCam detectors have substantial spatial structure in the pixel-to-pixel sensitivity variations on a variety of scales, which must be accurately calibrated and corrected via flat fields.  Early in the analysis of JADES images, it gradually became clear that the available ground-based flat fields for the LW channel contained correlated meso-scale errors in some regions of the detectors \citep{rieke23r, eisenstein2025}.  These appeared as spurious background structure in deep mosaics, especially those constructed from only mildly dithered data.
This ``weave'' pattern complicates source detection based on LW stacks (e.g., \brantspaper).  We therefore developed techniques to generate flat fields for the LW filters from the observed data themselves, by median-combining the available, homogeneously reduced, sparse field images after masking sources found in deep combinations of the available LW data.  While updated flat fields generated in a similar manner have been made available through the Calibration Reference Data System (CRDS),
we have continued to update and use our custom ``sky flats'', which have the advantage of being generated from data reduced through stage 1 in exactly the same way as images used for the released mosaics.
Their construction is described in detail in Appendix \ref{sec:sky-flats}.

For the SW channel we use the default flat fields available through CRDS as of \pmap\ (but see \citealt{Wu26} for discussion of potential SW flat field discrepancies).

\subsubsection{Reduction of $1/f$ Noise}
\label{sec:1overf}
The JWST near-infrared detectors incur significant $1/f$ or pink noise during readout leading to correlations between sequential pixel reads on a variety of scales \citep[see e.g.][for further discussion]{rauscher2012, schlawin2020, rauscher2024}.  While this noise is incurred in the individual data reads constituting ramps, it can lead to clear stripes in the horizontal/row (and vertical/column) directions in the resulting rate images.  A number of strategies to remove or mitigate this noise exist \citep[e.g.][]{bagley2023, schlawin2020, rauscher2024} operating either on the raw ramp data or on rate or even calibrated images, and designed for different kinds of data such as sparse fields or diffuse emission. We build on the techniques based on estimating and subtracting row and column mean or median amplitudes in each amplifier.

First, we construct a segmentation map of the calibrated exposures (the \texttt{\_cal.fits} files generated by by the stage 2 pipeline) and use this as a source mask.
For the wisp affected detectors (A3, A4, B3, and B4; see \S\ref{sec:exposure_bg}), we subtract a scaled, filter-specific single-component preliminary wisp template from the wisp susceptible regions after removal of an exposure-wide background estimated from source and wisp-free regions.
We then use \texttt{Background2D} from \texttt{photutils} to estimate and subtract a background on an $8\ \times 8$ point grid (i.e., a $7.7''$ scale for SW and $15.4''$ for LW), ignoring pixels in the source mask or clear outliers.
Next we fit a model for the $1/f$ noise to the residual flux in pixels that are not part of the source mask, reference pixels, or identified as large outliers. This model is of the form
\begin{eqnarray} \label{eqn:1overf}
\delta_{m, x, y} &= a_{m, y} + b_{x} + c_m + \epsilon
\end{eqnarray}
where $\delta_{m, x, y}$ is the residual flux in pixel $x, y$ belonging to amplifier $m$,
$a_{m,y}$ is the coefficient for the 512 pixels of row $y$ in amplifier $m$,
$b_{x}$ is the coefficient column $x$,
$c_m$ is a constant pedestal for amplifier $m$, and
$\epsilon$ is uncorrelated Poisson noise.
In total this model has $4\times 2044 + 2044 + 4 = 10,244$ parameters.  We fit using stochastic gradient descent within \texttt{tensorflow} on GPU-enabled nodes.  The resulting image of $\delta$ is subtracted from the calibrated image and recorded as the $1/f$ model prediction for each exposure.

Difficulties in this method arise when a masked source is a large fraction of the size of an amplifier, leaving $a_{m,y}$ poorly constrained, or when source outskirts are not sufficiently masked.  Bright diffraction spikes aligned with columns or rows may not be well masked by the segmentation and therefore bias the coefficients. Structure in the background smaller than the 256 pixel wide grid we use (e.g. due to persistence) may bias the coefficients.  The $1/f$ noise might be better fit and subtracted from countrate images or, indeed, from readout ordered ramp data before slopes are fit. Furthermore, the column- and row-wise constant model is an approximation to the true multi-frequency nature of the noise.

\subsubsection{Wisp, Persistence, and Background Subtraction}
\label{sec:exposure_bg}
Scattered light artifacts known as `wisps' can affect some SW detectors (A3, A4, B3, and B4), particularly at long wavelengths \citep[e.g.,][]{Robotham2023}. In one instance in GOODS-S pure parallel imaging particularly strong wisps also affected the LW images.

The wisps are largely stable in morphology in detector coordinates for a given filter.  Some variations are however present, which can lead to residual scattered light artifacts when using single parameter scaled wisp template images.  For this reason we have fit and subtracted linear combinations of multiple wisp templates from the affected areas of each image.  The derivation and application of these multi-component wisp templates is described in \citep{Wu26}.
Briefly, for wisp affected detectors and bands a linear combination of multiple templates is fit to the calibrated data in source-free wisp affected regions after subtraction of 1/f noise and a single median background value determined for the entire exposure. This yields a wisp model for each exposure, and an associated wisp flux uncertainty.

As described in \citet{eisenstein2023}, early JADES data were strongly affected by persistence \citep{leisenring2016} from a bright target observed just prior (the Trapezium cluster in the Orion nebula). In contrast to the small scale persistence from cosmic rays, wavefront sensing observations, or bright stars passing through the FOV during other observations, this persistence was large scale (affecting significant fractions of the A3, and B4 chips as well as $\sim 100 \times 1000$ pixel regions near the edges of B2 and B3, and a circular feature in A1). It was clearly visible (though fading) hours after the initial JADES observations. While the observations affected by Orion were the most severely impacted, we have found evidence for this large scale persistence in subsequent JADES and other program data. To mitigate this large scale persistence we constructed persistence templates for each of the affected detectors (A1, A3, B2, B3, and B4).  While these templates are approximate --- the timescale for the decay of persistence is likely pixel dependent, meaning that the morphology of the released flux will change with time --- they were found to improve the subtraction of this `detector background' in multiple instances.  A persistence model is generated for each exposure by masking sources, subtracting any wisp model, fitting and subtracting a 2D background, and then finding the amplitude of the template that best matches the data in unmasked, wisp-free regions. We note that some regions of the A3 chip are affected by both wisp and persistence, making template construction and fitting more involved.
The brightest large scale persistence artifacts were masked (\S\ref{sec:exposure-masking}).

The final background-subtracted calibrated exposures are generated by subtracting the $1/f$ noise prediction, any wisp model or persistence model for applicable detectors and bands, and finally a 2D background. This background is estimated with \texttt{photutils.background.Background2D} after masking sources and extreme outlier pixels.  For SCA exposures where a wisp or persistence template was fit this background is estimated on $16 \times 16$ box grid (128 pixel box size), median filtered, and then interpolated to the original pixel scale.  For other SCA exposures we use the original background estimated during $1/f$ fitting (\S\ref{sec:1overf}) and computed on a $8 \times 8$ box grid (box size of 256 pixels).

\subsubsection{Artifact masking}
\label{sec:exposure-masking}
We have visually inspected every exposure used in these DR5 mosaics and generated exposure-level masks for each image for which scattered light or other artifacts could not be adequately modeled and subtracted using the templates described above. In practice, this visual inspection is achieved by animating dither sequences with full size GIFs, as the artifacts largely stay fixed in detector coordinates while the physical scene moves in these coordinates.

In SW bands these masked artifacts include `claws', several instances of `dragon's breath', rare cases of strong cosmic-ray streaks that were not caught or persisted into subsequent exposures, instances of very strong large scale persistence from previous programs with high levels of background, and smaller scale persistence from bright stars in previous observations or from the wavefront sensing observations. In one deep pointing of the SAPPHIRES program in GOODS-N extremely strong wisps with a different morphology than typical were masked. 

The LW images were largely free of artifacts. As mentioned in \citet{eisenstein2023}, some of the 1180/Medium observations in GOODS-S were taken parallel to NIRSpec observations which suffered an electrical short.  The high amplitude and ringing pattern of the background rendered the LW data unusable, and those exposures were removed entirely from the processing. A small circular pattern that occasionally appeared in F277W at variable locations was also masked, as well as $\sim 200 \times 100$ pixel wide `blobs' at the edge of LW B module images in the 1210 and 3215 imaging.  Our stage 1 reductions of the \texttt{jw039900\_pa3558} LW data (from observations 17-20) yielded atypical background values and structure, and this data, comprising $\sim 4$ hours in each of 3 filters, was masked.  Finally, the extremely strong wisps in the deep GOODS-N pointing of the SAPPHIRES program were accompanied by scattered light in the LW bands that was also masked.

Some artifacts are difficult to see in the sequences of individual exposures, but may appear in stacks of mildly dithered exposures.  These include the faint edges of scattered light features, fading large scale persistence, or transient features near the edge of the detector. Such artifacts are identified and masked in intermediate mosaics after visual inspection (\S\ref{sec:submosaic-masks}).

\subsubsection{Astrometric Alignment}
\label{sec:astrometry}
Before images are combined to generate deep mosaics, they must be aligned with each other.  The approach of the JADES team is to compute and apply a single translation and rotation correction for all images of a given module and visit, under the assumption that the guider, fine-steering mirror, and distortion maps are at least as accurate as a correction that could be computed on an image-by-image basis in sparse fields.
To accomplish this we use module-level catalogs generated with \texttt{SourceExtractor} \citep{SourceExtractor, barbary2016} for every dither of two SW bands (when available), compute a shift and rotation relative to a global reference catalog (see below), and median the results before applying this single correction to all images in that module and visit.

Where we can check we have found this approach to work well with the distortion maps delivered to CRDS in February 2024, providing relative alignment to better than a few mas across different bands and dithers (\S\ref{sec:qa-align}) with only a few exceptions.  When substantial residuals were found for individual dithers ($\gtrsim 0.5$pixels) we compute and apply corrections on a per dither basis; this includes one repeated dither position in one visit of the 1210 program, as well as in much of the ancillary parallel program data taken in Cycles 1 and 2 (the PANORAMIC and BEACONS programs), often taken while not in fine guiding mode. For F162M (which is in the pupil wheel while F150W2 is in the filter wheel), we additionally adjusted the inter-SCA spacing for each module by fixed values relative to the default WCS assigned by the \pipe{} pipeline with \pmap.

The absolute reference catalog against which we compute these corrections is generated from a large number of individual F200W (or other SW band if F200W is not available) exposure-level catalogs. The true position of every reference star is fit along with a shift and rotation of every visit, as well as inter-module offsets.  The overlaps between different programs and different PAs are used to break the various degeneracies and explore any residual distortions, while the inclusion of a number of Gaia stars with known and proper-motion corrected positions is used to tie the entire catalog to the Gaia DR3 reference frame. More details are given in Appendix \ref{sec:refcat}.  This is different than previous JADES releases, where the reference catalogs were constructed from the CHaRGE HST mosaics (G. Brammer priv. comm.).

\subsection{Mosaicing}
\label{sec:mosaicing}
We use the \pipe\ pipeline to resample and combine images into mosaics.  This relies on the \texttt{drizzle} algorithm to redistribute flux from the input images into the output mosaics.  During this process we also use the multiple measurements of each location to identify outliers that have not otherwise been flagged.

\subsubsection{Mosaic Image Groups}
\label{sec:subregion}
When combining exposures in mosaics, we first group images by PID, epoch and position angle. We call these groups ``subregions''.  These different subregions are listed in Tables \ref{tbl:gs_subregions} and \ref{tbl:gn_subregions} for GOODS-S and GOODS-N, respectively. For bands in the the LW channel, we also group the images by module.

The diversity of programs used in the full mosaics leads to a wide range of position angles, which if included simultaneously in the imaging would lead to a complicated, spatially variable effective PSF with many diffraction spikes. The grouping of images by position angle (we allow about $\pm1\deg$ variation within a group) helps to simplify the final PSF and minimize its spatial variations across the subregion mosaic. As we will see, when there are multiple PAs covering a region, this also allows us to recover area under the diffraction spikes of bright stars.

The grouping by epoch allows us to detect transients \citep[e.g.][]{decoursey2025}.  It also allows us to identify high proper motion sources \citep[e.g. brown dwarfs][]{hainline2024b, hainline2026a}.  The grouping of the LW images by module was intended to allow mitigation of any differences in the effective response between the modules due to the different blue and red cutoffs of the ALONG and BLONG detector sensitivities.

The disadvantage of these image groupings is that fewer images are available for constructing a robust median against which outliers can be detected, and there is less diversity in sub-pixel phase and position angle with which to improve the PSF sampling and spatial resolution of the full mosaic.

\subsubsection{Outlier Detection}
\label{sec:outlier-detection}
Multiple dithers covering the same location on the sky can be used to identify outliers.  Such a procedure is implemented in the default \pipe\ pipeline as the \texttt{OutlierDetectionStep}, which compares individual exposures to a projected median image. For the bands in each channel with the poorest sampling of the PSF by the detector pixels, this step with default parameters was found to occasionally flag pixels in the cores of bright stars and compact objects.  This is because small changes in the pixel phase of the centroid can lead to large changes in pixel brightness.  We implemented a modified version of this step that uses custom band-by-band thresholds on a Laplacian filtered version of the median image.  These thresholds were determined from a suite of noiseless, model PSFs placed in the exposure with a variety of sub-pixel phase distributions.

This outlier rejection is different than was used in the processing of NIRCam images in previous JADES releases. We find that this new, less aggressive outlier detection can lead to flux increases up to $\sim 0.1$ mag in bright compact sources --- stars and nuclear dominated galaxies --- in the most undersampled bands (e.g., F070W, F090W, and F277W) with the amplitude of the effect decreasing towards the redder filters in each channel.
This issue affects the standard \pipe\ pipeline processed mosaics (D. Law, priv. comm.).

\subsubsection{Weighting and Resampling}
\label{sec:submosaic-weighting}
We use the \pipe\ pipeline resampling step to combine individual exposures in a given sub-region image group into mosaics onto a subset of the same pixel grid as the full mosaic.  This step relies on the \texttt{drizzle} algorithm. In most cases we first run the \pipe\ pipeline exposure background matching step (\texttt{jwst.skymatch.SkyMatchStep}), though the recovered offsets are extremely small given the background subtraction we have already done, and the step is disabled in rare cases where erroneous offsets led to entire exposures being flagged as outliers.

When combining individual exposures into a subregion mosaic, we weight them by their inverse read noise. While most of the exposures used in our mosaics are background noise dominated, we choose to use inverse read-noise variance weighting of the individual exposures as this is the only weighting option in the current \pipe\ pipeline that allows for different pixel weights for ramps shortened by detected jumps\footnote{We have developed a weight map for each exposure based on the combined read noise and Poisson variance in each pixel due to a uniform background and the valid number of groups in the ramp for that pixel. Use of these maps, which incorporate flat field and exposure time variations between pixels, will be implemented in future versions.}. Indeed, the read noise variance scales approximately as the inverse of the cube of the exposure time. We note that for most images in one filter of a subregion image group the exposure times and median background variance for a given band are generally comparable, such that the read noise weighting largely serves to severely down-weight any pixel with a shortened ramp.

For each of the GOODS-S and GOODS-N regions the subregion mosaics are resampled onto a WCS with the same pixel scale ($0.02999476''$ in GOODS-S and $0.0300021''$ in GOODS-N) and tangent point celestial coordinates ($\alpha=53.1227811076$, $\delta = -27.8051604556$ for GOODS-S and
$\alpha=189.22861342627$, $\delta= 62.2385675278$ for GOODS-N).
The coordinate system is ICRS. These pixel scales are close to the native SW channel detector pixel scale and were chosen to match empirical estimates of the archival HST image mosaic pixel scales.  Using this pixel scale for the LW filters as well dramatically simplifies aspects of the photometric processing.
Drawbacks of this choice include undersampling of the PSF in the bluer SW filters, and deconvolution noise in the highly oversampled LW filters.   Given the diversity of observing patterns (dither patterns and number of images) among the different input programs, we chose to keep the default \texttt{drizzle} pixel shrinkage parameter \texttt{pixfrac}$=1$ (i.e., no shrinkage). This affects the output PSF (\S\ref{sec:mpsf}).  In future versions we will explore additional pixel scales and pixel shrinkage parameters.

When generating these subregion mosaics we automatically split the output subregion image into a number of overlapping tiles of smaller area.  This tiling decreases memory usage and allows us to scale the same algorithms from single pointing mosaics to larger or deeper mosaics with hundreds of (potentially overlapping) contributing exposures.  The individual tiles are then snapped back together, discarding the overlapping regions.

\subsubsection{Mosaic Background Subtraction}
\label{sec:submosaic-bg}
For each tile of the subregion mosaics we estimate and subtract a 2D background.  This background is in addition to the exposure level background subtraction described in \S\ref{sec:exposure_bg}.  The method we use is similar to the method described in \citet{bagley2023}.  Briefly, the mosaic is first median smoothed with a thin ring filter of radius $\sim 2.4"$.  An iterative, multi-scale source detection defines a source mask. Each iteration convolves the image with a successively smaller Gaussian, detects successively smaller groups of connected pixels $>3\sigma$ above the background, and dilates the map of the resulting groups before the next iteration.
This combined source mask is applied before making a bi-weight estimate of the background in $10\times10$ pixel (0.3'' side) boxes, which is then $5\times5$ median filtered and spline interpolated to the original pixelization.
By design, the faint outskirts of large, bright galaxies are subsumed into the background.

\subsubsection{Additional Layers}
\label{sec:submosaic-layers}
We add several additional image layers to the default output of the stage 3 \pipe\ pipeline.
First, we compute per-pixel exposure time maps.  These are estimated from a simple sum of the exposure times of the images that contribute to a pixel, and thus do not account for relative weighting in the drizzling algorithm or shortened ramps (due to cosmic rays) for individual pixels in the contributing images.  These exposure times are stored in the \texttt{EXP} extension.
Next, we record the number of exposures that contribute to each pixel in the \texttt{NIM} image extension.  This number can be useful in evaluating the robustness of image features; empirically we find that compact, bright, single band artifacts are much more likely when \texttt{NIM}~$ \leq 3$.  This layer is also used to record mask information at the subregion level (\S\ref{sec:submosaic-masks}).  Both of these layers are propagated to the final, combined mosaics for GOODS-N and GOODS-S.

\subsubsection{Subregion masking}
\label{sec:submosaic-masks}
\begin{figure*}
\begin{center}
\includegraphics[width=\textwidth]{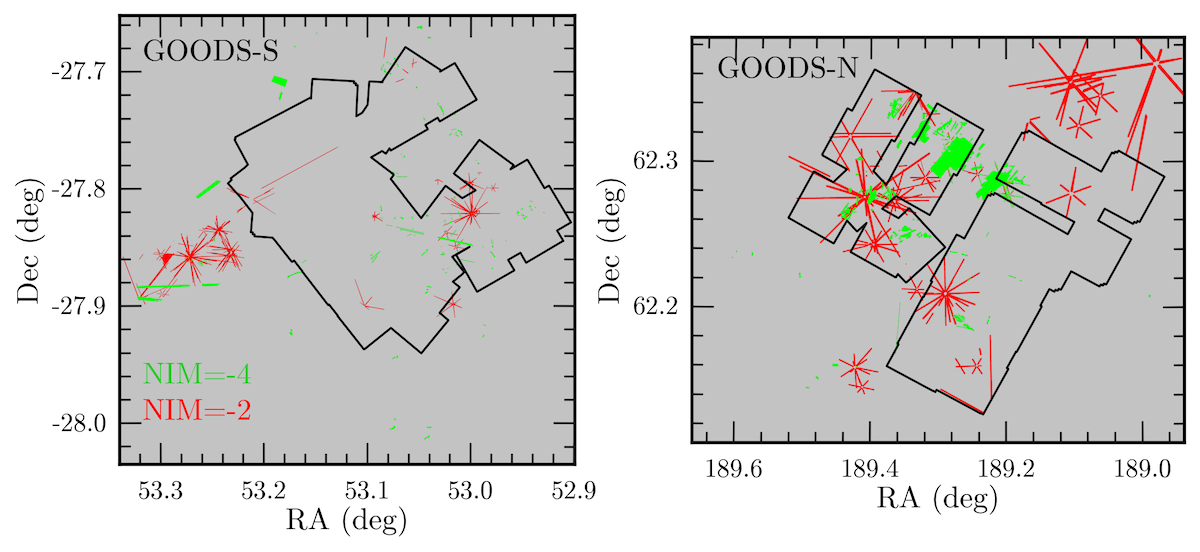}
\caption{Subregion mosaic masks.  These maps show the union of all artifact (\texttt{NIM}$=-4$; green) and diffraction spike masks (\texttt{NIM}$=-2$; red) over all subregions and bands.
\label{fig:nim_layer_mask}}
\end{center}
\end{figure*}

After mosaics for the subregion image groups are made, we visually inspect each one and construct pixel masks that can be used to censor features during the full mosaic construction. These masks are added as negative numbers to the \texttt{NIM} layer. The masked features include artifacts that were not captured during the exposure level masking, often because they were too faint, or because it is easiest to see them in cross band comparisons (e.g. via RGB images). Pixels masked due to these clear artifacts are assigned a value of \texttt{NIM}~$=-4$, and are removed from consideration when combining subregion mosaics into the full mosaic.  Figure \ref{fig:nim_layer_mask} shows the union of these artifact masks across all subregions and bands for GOODS-S and GOODS-N.

\begin{figure*}
\includegraphics[width=\textwidth]{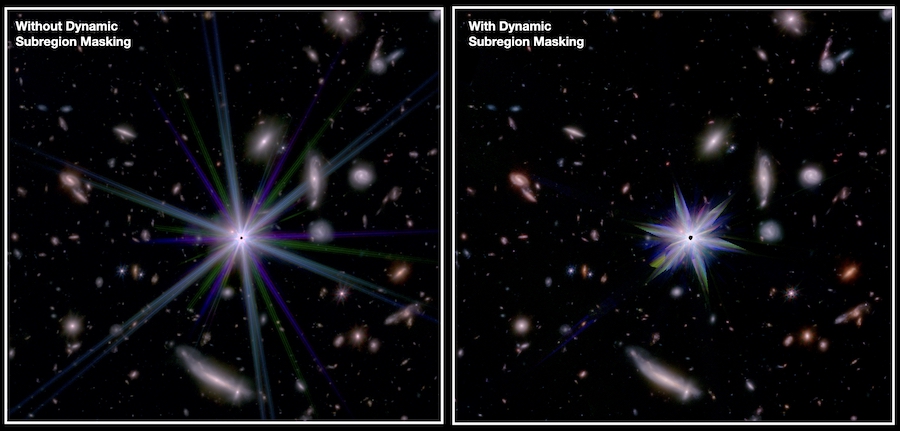}
\caption{Example of the effect of diffraction spike masking in the subregion mosaics. {\it Left}: A color image constructed from mosaics without masking of the diffraction spikes at the subregion level.  Note the diffraction spikes at multiple PAs and in different bands (colors).  {\it Right}: Same, but with subregion masking.
\label{fig:diffraction_spike}}
\end{figure*}

These pixel masks are also used to mark diffraction spikes around bright stars, and are generated by hand through visual inspection.  Pixels masked in this way are assigned a value of \texttt{NIM}~$=-2$, and are only used in the full mosaic if no other subregion has a valid (\texttt{NIM}$>0$) pixel in that band.  Otherwise, they do not contribute to the full mosaic. The union of all diffraction spike masks is shown in Figure \ref{fig:nim_layer_mask} for GOODS-S and GOODS-N. With this scheme, we can effectively remove the diffraction spikes when there are supporting data taken at a different PA with no diffraction spike, as demonstrated in Figure \ref{fig:diffraction_spike}.  However, this can lead to `orphan' diffraction spikes far from the star if the supporting data run out, and, for close stars, the diffraction spike masks at different PAs can interact. For a small subset of bright stars in a subset of filters, we have forced the censoring of their diffraction spikes from the images by setting \texttt{NIM}~$=-4$ in their masks.

\subsubsection{Full Mosaic Generation}
\label{sec:full_mosaic}
Combining the different subregion mosaics for each band into a full mosaic for the GOODS-S or GOODS-N regions does not require further resampling, as they were already sampled onto a pixel grid with a common tangent point and pixel scale.  The full GOODS-N mosaics are $40600 \times 33400$ pixels (20.3$'$ by 16.7$'$), while the GOODS-S mosaics are $46700 \times 46000$ pixels (23.35$'$ by 23$'$).
The size of the mosaics is chosen to encompass the JADES GTO data, the full subregion of any data that overlaps JADES (excluding some SW-only SAPPHIRES or shallow 3-band POPPIES data which extends off the northwest edges of the GOODS-N mosaic), and very nearby deep, multiband fields provided by PIDs 1283, 2079, 6511, and 2514.

When combining the different subregion mosaics into the full mosaic we weight each input subregion mosaic by its \texttt{WHT} map after re-normalizing the \texttt{WHT} map by the inverse of the median Poisson and read noise variance in background regions.  This gives a relative weight to each pixel within a contributing subregion mosaic that accounts for shortened ramps, but the relative weights \emph{between} different subregion mosaics are inversely proportional to the average background plus read noise variance.  These weights are used to propagate the \texttt{SCI}, \texttt{ERR}, and \texttt{WHT} layers to the full mosaic.
The \texttt{EXP} and \texttt{NIM} layers are accumulated while accounting for subregion mosaic masks.  We do not attempt any additional outlier rejection during this combination, so as to preserve transients.  However, this does mean that a rare set of artifacts make it into the full mosaics (\S\ref{sec:remaining}), particularly from subregions that don't have enough dithers to robustly detect pixel-level issues in earlier processing.

\begin{figure*}
\includegraphics[width=\textwidth]{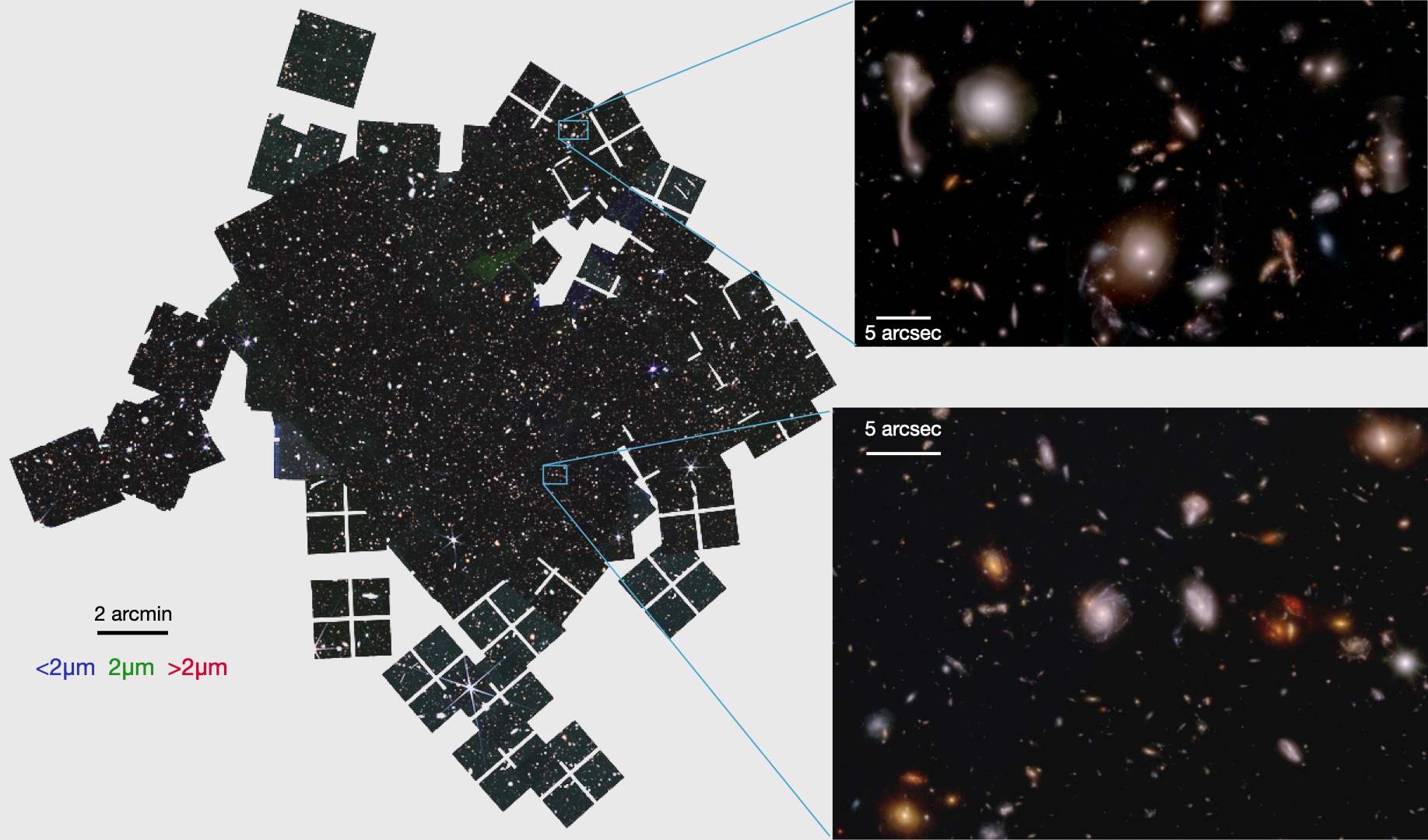}
\caption{RGB image of the GOODS-S field, including the combined mosaics of several bands in each color channel.  The green channel includes F200W and F210M data.  Only pixels with valid data in all of the channels are shown.  Panels on the right show selected regions at higher resolution.
\label{fig:gs_mosaic_rgb}}
\end{figure*}

\begin{figure*}
\includegraphics[width=\textwidth]{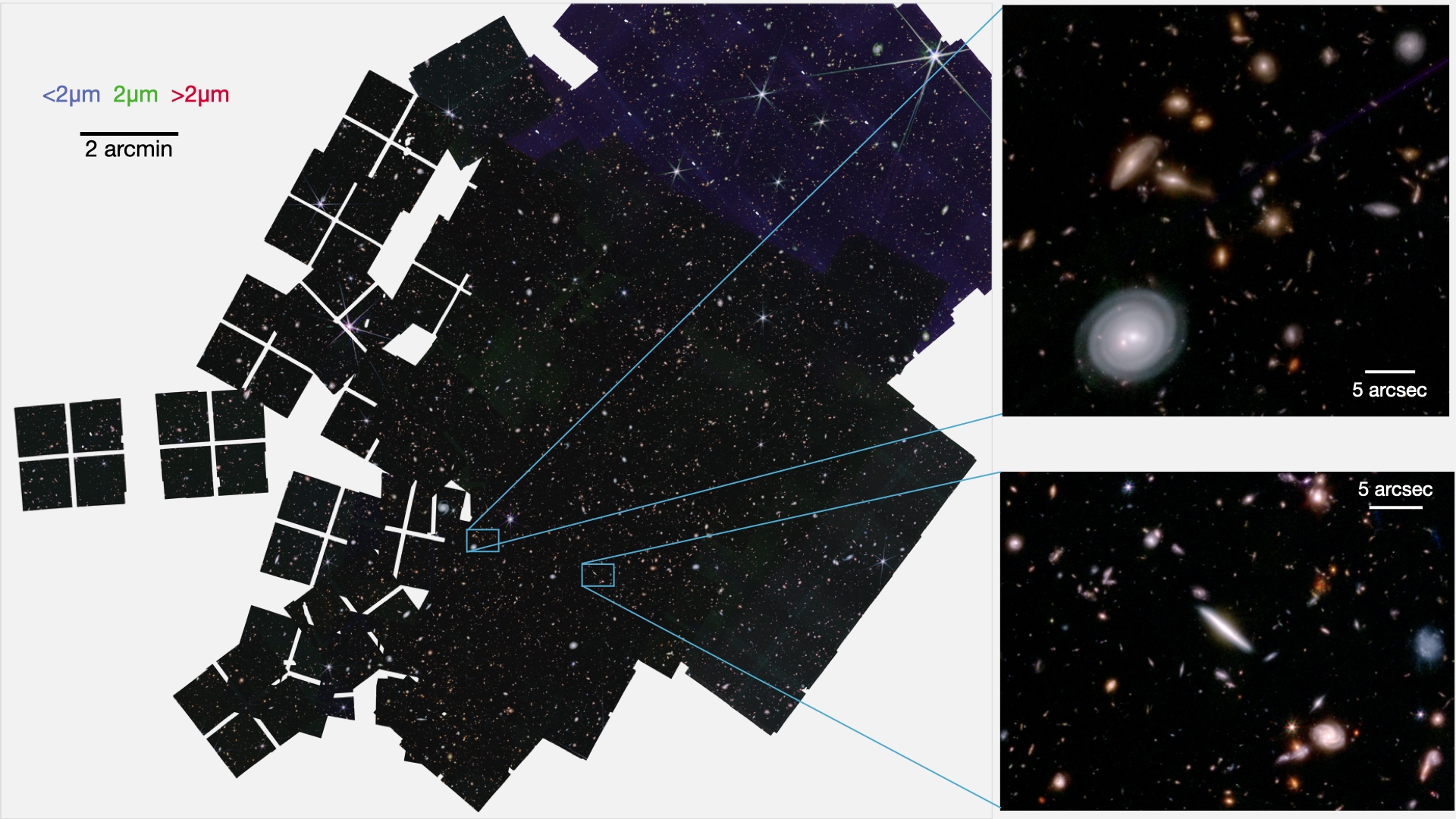}
\caption{RGB image of the GOODS-N field, including the combined mosaics of several bands in each color channel. The green channel includes F200W and F210M data.  Only pixels with valid data in all of the channels are shown.  Panels on the right show selected regions at higher resolution.
\label{fig:gn_mosaic_rgb}}
\end{figure*}

\begin{figure*}
    \includegraphics[width=\textwidth]{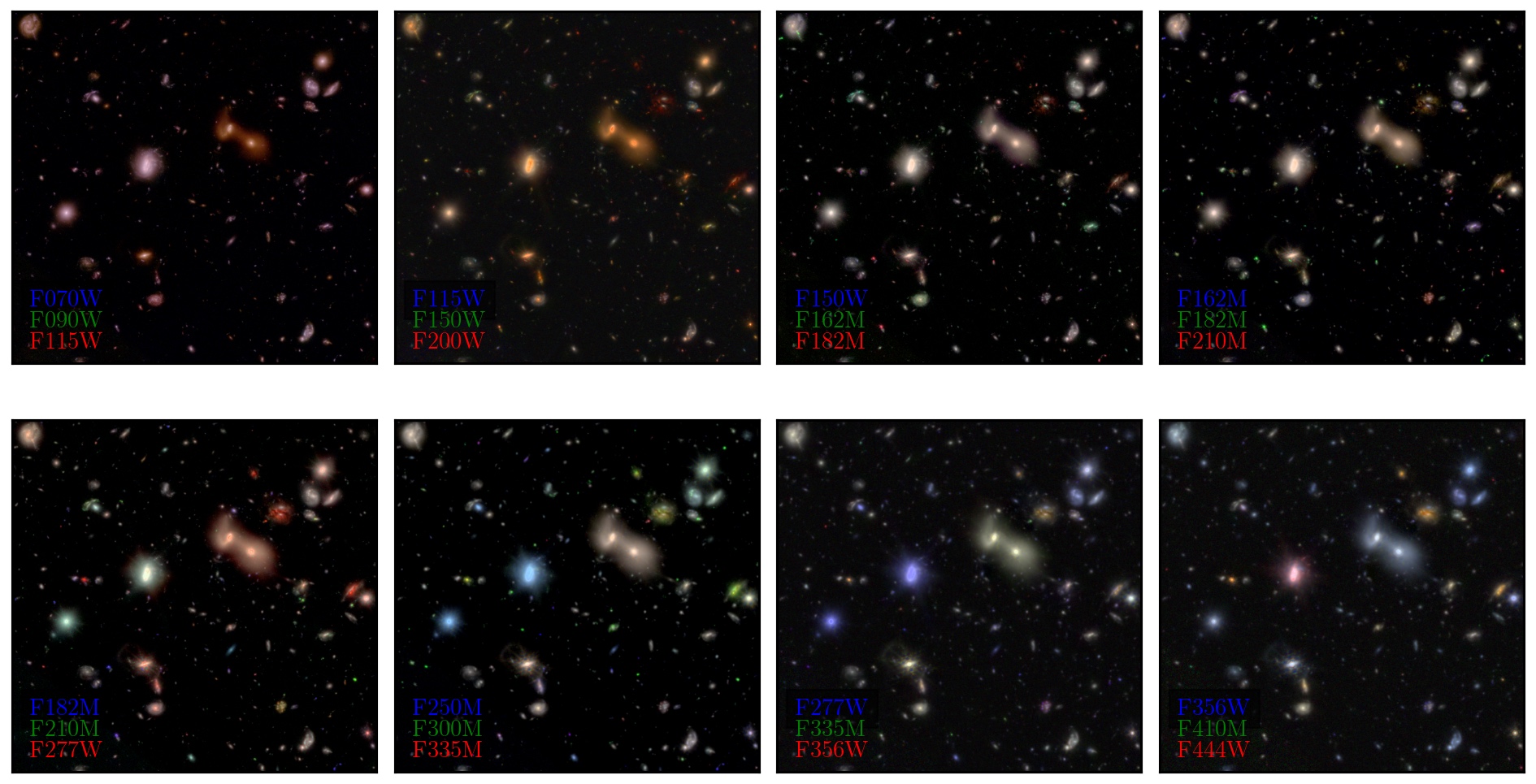}
    \caption{Multi-band coverage in the JOF, showing RGB images of the same region, where the channels are comprised of sequential bands as noted in each panel. The cutouts are 45''$\times$43''.
    \label{fig:multiband}}
\end{figure*}

In Figure \ref{fig:gs_mosaic_rgb} we show RGB images of the entire GOODS-S mosaic, with some interesting regions expanded to show more detail.  We show the GOODS-N RGB in Figure \ref{fig:gn_mosaic_rgb}. In Figure \ref{fig:multiband} we show RGB images of a single region in the JOF constructed from 8 different filter combinations covering 14 different filters, including 7 medium band filters.  Emission-line galaxies are easily identified as mono-color (red, green, or blue) sources in these images.

\subsubsection{Bithash Image}
\label{sec:bithash}
A challenge facing the JWST scientific community involves tracking the origin of all data contributing to the composite mosaics and catalog products generated from many partially overlapping programs, so that appropriate credit can be provided to the programs contributing data to scientific result.
To help clarify which JWST programs contribute to the data products, we have constructed a 32-bit integer
``bithash''
image that encodes the spatial coverage of the contributing programs to our composite mosaics. Each JWST PID contributing to our JWST filter mosaics is assigned a unique bit $b\in[0,30]$, and sometimes more than one to help encode epoch information.
We create a bithash image of the size of our mosaic, and for every pixel a given JWST program covers (in any filter) we add $1<<b$ to the bithash image at that location.
This bithash is propagated to the source catalogs presented in \brantspaper; for each source in the catalog, the value of the bithash image at the source location is recorded in its \texttt{PID\_HASH} field.
The bithash image includes an HDU with a list of the JWST programs corresponding to the bit values, allowing for a pixel or object's \texttt{PID\_HASH} to be decoded, but these bit values are also reported in Tables \ref{tbl:gs_subregions} and \ref{tbl:gn_subregions} for convenience. Bits currently run from 0 to 30.  The JADES GTO programs have bits 0 to 7 as well as 9, while the JADES affiliated programs are bits 8, 10, 12, 16, and 22.

\subsection{Model PSFs}
\label{sec:mpsf}
It is important to quantify the PSF of the resulting mosaics for use in inferring the sizes of objects, for the calculation of aperture corrections to aperture photometry \citep[e.g.][]{robertson2023}, and for creating multiband PSF-matched images through image convolution (e.g, \brantspaper{}). We have constructed model PSFs for each sub-mosaic, using modified versions of procedures described in \citet{ji2024a}
which we release along with the images themselves.
An advantage of the model PSFs is that they are effectively noiseless, even at large angular scales, and can be used even when there are not enough suitable observed stars to determine the effective PSF empirically.

The PSF of individual JWST/NIRCam images can be modeled using the \webbpsf program, which propagates light through the telescope and detectors using Fourier methods and periodic in-flight measurements of optical path distortions derived from the wavefront sensing data.  Efforts to improve the fidelity of \webbpsf \citep[formerly \texttt{WebbPSF}][]{perrin25} have been ongoing, including the effects of high in-flight pointing stability and detector effects such as charge diffusion and inter-pixel capacitance. The NIRCam pixels typically undersample the PSF, and the mosaicing process can therefore alter the final PSF from the individual exposure level PSFs in a way that depends on the dither pattern and the parameters of the \texttt{drizzle} algorithm.

To incorporate the effects of mosaicing in the PSF model, we construct mock exposures where the flux layer of the astrometrically aligned images is replaced with model point-source images constructed with \webbpsf versions 1.2.1 or 1.5.0 (for data taken after Nov 2024).  We used default values for the charge diffusion approximation but disabled the IPC effect as we applied an IPC correction to our data. These point sources are placed at consistent on-sky locations, defined by the HEALPIX grid (with \texttt{NSIDE}$=2^{12}$, leading to $\sim 6$ fake stars per module).  These mock images are then propagated through the same mosaicing process used for the science images. In principle this method can be used to characterize the mosaic PSF at any location on the sky, but we extract an effective PSF for each sub-region mosaic using the \texttt{EPSFBuilder} methods from
the \texttt{photutils} package with the known input centers for the injected point sources.
This method does not include the effect on the output PSF of any astrometric alignment errors between exposures in the same band and subregion.
We refer the reader to \brantspaper{} for more details and an application of these mPSFs to aperture corrections and common PSF images.

\begin{figure}
\begin{center}
\includegraphics[width=0.48\textwidth]{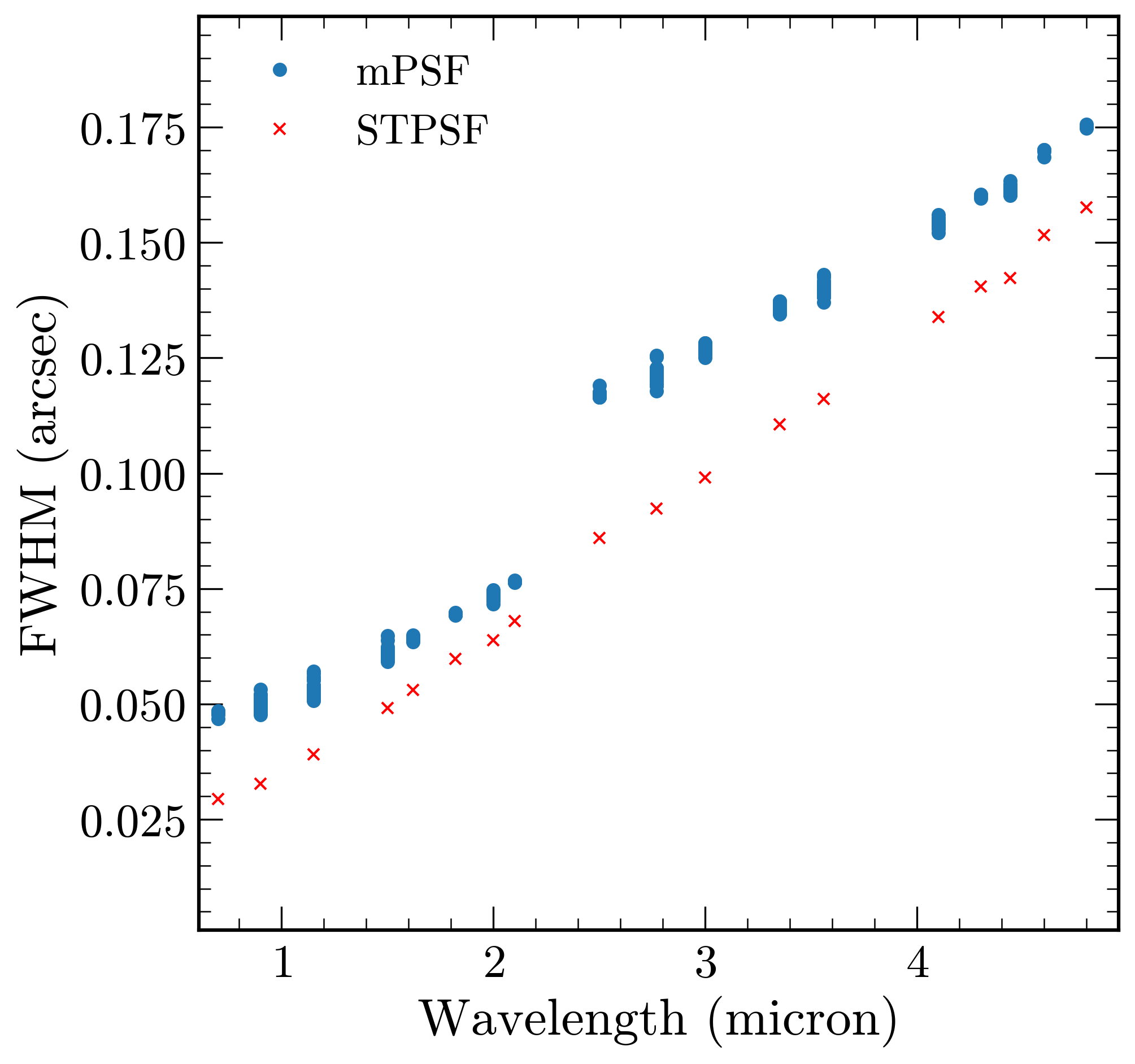}
\caption{The Full Width at Half-Maximum (FWHM) of the model PSFs for every subregion, as a function of wavelength and compared to the native exposure level PSF FWHM. The latter is measured in the same way from PSF images constructed with \webbpsf.
\label{fig:mpsf_fwhm}}
\end{center}
\end{figure}

In Figure \ref{fig:mpsf_fwhm} we show the full width at half-maximum (FWHM) of the  model PSFs in every subregion, as a function of wavelength. These are measured using the \texttt{photutils.psf.fit\_fwhm} method, which takes into account undersampling of the PSF by large pixels.  We also show the FWHM of the native, exposure level PSFs constructed with \webbpsf{} and measured in the same way as for the model PSFs.  We find that the cores of the model mosaic PSFs are broader than the exposure level PSFs, consistent with \citet{ji2024a}.  There is scatter in the model PSF FWHMs that likely arises from different dither patterns.

\section{Image Quality}
\label{sec:image_quality}

\subsection{Astrometric Alignment}
\label{sec:qa-align}
The absolute astrometric accuracy of the full mosaics are ultimately tied to the accuracy of the reference catalog.  See \S\ref{sec:astrometry} and Appendix \ref{sec:refcat} for details of the reference catalog construction, which is ultimately tied to the positions of \emph{Gaia} DR3 stars.

\begin{figure*}
\includegraphics[width=\textwidth]{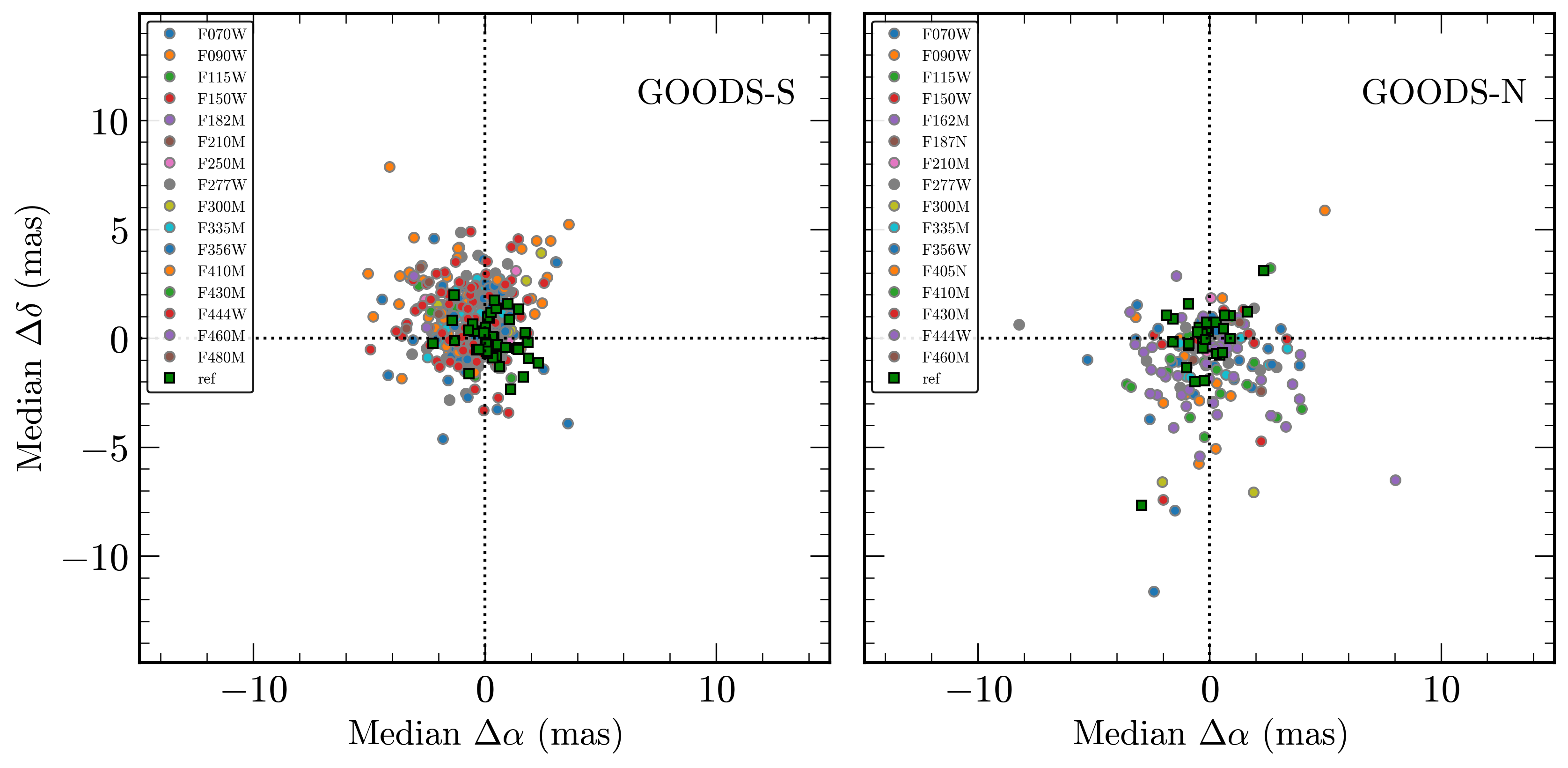}
\caption{Median alignment offsets between bands for every subregion in GOODS-S ({\it left}) and GOODS-N ({\it right}).  Each circle shows the median offset for each subregion between source positions measured in a reference band (usually F200W or an adjacent SW band if that is not available) and another band, indicated by the point color.  There are separate points for each subregion.  The offset median between the reference band positions and the reference catalog positions are shown as green squares.
\label{fig:band_alignment_median}}
\end{figure*}

Here we quantify the internal alignment accuracy of the mosaics in several ways.  First, for each subregion we compare centroid positions measured from the mosaic in each band to those measured in a reference band (usually F200W).  These measurements use the \texttt{XWIN\_IMAGE} and \texttt{YWIN\_IMAGE} windowed centroids computed by \texttt{SourceExtractor} and propagated through the subregion mosaic WCS. This provides an estimate of the degree to which the different bands are misaligned with each other in each subregion.  The median values for each band of each subregion are shown in Figure \ref{fig:band_alignment_median} for GOODS-S and GOODS-N.  We also include in this figure an estimate of the median offset of the reference-band positions from the reference catalog.

Scatter about these median offsets arises due to random measurement error on the centroid positions (increasing for fainter sources) and due to spatial patterns in the offsets, which may be caused by either misalignment of different visits within a subregion or from residual distortion errors within the detectors (see Appendix \ref{sec:refcat} for further discussion of these residual distortions).

\begin{figure*}
\includegraphics[width=\textwidth]{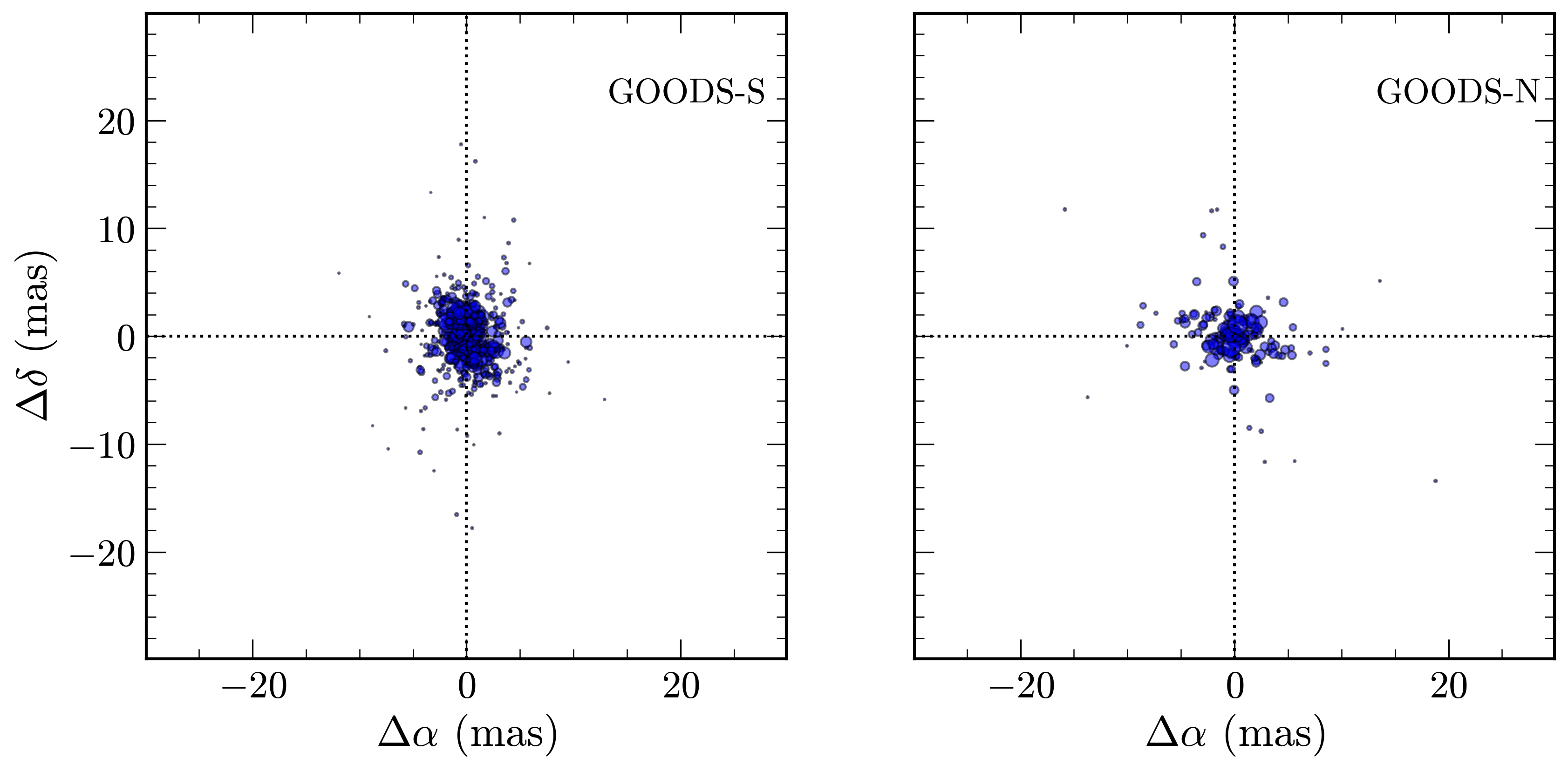}
\caption{Median alignment offsets between overlapping subregions in GOODS-S ({\it left}) and GOODS-N ({\it right}).  Each circle shows the median offset between positions measured in two subregions in a reference band (usually F200W or an adjacent SW band if that is not available), with the point size proportional to the number of matches (and roughly the overlap area).
\label{fig:region_alignment_median}}
\end{figure*}

Next, we compare the measured positions of sources in the reference band of each subregion to the measured position of matched sources in any overlapping subregion.  This quantifies the consistency of the alignment of the subregions across the field. The medians of these offsets are shown for GOODS-S and GOODS-N in Figure \ref{fig:region_alignment_median}.  In many cases the fields only overlap slightly, and there are only a few objects in common at an edge.  While the agreement is generally good, this comparison reveals several subregions that are in tension, with offsets These tend to be subregions with small overlap at differing position angles where there are not many interlocking observations.  Of the 341 subregion pairs with more than 20 matched objects, 280 (82\%) have $< 5$ mas median offsets while 12 have $>10$ mas offsets (but all are $<20$ mas). These 12 nearly all involve one or two pure parallel subregions; the exception is the \texttt{jw011800\_medium\_obs223}/\texttt{jw012860\_dec23} pair with an 18 mas offset in RA in an overlap area of $0.34$ arcmin$^{2}$.

\begin{figure*}
\includegraphics[width=\textwidth]{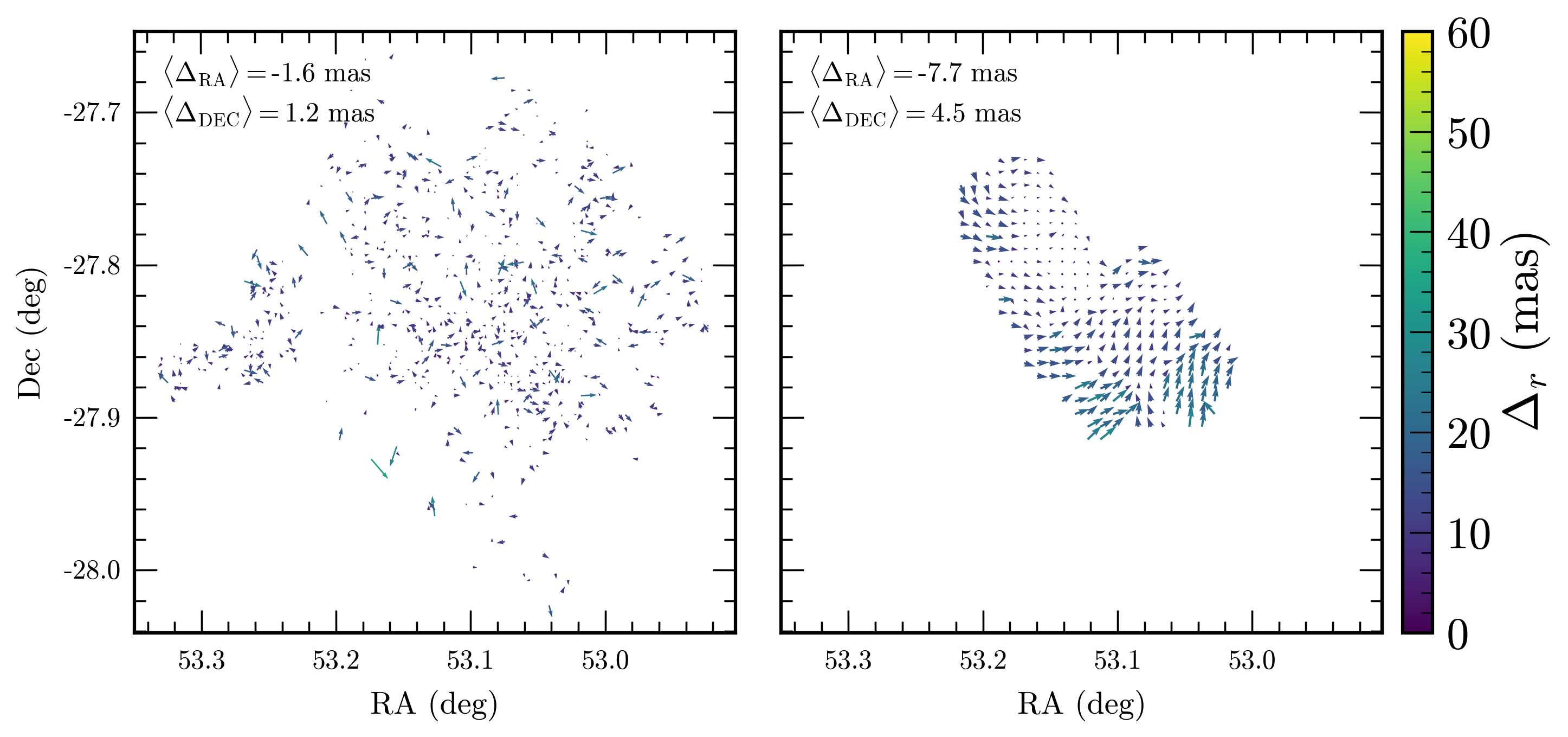}
\caption{Median offsets in 30''$\times30$'' blocks in GOODS-S of the DR5 catalog positions (\brantspaper) based on the mosaics presented in this paper from the JWST-based astrometric reference catalog ({\it left}) and from the previous data release based on HST positions ({\it right}).  We require at least 4 matched objects in each bin.
\label{fig:gs_mosaic_alignment}}
\end{figure*}

\begin{figure*}
\includegraphics[width=\textwidth]{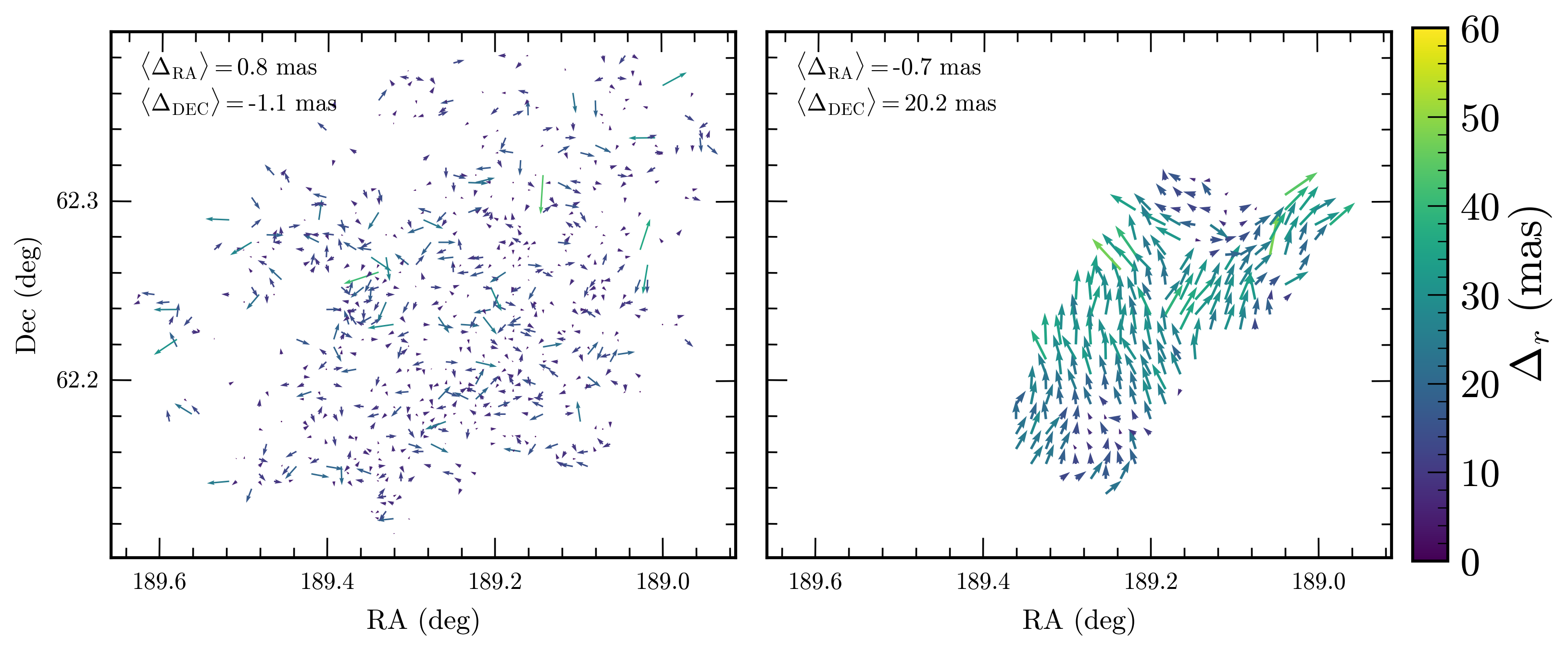}
\caption{Median offsets in 30''$\times30$'' blocks in GOODS-N of the DR5 catalog positions (\brantspaper) based on the mosaics presented in this paper from the JWST-based astrometric reference catalog ({\it left}) and from the previous data release based on HST positions ({\it right}).  We require at least 4 matched objects in each bin.
\label{fig:gn_mosaic_alignment}}
\end{figure*}

Finally, we can compare positions measured from the full mosaics by \brantspaper{} to the astrometric reference catalog.  The results are shown in Figure \ref{fig:gs_mosaic_alignment} for GOODS-S and in Figure \ref{fig:gn_mosaic_alignment}.  These figures also show the offsets from the previous JADES data release catalogs, which used a different alignment procedure based on HST-derived reference catalog.  A notable difference from the previous releases is an average 20 mas shift in declination in GOODS-N.

\subsection{Depth}
\label{sec:depth}
Due to the many different programs included in the full mosaics, varied dither patterns, and artifact masking, the depth of the final combined mosaics is often highly spatially variable. The estimation of aperture photometry uncertainties on a pixel-by-pixel basis, accounting for correlated noise, is described in detail in \brantspaper.
Briefly, low-resolution images of the estimated local uncertainty are constructed from the RMS of fluxes in nearby empty apertures for a variety of aperture sizes and for every band.  These are used to fit a separate power-law model in each band for the uncertainty as a function of aperture size, the combination of subregions as given by a filter-specific bithash image (\S\ref{sec:bithash}), and the propagated single pixel background noise estimate (as given by the \texttt{WHT} image).  This model is then used to compute the estimated uncertainty in a $r=0.1"$ aperture for every pixel in each filter.

\begin{figure*}
\includegraphics[width=\textwidth]{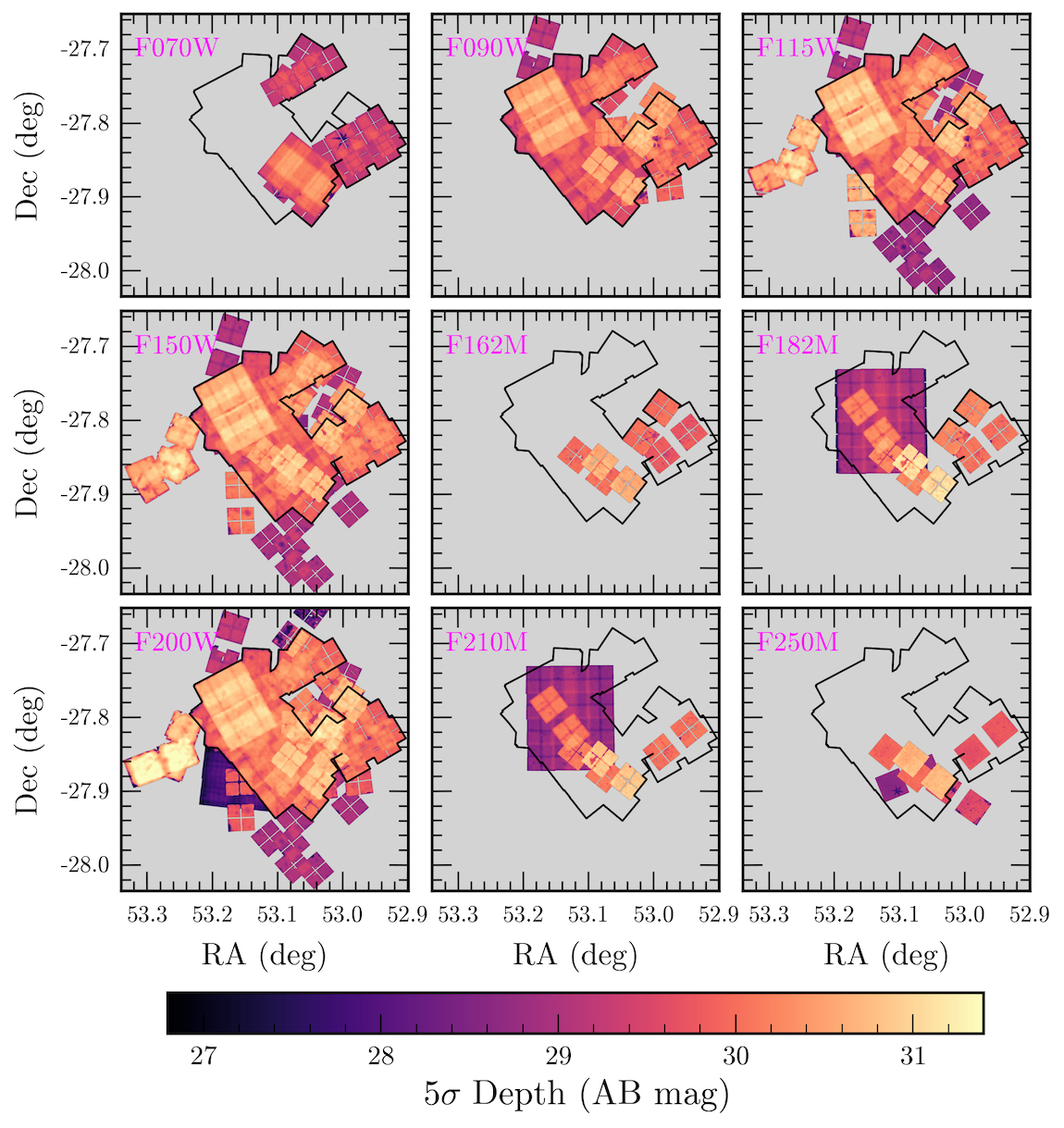}
\caption{Maps of the $5\sigma$ AB magnitude depth in circular apertures $r=0.1''$ for the SW channel filters combined mosaics in GOODS-S. These are corrected for aperture losses appropriate for a point source. The LW filter F250M is also included here.
Thin black lines indicate the footprint of the JADES GTO 8-band imaging.
\label{fig:gs_sw_depthmap}}
\end{figure*}

\begin{figure*}
\includegraphics[width=\textwidth]{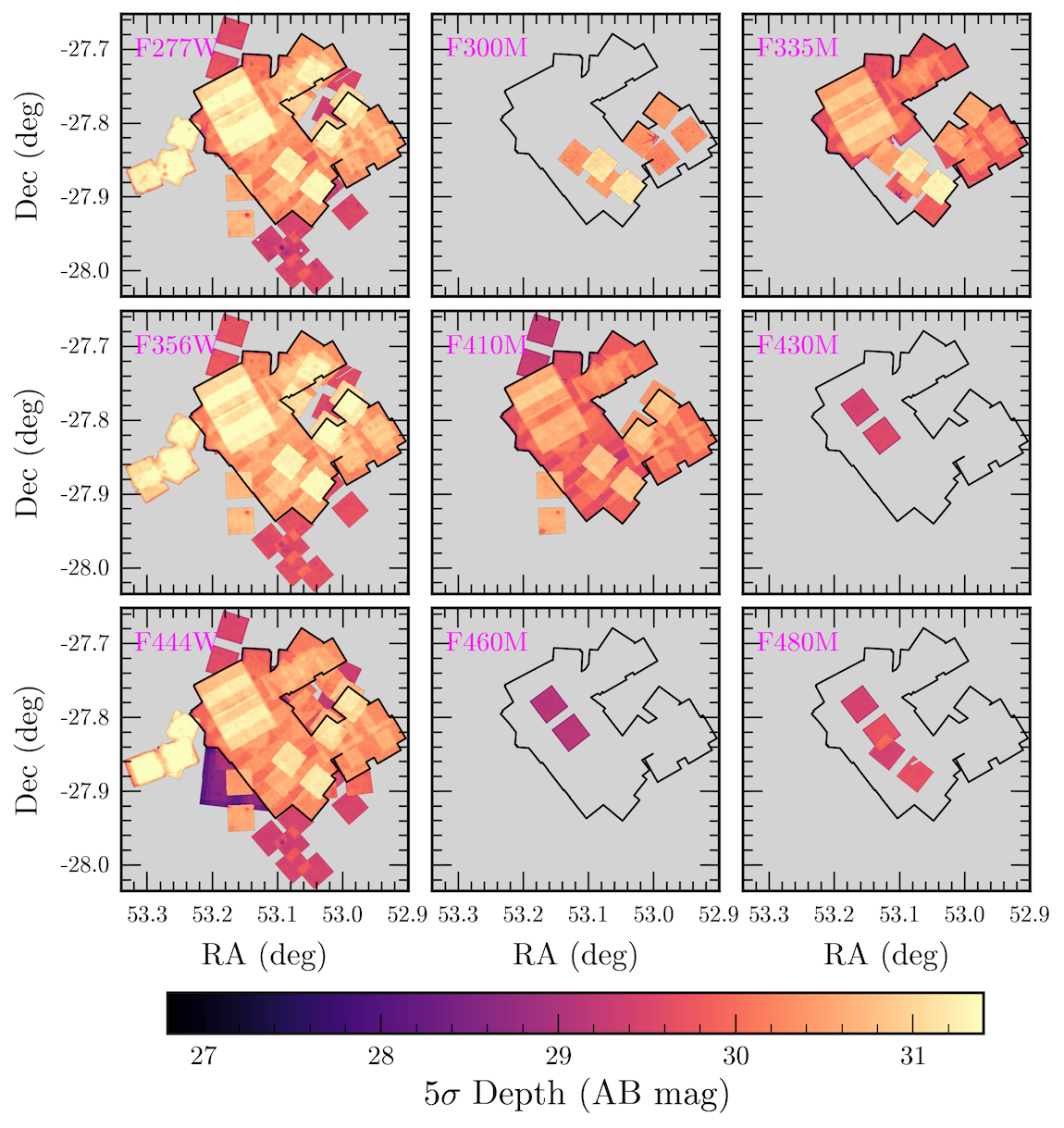}
\caption{Maps of the $5\sigma$ AB magnitude depth in circular apertures $r=0.1''$ for the LW filters combined mosaics in GOODS-S, excluding the F250M filter shown in \ref{fig:gs_sw_depthmap}.
These are corrected for aperture losses appropriate for a point source.
Thin black lines indicate the footprint of the JADES GTO 8-band imaging.
\label{fig:gs_lw_depthmap}}
\end{figure*}

\begin{figure*}
\includegraphics[width=\textwidth]{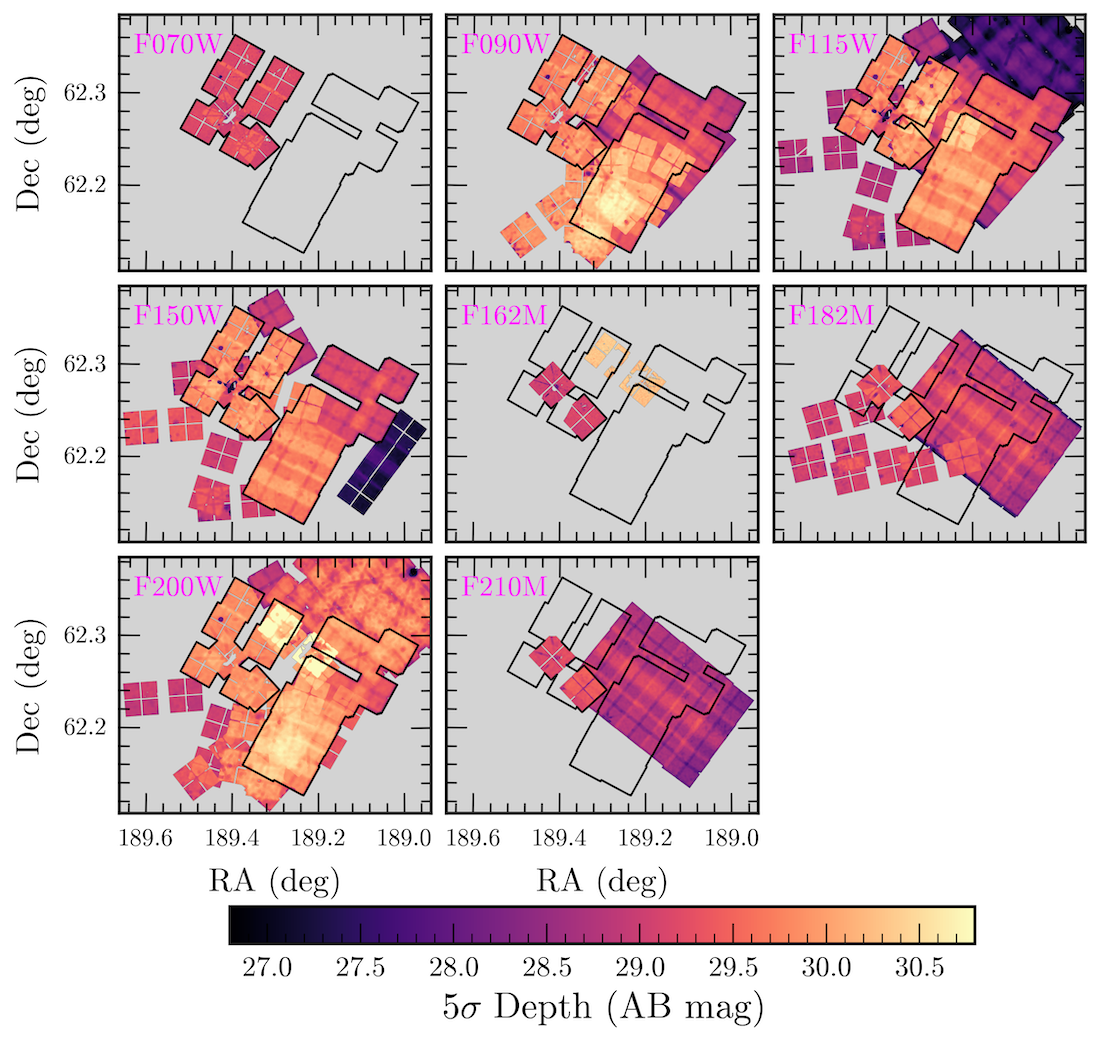}
\caption{Maps of the $5\sigma$ AB magnitude depth in circular apertures $r=0.1''$ for the SW channel filters combined mosaics in GOODS-N.
These are corrected for aperture losses appropriate for a point source.
Thin black lines indicate the footprint of the JADES GTO 8-band imaging.
\label{fig:gn_sw_depthmap}}
\end{figure*}

\begin{figure*}
\includegraphics[width=\textwidth]{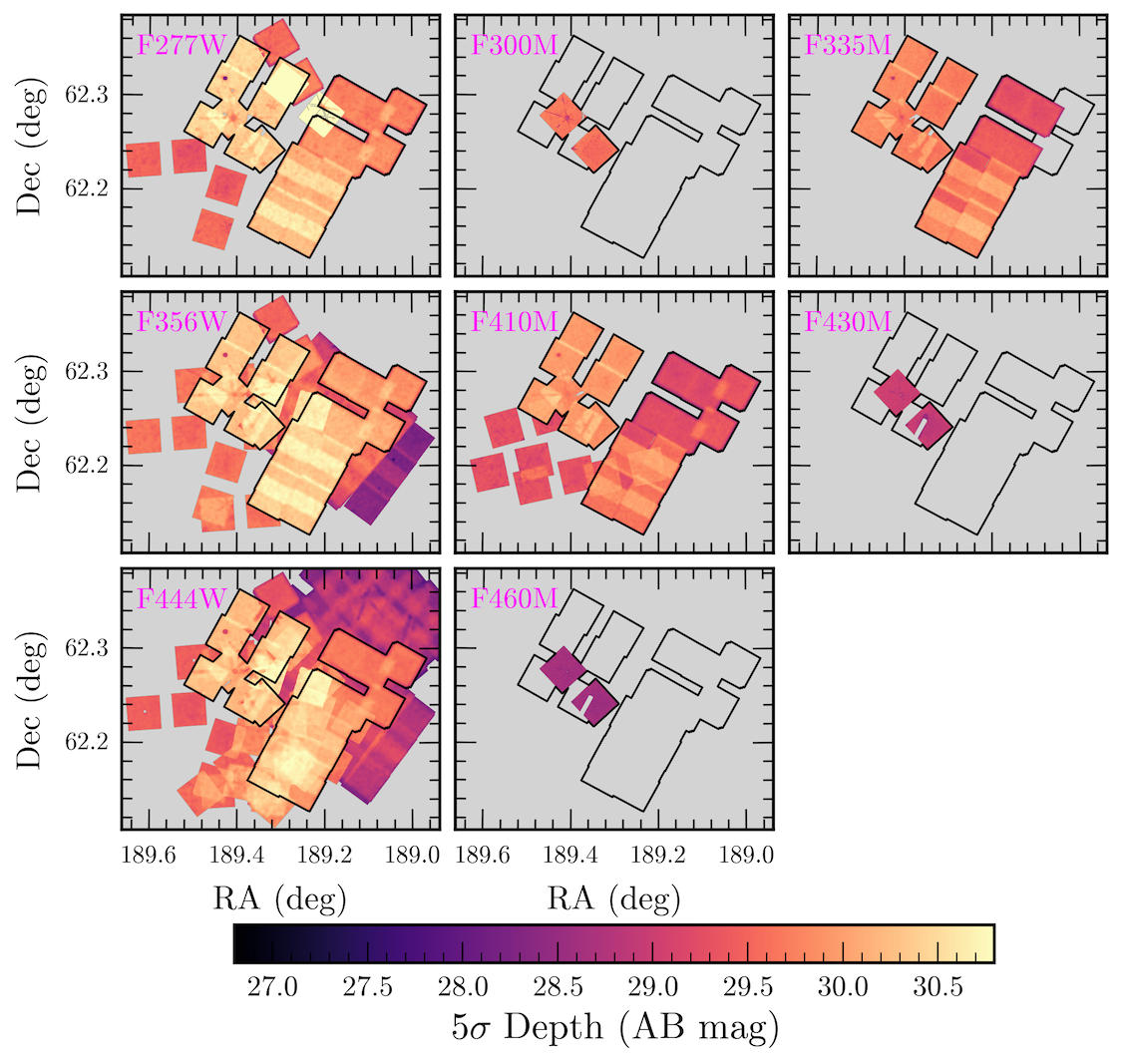}
\caption{Maps of the $5\sigma$ AB magnitude depth in circular apertures $r=0.1''$ for the LW channel filters combined mosaics in GOODS-N.
These are corrected for aperture losses appropriate for a point source.
Thin black lines indicate the footprint of the JADES GTO 8-band imaging.
\label{fig:gn_lw_depthmap}}
\end{figure*}

\begin{figure*}[h]
\includegraphics[width=\textwidth]{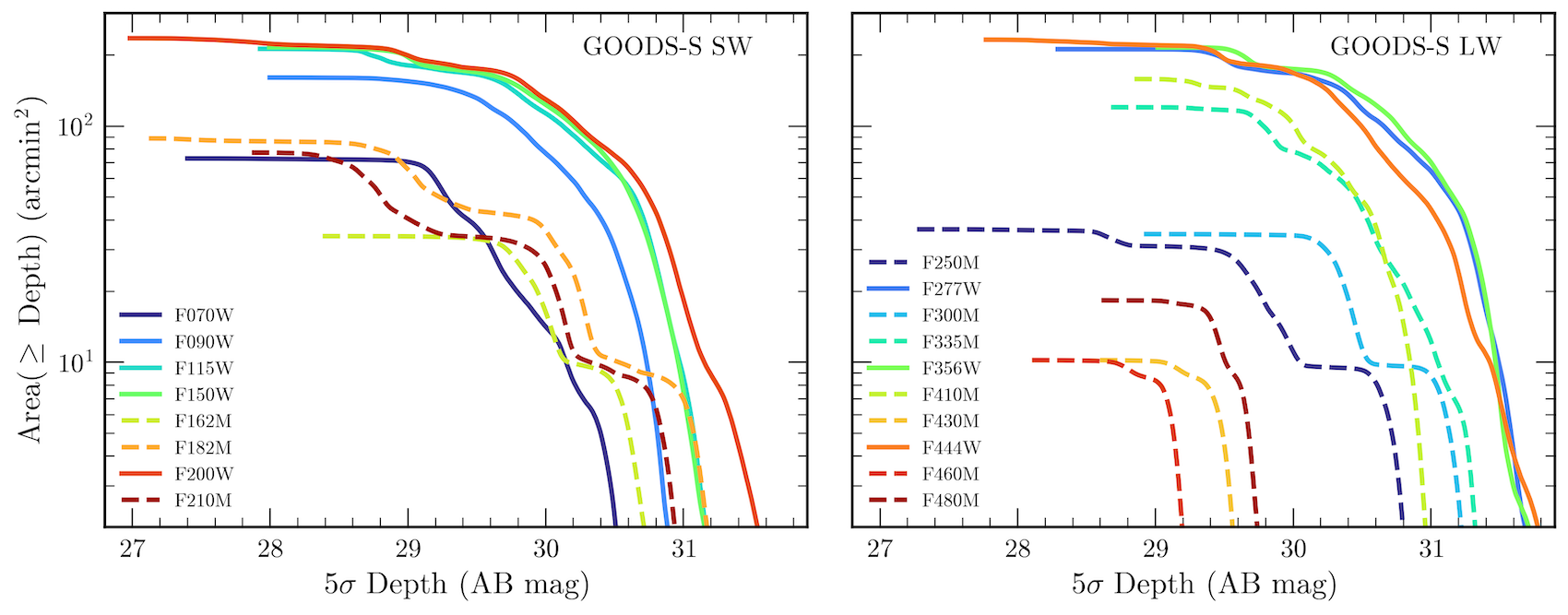}
\caption{Cumulative area in GOODS-S as function of the $5\sigma$ depth in a $r=0.1''$ aperture (corrected for aperture losses appropriate for a point source) for all filters. \label{fig:gs_depth_curves}}
\end{figure*}

\begin{figure*}[h]
\includegraphics[width=\textwidth]{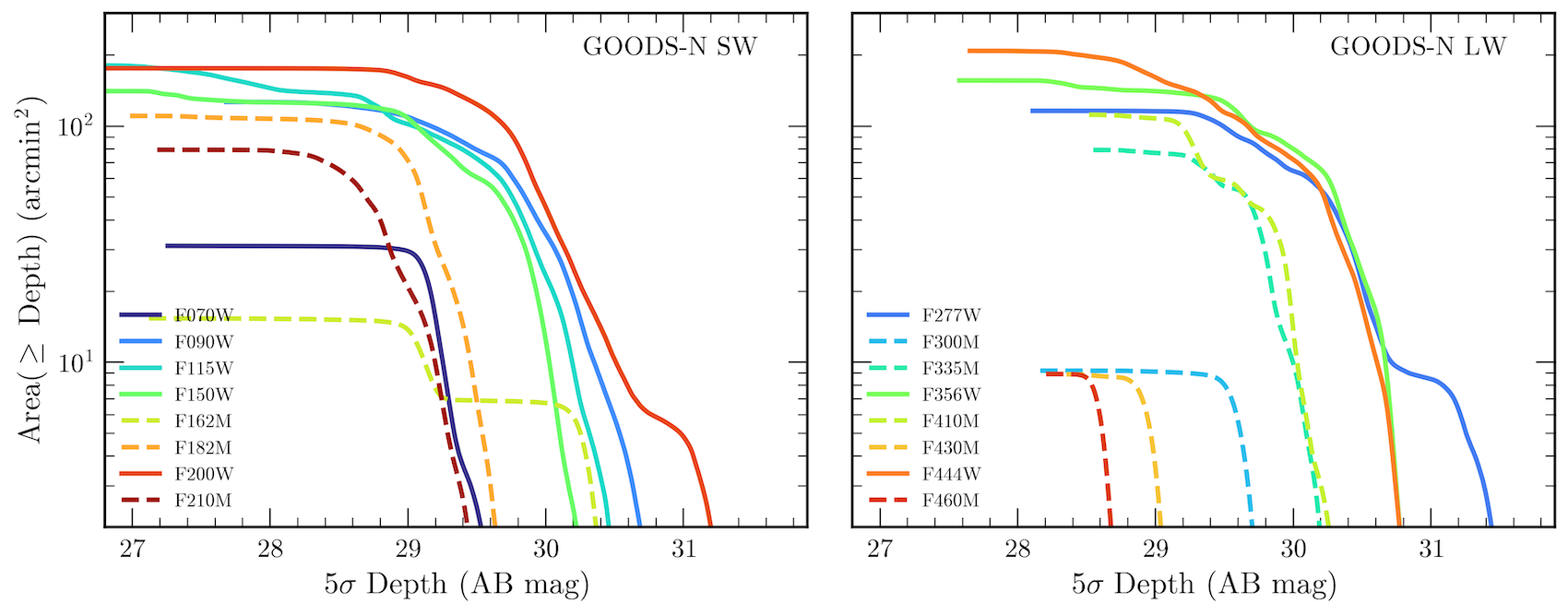}
\caption{Cumulative area in GOODS-N as function of the $5\sigma$ depth in a $r=0.1''$ aperture (corrected for aperture losses appropriate for a point source) for all filters.
\label{fig:gn_depth_curves}}
\end{figure*}

In Figures \ref{fig:gs_sw_depthmap} and \ref{fig:gs_lw_depthmap} we show maps of the $5\sigma$ AB magnitude depth for $r=0.1''$ apertures in the GOODS-S SW and LW channels, respectively.
These magnitudes include an aperture correction appropriate for a point source.
Similar maps are shown for GOODS-N in Figure \ref{fig:gn_sw_depthmap} and Figure \ref{fig:gn_lw_depthmap}.  Finally, in Figures \ref{fig:gs_depth_curves} and \ref{fig:gn_depth_curves} we show curves of the cumulative area imaged in GOODS-S and GOODS-N respectively as a function of depth for each filter.

\subsection{Remaining Issues and Caveats}
\label{sec:remaining}
While the resulting images are generally clean and a clear improvement over past JADES releases, there are some rare flaws of which users should be aware.

\subsubsection{`Hot' Pixels}

Some of the subregions have only 2 dithered exposures, particularly around the edges of the chips, or in areas subject to substantial large scale artifact masking.  In such regions, unmasked hot pixels can leak through the outlier rejection.  As we do not do outlier rejection between subregions, such outliers will print through to the full mosaic, although often these shallow regions are heavily downweighted in the coaddition.  The pixel outliers in the full mosaic are therefore rather rare, but some do exist.  As the \brantspaper\ detection catalog is based on a LW stack, SW hot pixels don't create fake objects.  LW hot pixels are noticeably more compact than the PSF, are typically isolated to a single band, and can be easily spotted in thumbnail images.

\subsubsection{Large-scale artifacts}

Our visual inspection has caught and removed many of the scattered light features, but a few remain.  There also can be mismatches of the background at the edges of chips and the edges of subregions, which will cause features with a straight edge.  False signals can often be diagnosed by comparing F200W to F277W imaging, because false features almost never jump between the SW and LW arms.  When an area is covered by multiple subregions, this also is a powerful diagnostic.

We stress that not everything strangely extended is false.  We have sometimes gone to inspect suspected scattered light only to find a real galaxy tidal feature, found in multiple subregions and in both SW and LW.  These images routinely display extended tidal shells and tails.

We have tried to subtract or mask the residual large-scale persistence but some patches remain, usually near the edge of the A3 or B4 chips in F090W.  One should therefore be suspicious of low-surface brightness streaks in F090W that don't show up in redder filters.

There can be cosmetic issues when the depth changes rapidly, such as when a small patch of a deep subregion has been masked, leaving only a shallow backing subregion.

\subsubsection{`Weave'}
Compared to past data releases, we have substantially reduced the patterns induced by correlated meso-scale structure in the LW flat fields by using our new sky-flats.  However some structure remains, particularly in the medium bands of the \texttt{jw032150} and \texttt{jw012870} subregions taken in parallel to deep NIRSpec observations.  While this is below the level that would create fake objects in the \brantspaper\ catalog, in some regions and bands we haven't quite reached the photon-limited statistical error on meso-scales.
We estimate the amplitude (peak-to-valley) of the effect on scales of 0.3'' to be $\lesssim2$\% of the background surface brightness in the F250M, F300M, and F335M bands for the affected subregions (i.e., $\lesssim \pm 0.001 {\rm MJy}/{\rm sr}$).  The amplitude is much less in the more common wide and medium bands.

\subsubsection{Background Oversubtraction}
Our algorithms are tuned to perform well on smaller, fainter objects.  Large bright galaxies (or occasionally sets of large galaxies) do drive oversubtraction of the background, which will in turn bias the photometry of these galaxies.  This oversubtraction happens both because of the $1/f$ corrections and because of the meso-scale background subtraction at the exposure and subregion mosaic level.  In the SW channel, each amplifier of the SCA is only about 16$''$ wide, so galaxies approaching this size will cause oversubtraction.

Oversubtraction is visible as negative flux regions around bright galaxies in the single-band images.  Galaxies need to be larger than 5-8$''$ to display this. Galaxies with sizes $\gtrsim 2''$ may have their outskirts subsumed into the background subtraction (but not exhibit negative halos). In cases where the galaxy spans a SW amplifier boundary, additional artifacts associated with the boundary can sometimes be seen.  We stress that large galaxies are rare in the images; the vast fraction of the mosaic does not show this effect, despite the very low noise level.

\subsubsection{Variability within multi-epoch stacks}

While the multi-year extent of this data set is an opportunity for time-domain astronomy, it also means that that there is a small chance that variability can affect the reported spectral energy distributions of objects in the stacked catalogs.  Most obviously, if an object varies in flux, those filters observed in one year will be reported as having different flux than those observed in another year.  Further, some filters may have been observed in multiple years and others not, leading to a similar issue.  To give a particularly perverse example, object 386437 at position 53.1065979 --27.7425880 was a supernova that occurred in cycle 1, but fell in the intra-module SW chip gap so only had LW data.  The area was observed in both SW and LW in cycle 2, after the supernova had faded, so the resulting object presents in the stacked catalog as a compelling F200W dropout, with flux only in the LW bands.  We conclude that variability hypotheses should be considered \citep{decoursey2025} when diagnosing very unusual spectral energy distributions (SEDs).

\subsubsection{Astrometric uncertainties in isolated regions}

The astrometric solution is based on overlaps of multiple pointings.  There are a few modules in the extreme edges of the full field that lack that overlap and therefore have less certain astrometry, based on a few Gaia stars, priors from the HST mosaic, or our model of the inter-module astrometric separation combined with the tie-down of the other module.  We have less confidence in the astrometric precision in these stand-alone regions.

\subsubsection{Diffraction spikes}

We have masked diffraction spikes from the full mosaic only when multiple position angles in a filter allowed an uncontaminated view.  However, this decision is on a filter-by-filter basis, so one can have cases where a diffraction spike appears only in some filters, or when a spike is masked in some region and then reappears further away.  Of course, these long linear features point toward bright stars (sometimes off the field).  Science applications should be careful to consider the impact of the remaining diffraction spikes.

Because of the variations in position angles, alternative handling of the diffraction spikes would be best done in the subregion mosaics, before the coaddition.

\section{Summary \& Conclusions}\label{sec:conclusions}

We have presented the NIRCam imaging for the fifth data release from the JADES collaboration, incorporating 1253 hours of NIRCam imaging in GOODS-S and GOODS-N.  This includes the full NIRCam imaging from the JADES GTO project, but also substantial imaging from 19 other programs, five of which were affiliated with JADES.

The resulting mosaics present some of the deepest near-infrared images yet obtained, utilizing a pipeline built from several years of preparation and then engagement with on-orbit data.  The union of the mosaics cover 469 arcmin$^2$ with at least one filter and 250 arcmin$^2$ with at least eight.  The data have been carefully vetted, and we have carefully addressed many issues diagnosed in our three earlier data releases.  The paper has described many of the algorithms now in use and presents our key validation test.  We describe custom steps in our reduction, including crosstalk correction, $1/f$ noise fitting and removal, wisp and persistence template fitting and removal
(but see \citealt{Wu26} for details)
, custom LW sky-flats, and a mosaic outlier rejection step designed to preserve the cores of compact, unsaturated sources.

The JADES program is the largest program yet conducted by JWST, and JADES itself contributes 578 hours of SW imaging to these mosaics.  Affiliated programs contribute another 211 hours.  But importantly another 464 hours come from other programs, testament to the great interest in these fields and the substantial contributions that many groups have made.  Of course, JWST is just a recent observer of the GOODS fields, and we hope that these comprehensive mosaics provide a compelling opportunity to leverage the years of multi-wavelength imaging and spectroscopy focused on these two fields.

\bigskip
This research was funded through the JWST/NIRCam contract to the University of Arizona (NAS5-02105), and JWST program 3215. DJE is supported as a Simons Investigator and by JWST/NIRCam contract to the University of Arizona. Support for program 3215 was provided by NASA through a grant from the Space Telescope Science Institute, which is operated by the Association of Universities for Research in Astronomy, Inc., under NASA contract NAS 5-03127. ST acknowledges support by the Royal Society Research Grant G125142. SA acknowledges grant PID2021-127718NB-I00 funded by the Spanish Ministry of Science and Innovation/State Agency of Research (MICIN/AEI/ 10.13039/501100011033). WMB gratefully acknowledges support from DARK via the DARK fellowship. This work was supported by a research grant (VIL54489) from VILLUM FONDEN. AJB and AJC acknowledge funding from the ``FirstGalaxies" Advanced Grant from the European Research Council (ERC) under the European Union’s Horizon 2020 research and innovation programme (Grant agreement No. 789056). AJC gratefully acknowledges support from the Cosmic Dawn Center through the DAWN Fellowship. The Cosmic Dawn Center (DAWN) is funded by the Danish National Research Foundation under grant No. 140. ECL acknowledges support of an STFC Webb Fellowship (ST/W001438/1). Funding for this research was provided by the Johns Hopkins University, Institute for Data Intensive Engineering and Science (IDIES). RM acknowledges support by the Science and Technology Facilities Council (STFC), by the ERC through Advanced Grant 695671 ``QUENCH'', and by the UKRI Frontier Research grant RISEandFALL. RM also acknowledges funding from a research professorship from the Royal Society. PGP-G acknowledges support from grant PID2022-139567NB-I00 funded by Spanish Ministerio de Ciencia e Innovaci\'on MCIN/AEI/10.13039/501100011033, FEDER, UE. JAAT acknowledges support from the Simons Foundation and JWST program 3215. Views and opinions expressed are however those of the authors only and do not necessarily reflect those of the European Union or the European Research Council Executive Agency. Neither the European Union nor the granting authority can be held responsible for them. The research of CCW is supported by NOIRLab, which is managed by the Association of Universities for Research in Astronomy (AURA) under a cooperative agreement with the National Science Foundation.

This work is based on observations made with the NASA/ESA/CSA James Webb Space Telescope. The data were obtained from the Mikulski Archive for Space Telescopes at the Space Telescope Science Institute, which is operated by the Association of Universities for Research in Astronomy, Inc., under NASA contract NAS 5-03127 for JWST. JADES DR5 includes NIRCam data from JWST programs 1176, 1180, 1181, 1210, 1264, 1283, 1286, 1287, 1895, 1963, 2079, 2198, 2514, 2516, 2674, 3215, 3577, 3990, 4540, 4762, 5398, 5997, 6434, 6511, and 6541. The authors acknowledge the teams of programs 1895, 1963, 2079, 2514, 3215, 3577, 3990, 6434, and 6541 for developing their observing program with a zero-exclusive-access period. The authors acknowledge use of the {\it lux} supercomputer at UC Santa Cruz, funded by NSF MRI grant AST 1828315. This research made use of \texttt{photutils}, an \texttt{astropy} package for detection and photometry of astronomical sources \citep{bradley2025a}.

\vspace{5mm}
\facilities{JWST, HST}

\software{JWST Calibration Pipeline, \texttt{STPSF}, \citep{perrin25}, \texttt{FitsMap} \citep{hausen2022b}, \texttt{astropy} \citep{2013A&A...558A..33A, 2018AJ....156..123A}, \texttt{photutils} \citep{bradley2025a}, \texttt{SourceExtractor} \citep{SourceExtractor}, \texttt{sep} \citep{barbary2016}}

\appendix

\section{Data Model of the Mosaic Image Files}
\label{app:data_release}

We have produced separate multi-extension FITS files for each of the available NIRCam filters for the GOODS-S and GOODS-N regions. These files are the result of all reduction steps described in Section~\ref{sec:pipeline}. They are available and documented at MAST via \dataset[10.17909/8tdj-8n28]{\doi{10.17909/8tdj-8n28}}. The mosaics have five image extensions, which we summarize here.

\begin{itemize}
    \item \texttt{SCI}: The resampled pixel values, in units of MJy/sr.

    \item \texttt{ERR}: The resampled uncertainty estimates, same units as \texttt{SCI}.  These uncertainty estimates include read noise, background and source Poisson noise, and propagated uncertainties from the flat fields.  They do not include correlated pixel uncertainties induced by the resampling.

    \item \texttt{WHT}: Estimate of the pixel uncertainty excluding source Poisson noise or correlated noise, suitable for relative weighting of each pixel in subsequent coaddition.

    \item \texttt{EXP}: The sum of the exposure time of every pixel that contributed to the mosaic pixel, not accounting for ramps shortened due to detected jumps.

    \item \texttt{NIM}: The number of individual exposures that contributed to this pixel.
\end{itemize}

Key header information includes the \texttt{FILTER} and WCS keywords.
The data model of the subregion mosaics is identical.
In the subregion mosaics \texttt{NIM} values of $-2$ and $-4$ are used to denote areas affected by diffraction spikes or artifacts, respectively.
There are additional header keywords in the subregion mosaics that can be used for relative weighting of the mosaics, that indicate the mean epoch and PA, and for LW that identify the module/SCA of the subregion mosaic.
The data volume of the subregion mosaics totals 2 TB.

The ``bithash'' image (\S\ref{sec:bithash}) has the following structure:
\begin{itemize}
    \item \texttt{PRIMARY}: Empty
    \item \texttt{BITHASH}: An array of same size as the full mosaics giving 32-bit integers that encode the programs that contributed data in any band to that pixel.
    \item \texttt{DECODER}: A FITS binary table giving the mapping between subregion name and bit.
\end{itemize}

\section{Astrometric Reference Catalog}\label{sec:refcat}

The data set in the GOODS fields is now sufficiently large, with overlapping pointings of varying position angle, to offer a very effective route to solve for the astrometry of interlocking observations.  We use this to construct a NIRCam-based Gaia-registered reference catalog of objects for the registration of the individual visits.

As described in \S\ref{sec:astrometry},
we start with an initial processing of the visits
to find well-detected compact objects and derivve initial astrometric solutions that match
the objects in each visit to external catalogs from HST.  For the remaining work, we limit ourselves to detections 
between 70 and 5000 nJy,
signal-to-noise ratio of at least 20, and compact morphology smaller than 5 pixels; this typically yields around 
100 objects per module.  We project spherical coordinates to a single tangent plane centered on the middle of
the GOODS field; at present, the data set is compact enough (about 10$'$
radius) that the radial distortion of the tangent plane is below our tolerances.
We then  
find matches between these detections in different exposures to define 
unique objects, using a friends-of-friends algorithm with 0.1$''$ linking.

We will solve for a refined astrometric solution by introducing parameters for the translation and rotation
for each module in each exposure as well as the unknown true position
of each unique object.  We optimize these parameters using a loss function that is a softened least
squares of the residual of the detected position relative to the true 
position.  In detail, we use $L = c^2/(1+c^2/s^2)$ where $c^2 = [(\Delta x)^2 + (\Delta y)^2]/\sigma^2$, for relative positions $\Delta x$ and $\Delta y$, a tolerance $\sigma$, and a softening $s$.
We use $s=4$, while $\sigma$ starts at 15 mas and then drops to 7.5 mas
after some burn-in.  The softening implies that outliers are capped at a 
$4\sigma$ penalty, while still allowing a non-zero derivative for the 
optimizer.

We then tie this to the Gaia reference frame by adding another
``exposure'' that is simply the Gaia catalog; this exposure is
allowed no freedom of rotation or translation.  Astrometry of the
Gaia stars in the NIRCam exposures was done by a separate PSF-fitting
code.  These stars are typically saturated in their cores, but the
diffraction spikes give very accurate positions.  We observe only
1-2 mas of variation in the measured position from exposure to
exposure in a dither sequence.  Proper motions of the Gaia layer
are corrected to the mean epoch of the exposures, and the proper
motion relative to that mean epoch is applied to the measured
position in each exposure, so that the multiple detections of each
Gaia star are on the same epoch.  Gaia stars are treated as other 
objects in the loss function, save that the loss penalty is multiplied
by a boost that starts at 9 and increases by a factor of 16 as the 
solution converges.

We further augment the loss function to favor a consistent separation
and rotation between the two modules.  In each module, we pass three 
reference pixel locations through the WCS solution and use this triplet
of positions to define the relative 2-d separation and rotation between
the pair of modules.  These three parameters are defined separately 
for each filter, as the small optical wedge in the filters causes the
astrometric separation of the modules to vary.  We add these parameters
to the fit, and adopt a quadratic loss with an 8 mas error.  Here, the
rotation angle is converted to a distance at 1 arcminute radius, roughly
the size of the module.

This interlocking of the two modules is useful in two major ways.  First,
it allows observations of varying position angle to lock down the
mosaic.  For example, the JADES Prime mosaic was observed at position angles
from 298$^\circ$ to 321$^\circ$, while FRESCO was observed at 0$^\circ$;
the interlocking rectangles provides a strong fabric.  Second, it assures
that outlying pointings where one module is not overlapping other data
can benefit from the tight registration of the module that is overlapping
the rest of the mosaic, as the solution will pull to the intermodule prior
unless overruled by a Gaia star.

We do some cleaning of the input detection lists.  For example, only the
brightest source within 2$''$ is retained, as we want to avoid percolation
between close detections in the input list.  We also limit the Gaia stars
to $G<19.8$ and keep only stars with measured proper motions less than 
0.5 mas/year.

In GOODS-S, we use 1398 exposures, mostly in F200W but using F210M or F115W if
that is not available for a visit.  From over 200,000 input detections,
we find 11,436 unique objects with multiple detections, connecting to about 147,000 individual detections.  These are augmented by 477 detections of 70 Gaia stars.

We then optimize using optax, solving for about 27,000 parameters:
the 2-d positions of the unique objects, the 3 astrometric
parameters per exposure, and the 3 intermodule parameters per filter.
We start with a burn-in, then mark as inactive any detections that 
have residuals more than 30 mas.  This affects only about 3\%  of
detections.  We then restart from the original positions and repeat
the burn-in.  After this, we halve the $\sigma$ in the loss function,
run further, and then increase the Gaia boost factor, and run further.
The optimization appears to be very well converged.

We have three main checks on the astrometric performance.  The first
is the scatter between positions for a single object, where we take the
median position within each dither sequence and then compare between 
observations/visits.  This appears to saturate at around 5 mas of 2-d
scatter, for brighter objects.
\begin{figure}
    \noindent
    \includegraphics[width=4in]{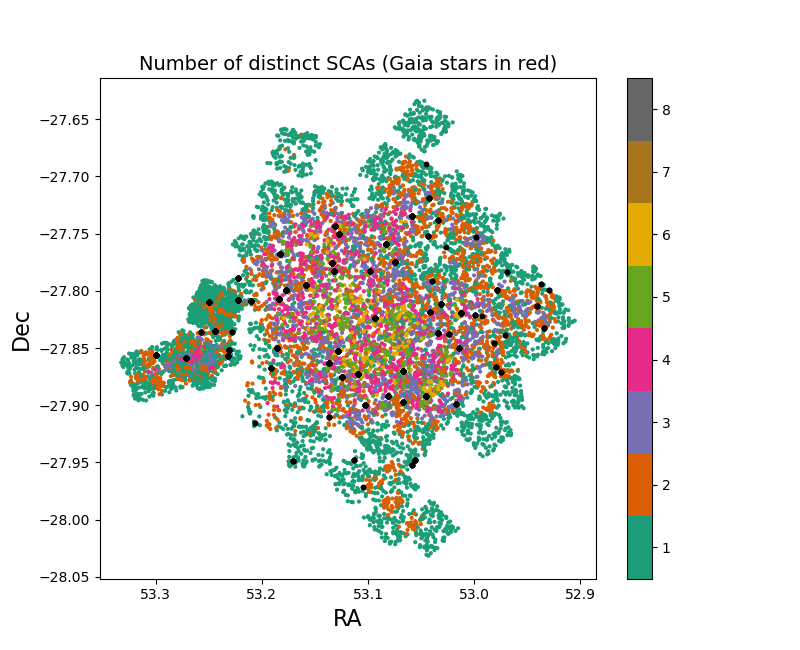}\hspace{-24pt}
    \includegraphics[width=4in]{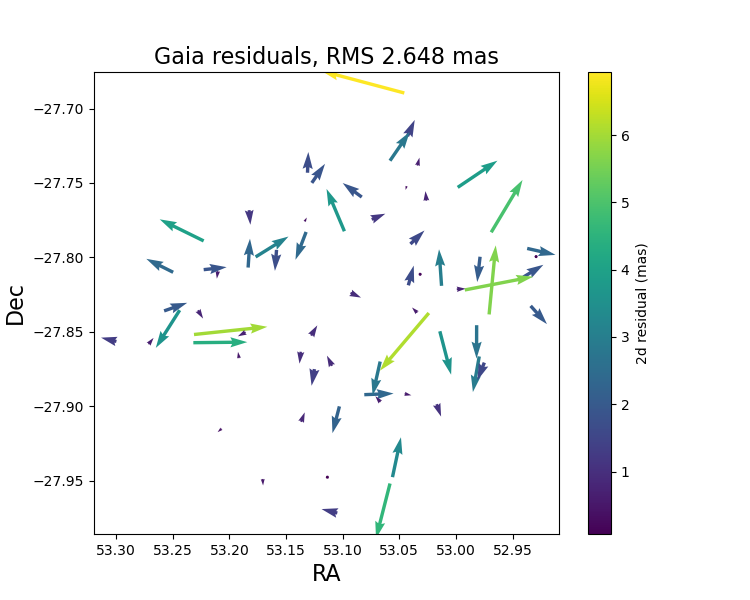}
    \caption{\label{fig:gaia}%
    (left) The locations of the unique objects in the GOODS-S field colored by the number of distinct SW SCAs on which they are detected.  This is a good proxy for how much interlocking leverage is present, as this indicates that the visits are well offset from one another.  In the center of the field, the interlocking is very dense, with a couple hundred stars falling on 6 separate SCAs and a handful on all 8.  The Gaia stars used in the optimization are shown in black.
    (right) The astrometric residuals of the Gaia stars, comparing the final median position in the NIRCam detections to the Gaia catalog, after correcting proper motion to a common epoch.  The color is the vector norm in milliarcseconds.  The rms residuals over the field is only 2.6 mas.  }
\end{figure}

\begin{figure}
    \noindent
    \includegraphics[width=4in]{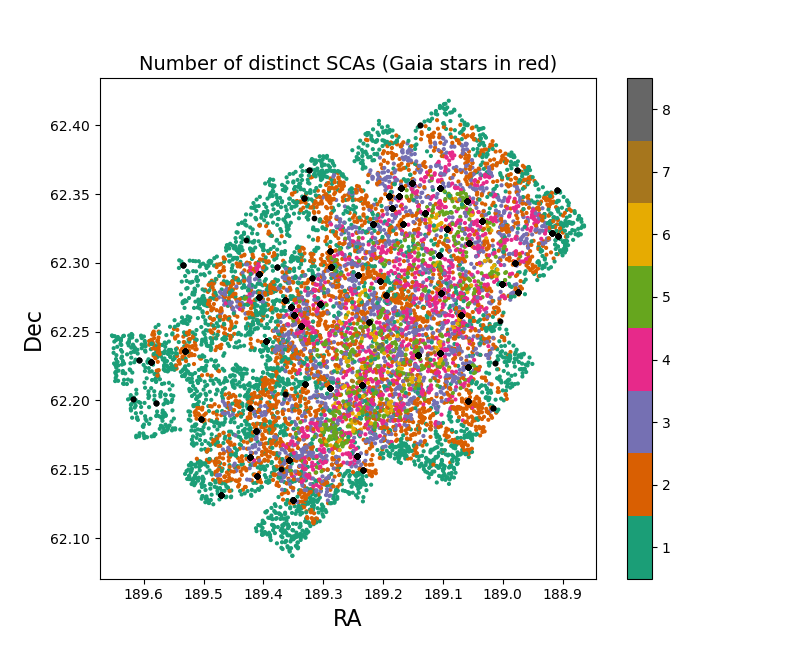}\hspace{-24pt}
    \includegraphics[width=4in]{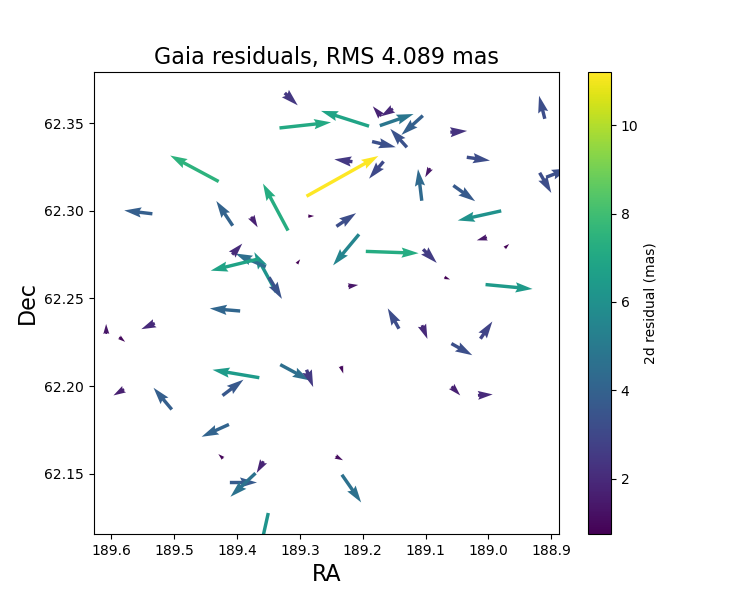}
    \caption{\label{fig:gaia_GN}%
    As Fig.~\ref{fig:gaia}, but for GOODS-N.}
\end{figure}

The second check is the offset of the Gaia stars from their catalog
location.  This is 2.6 mas of 2d residual and notably, it is tighter 
in the middle of the mosaic where we have the most cross-linking between
exposures.  The total variance is clearly driven by the northwest 
portion of the mosaic as well as the eastern region, where we have only
limited overlap of pointings.
There is a mild hint that the northwest edge of the mosaic
has a coherent distortion, while the eastern region looks more random.

In practice, we find that the intermodule separations of the final fitted
location of the modules have a scatter of 3-5 mas, depending on filter, 
with rotational scatter of 1-2 mas.  This means that the interlocking 
solution has constrained the separations mildly better than our prior 
of 8 mas.  We experimented with tighter priors, locking the modules 
together to 1 mas, and this did create substantially larger residuals 
in our Gaia fits.  In particular, this strongly increased the coherent
residuals in the northwest edge, as if the intermodule separation favored 
by the bulk of the interlocked exposures at position angles from 298 to 360
degrees were systematically different than those needed to fit the positions
in northwest region that was created by parallel observations at position
angles closer to 50 degrees, i.e., several months later.  We do not
mean that the effect is due to the change in position angle, but
it seems to correlate with this period of time.  This surely needs more
study; our main conclusion is that the inter-module separation seemingly
varies at the 5 mas level.

The third check is the residual distortion maps.  Here we consider only
those objects that have detections on multiple shortwave detectors, requiring 
a substantial change in pointing, and compute the median of the residuals
of detections falling a given portion of the detector.  We produce a 16 by 16
grid within each detector.  These distortion maps show a variation of no worse
than 2.5 mas rms (2d) within each detector.  In detail, we estimate that this
is due to a 1 mas per dimension random noise and a small remaining affine 
distortion.  Chips A2 and A3 have the largest coherent trends, but even
these are only 5 mas edge-to-edge.  We have not attempted to apply these
residual distortions, but they likely indicate that further improvements
in the distortion maps are possible.  At this point, we regard the smallness
of the coherent residuals in the stack as an indication that process of 
trusting the input distortions and then interlocking the modules with only
translations has validated the initial trust.

It is difficult to be highly quantitative about the accuracy of the astrometric
solution, because we are putting a high loss penalty on the Gaia stars, which 
of course pulls the solution to match them.  Before the penalty is increased 
by a factor of 16, the Gaia 2d residuals are 4.4 mas, but again this is 
dominated by the northwest edge and eastern extension, with the center of the
mosaic much quieter.  We suspect that the accuracy is at or below
3 mas in the center, with some indication of a coherent error of a
few mas in the northwest.  But we have not yet conducted trials
omitting some Gaia stars to use as tests of accuracy.

\section{Sky Flats}
\label{sec:sky-flats}
The NIRCam detectors have substantial spatial structure in the pixel-to-pixel sensitivity variations on a variety of scales, which must be accurately calibrated. Errors in the flat fields can introduce artifacts in the images (on the scale of the flat field error times the background level) and affect the precision and accuracy of measured fluxes, particularly for very faint sources in deep images. Earlier JADES data releases \citep{rieke23r, eisenstein2025} made use of sky flats generated from JADES imaging as well as other public datasets in order to mitigate issues with the LW ground flats.  Subsequent investigation  revealed substantial background structure (on the 10-20 pixel scale, $\sim 0.3$'') for deep data taken with certain dither patterns, especially those patterns optimized for deep NIRSpec observations taken as primary. This was traced to remaining spatially correlated errors in the flat fields, which combined with the dither pattern created structure in the mosaics at the level of $\sim 2$-$5$\% of the background spanning the entire detector.  This pattern was most prominent in the medium bands of the 3215 program, but also present in the wide and medium bands of the 1287 program, and persisted when using the sky-flats delivered to CRDS as part of \texttt{jwst\_0956.pmap}.

We therefore reconstructed LW sky-flats using a larger dataset available at that time.   These include the JADES data in both GOODS-N and GOODS-S and images from many of the additional programs in these fields described above and obtained before Jun 2024, as well as images from program 1345 \citep[The Cosmic Evolution Early Release Science Survey (CEERS), PI:Finkelstein][]{finkelstein2023}.  We favored these data as they are of relatively sparse fields lacking large-scale diffuse structure.  This simplifies the source detection and masking.  They are also well matched to the typical exposure times and readout patterns of JADES, and are not read-noise dominated.  All of these images were processed through the same version of the stage 1 pipeline described in \S\ref{sec:stage1}. To construct the sky flats we first created source masks for every exposure.  When available (e.g., for GOODS-S and GOODS-N), these source masks were derived from segmentation maps constructed with combined LW mosaics from a previous JADES data version. For other fields we produced subregion-level stacks of all LW data available, and created segmentation maps from these stacks which were then re-projected to the individual exposures.  In all cases the segmentation maps were dilated to mitigate flux from the outskirts of the detected galaxies. We also masked any measured pixel value with any DQ bit set in the rate images (including jumps) an applied any artifact masks (\S\ref{sec:exposure-masking}). We then divided each count-rate image by the median value in the remaining pixels of that image, and constructed the sigma-clipped median value across rate images for each pixel.

We estimated the uncertainty of the flat field value for each pixel from the scatter across background-normalized rate images.  We also constructed flat fields with parts of the data to investigate temporal variability. Pixels that were abnormally sensitive or insensitive were marked as invalid, as well as those with abnormally large scatter, or which were masked in a substantial fraction of the rate images.  The marked pixels comprise $\sim2.4$\% and $\sim 1.7$\% of pixels in the A and B modules, respectively.

In some medium bands the aggressive masking led to insufficient data to reliably estimate the sensitivity of every pixel.  However, comparison of the wide band sky-flats to each other and to the medium band sky-flats for the pixels where there were enough data suggested that for most medium bands there was little difference (slope of 1.00 and approximately 1\% scatter).  We therefore substituted the nearest wavelength wide band flat field for the flat fields in F250M, F300M, and F335M as well as the B module flat fields F430MB, F460MB, and F480MB, adding an additional 1\% in quadrature to the flat field uncertainty. For F410M there were sufficient data to directly construct a sky-flat, while for A module the redder medium bands (F430MA, F460MA, and F480MA) displayed substantial large scale structure in the ratio to the F444WA sky flat.
For these filter/module combinations we fit a 2-dimensional polynomial to the ratio of the medium band to F444WA sky-flat in valid pixels, and then multiplied these polynomials by the full F444WA sky flat to produce the medium band sky-flats. Reductions of the deep, single-pointing, NIRSpec parallel data with the resulting flat fields resulted in a substantial reduction of the residual background pattern.

The origin of the remaining differences between flat fields on small scales is unclear.  While the amplitude of the differences is small ($\sim 1$\% RMS) there is clear spatial structure in the ratios.
This may indicate that further improvements in the wide-band flat field determination is possible, or that the use of wide-band sky-flats for the rare medium bands introduces structured errors.
The origin of the large scale gradients of the F430M, F460M, and F480M A module sky-flats with respect to F444W sky-flats are similarly unclear.
We used data from the calibration program 1476 (PI: Boyer) which took multiple images of an LMC field shifted across the detectors to measure the response of the detector to point sources independently of the sky-flats, but there were not enough independent measurements to draw strong conclusions.

\bibliography{ms}

\end{document}